\documentclass{emulateapj}
\usepackage[optb]{optional}

\def\lsim{\mathrel{\rlap{\lower 4pt \hbox{\hskip 1pt $\sim$}}\raise 1pt\hbox {$<$}}}
\def\gsim{\mathrel{\rlap{\lower 4pt \hbox{\hskip 1pt $\sim$}}\raise 1pt\hbox {$>$}}}

\shorttitle{Standing accretion shock instability and neutrino-driven explosions}
\shortauthors{Marek and Janka}

\begin{document}

\title{Delayed neutrino-driven supernova explosions aided by the standing
       accretion-shock instability}

\author{A.\ Marek\altaffilmark{1}
        and H.-Th.\ Janka\altaffilmark{1}}
\email{amarek@mpa-garching.mpg.de}
\email{thj@mpa-garching.mpg.de}
\altaffiltext{1}{Max-Planck-Institut f\"ur Astrophysik,
       Karl-Schwarzschild-Str.~1, D-85748 Garching, Germany}

\begin{abstract}
  
  We present two-dimensional hydrodynamic simulations of 
  stellar core collapse and develop the framework for a detailed 
  analysis of the energetic aspects of neutrino-powered supernova 
  explosions. Our results confirm that the neutrino-heating
  mechanism remains a viable explanation of the explosion of a wider 
  mass range of supernova progenitors with iron cores, but the explosion
  sets in later and develops differently than thought so far. The
  calculations were performed with an energy-dependent treatment of the
  neutrino transport based on the ``ray-by-ray plus'' approximation,
  in which the neutrino number, energy, and momentum equations are
  closed with a variable Eddington factor obtained by iteratively
  solving a model Boltzmann equation. We focus here on the evolution
  of a 15$\,M_\odot$ progenitor and provide evidence that shock revival
  and an explosion are initiated at about 600$\,$ms after core bounce, 
  powered by neutrino energy deposition. This is significantly later
  than previously found for an 11.2$\,M_\odot$ star, for which we also
  present a continuation of the explosion model published by Buras et al.
  The onset of the blast is fostered in both cases
  by the standing accretion shock instability
  (SASI). This instability exhibits highest growth rates for the 
  dipole and quadrupole modes, which lead to large-amplitude bipolar 
  shock oscillations and push the shock to larger radii, thus
  increasing the time accreted matter is exposed to neutrino heating
  in the gain layer. As a consequence, also convective overturn behind 
  the shock is strengthened, which otherwise is suppressed or damped 
  because of the small shock stagnation radius. When the explosion
  sets in, the shock reveals a pronounced global deformation with a 
  dominant dipolar component. In both the 11.2$\,M_\odot$ and 
  15$\,M_\odot$ explosions long-lasting equatorial downflows supply 
  the gain layer with fresh gas, of which a sizable fraction is 
  heated by neutrinos and leads to the build-up of the explosion 
  energy of the ejecta over possibly hundreds of milliseconds. 
  A ``soft'' nuclear equation of state that
  causes a rapid contraction and a smaller radius of the forming 
  neutron star and thus a fast release of gravitational binding 
  energy, seems to be more favorable for the development 
  of an explosion. Rotation has the opposite effect because in the
  long run it leads
  to a more extended and cooler neutron star and thus lower neutrino
  luminosities and mean energies and overall less neutrino heating.
  Neutron star g-mode oscillations, although we see their presence,
  and the acoustic mechanism play no important role in our simulations. 
  While numerical tests show that our code is well able to follow also
  large-amplitude core g-modes if they are instigated, the amplitude
  of such oscillations remains small in our supernova runs and 
  the acoustic energy flux injected by the ringing neutron star is
  minuscule compared to the neutrino energy deposition.
  
\end{abstract}


\section{Introduction}

The physical processes that start the explosion of massive stars 
are still not well understood, although this is of crucial importance
for predicting supernova and remnant properties, nucleosynthesis
conditions and yields, and the observable signals from supernovae
like neutrinos and gravitational waves. Neutrinos are thought to be
the main agent of energy transport and energy loss from collapsing 
stars, an expectation that was spectacularly verified by the detection
of two dozen neutrinos in connection with Supernova~1987A.  
Colgate \& White (1966) suggested already 40 years ago
that the intense flux of neutrinos radiated from the nascent neutron 
star might deposit the energy needed to reverse the stellar collapse.
The detailed physics of this neutrino-driven mechanism, however,
was worked out only later, based on numerical results of Wilson (1985).
Bethe \& Wilson (1985) showed that neutrinos deposit their energy 
behind the stalled supernova shock front mainly by the absorption
of $\nu_e$ and $\bar\nu_e$ on free neutrons and protons, respectively,
which are the most abundant nuclear species in the postshock layer.
Provided the neutrino luminosities are sufficiently large,
this energy input was found to be so strong that the stagnating
shock can be ``revived'' and thus accelerates outward to propagate
through the overlying, still collapsing layers of the star.

The viability of this neutrino-heating mechanism has recently 
been demonstrated for stars near the lower mass end of supernova
progenitors for stiff as well as soft nuclear equations of state
(Kitaura, Janka, \& Hillebrandt\ 2006; Janka, Marek, \& Kitaura 2007;
Janka et al.\ 2008). The investigated progenitor with a main-sequence
mass of 8.8$\,M_\odot$ (Nomoto 1984, 1987) can be 
considered as representative of the $\sim\,$8--10$\,M_\odot$ range.
The cores of such stars consist of oxygen, neon, and magnesium 
instead of iron and possess an extremely steep density gradient
at their surface. The latter fact leads to a very rapid 
decrease of the mass accretion rate onto the forming neutron star,
enabling the stalled shock to continue its expansion. This creates
favorable conditions for neutrino energy deposition, thus allowing
a neutrino-driven baryonic wind to be launched,
which is sufficiently powerful  
to eject the gravitationally loosely bound outer layers of the  
star. The new results confirm qualitatively previous simulations
by Mayle \& Wilson (1988), although the more
sophisticated treatment of the neutrino transport in the new
models, which were computed with the {\sc Prometheus-Vertex} code, 
leads to the predictions of a lower
explosion energy and a lower neutron excess in the ejecta.

In case of more massive progenitors, Wilson \& Mayle (1988, 1993)
could obtain explosions only by assuming that the neutrino luminosities
and thus the neutrino energy deposition behind the stalled shock were
boosted by enhanced neutrino transport in the neutron star. According to
Wilson \& Mayle this could happen, e.g., because of the presence
of neutron-finger instabilities below the neutrinosphere.
The existence of neutron-finger unstable conditions, however,
was repudiated on grounds of a detailed analysis of the neutrino 
transport conditions in supernova cores (Bruenn \& Dineva 1996).
Nevertheless, numerical simulations (Buras et al.\ 2006b; 
see also Dessart et al.\ 2006)
showed that Ledoux convection occurs inside the proto-neutron star and
later than about 200$\,$ms after bounce leads to accelerated lepton 
number and energy losses because of significantly increased muon and
tau neutrino luminosities. The fluxes of $\nu_e$ and $\bar\nu_e$,
which are mostly responsible for the shock heating, however,   
hardly change on a post-bounce timescale of some 
hundred milliseconds, and the mean energies of the radiated 
neutrinos and antineutrinos of all flavors even decrease compared
to models that ignore convection.

A potentially stronger multi-dimensional effect that is clearly
supportive for delayed shock revival, was discovered to be convective
overturn in the layer behind the stagnant shock. In this region neutrino
heating tends to create a negative entropy gradient, because the
energy deposition is maximal just outside of the gain radius  
(Herant et al.\ 1994; Burrows, Hayes, \& Fryxell 1995, 
Janka \& M\"uller 1996). Cooler matter that is accreted by the
shock is carried in narrow downdrafts closer to the gain radius,
where it readily absorbs energy from the neutrinos. At the same
time, heated matter with its higher entropy becomes buoyant and 
starts rising and expanding, thus reducing energy loss by the
reemission of neutrinos and pushing the shock farther out. 
Convective overturn therefore improves the efficiency
of the neutrino energy transfer because it allows more
matter to be exposed to strong heating near the gain radius. 
Effectively this means that the mass in the gain layer grows and
that gas stays there for a longer time. This is reflected by an 
increase of the mean timescale $\tau_\mathrm{adv}$ needed by
accreted matter to move inward from the shock to the gain radius.
A larger value of $\tau_\mathrm{adv}$ implies that the ratio
of the advection timescale to the neutrino-heating timescale,
$\tau_\mathrm{heat}$, gets closer to unity, which is a necessary
condition for a model to approach an explosion (see the analysis
in Buras et al.\ 2006b; Thompson, Quataert, \& Burrows 2005;
Murphy \& Burrows 2008).

Convection in the flow between shock and gain layer, however, 
is subject to strong damping because the gas is falling rapidly
and the time for seed perturbations to grow is therefore short
until the gas crosses the gain radius and reaches
the Ledoux-stable cooling region (Foglizzo, Scheck, \& Janka 2006).
Convective activity is therefore {\em not} able to develop in
the gain layer despite the negative entropy gradient there, unless
the initial density perturbations are sufficiently large or the ratio
of advection timescale to convective growth timescale,
$\tau_\mathrm{conv}$, exceeds
a critical value (for more details, see Scheck et al.\ 2008). 
Two-dimensional (2D) supernova simulations (Buras et al.\ 2006a,b)
with the {\sc Prometheus-Vertex} code, employing a sophisticated
energy-dependent description of the neutrino transport, indeed 
show that
convective overturn in the hot-bubble region becomes much less 
vigorous than in previous multi-dimensional models with a grey
neutrino diffusion scheme (e.g., Herant et al.\ 1994;
Burrows et al.\ 1995; Fryer 1999; Fryer \& Warren 2002, 2004),
mainly because the accretion shock in the new models lingers at 
rather small radii and the correspondingly high infall velocities 
in the postshock region are unfavorable for the growth of convection.
Scheck et al.\ (2008), performing parametric hydrodynamic
studies of the post-bounce accretion phase in collapsing stellar cores, 
found that convection is suppressed when (i) the neutrino heating 
is too weak to produce a steep entropy gradient behind the shock,
or (ii) the rapid contraction of the nascent neutron star leads
to shock retraction and thus causes very short infall
timescales of the gas between shock and gain radius.

In such a situation, however, the standing accretion shock
instability (SASI; Blondin, Mezzcappa, \& DeMarino 2003), 
which is a generic hydrodynamic 
instability of the shocked accretion flow to non-radial 
deformation modes, can exhibit particularly large
growth rates (cf.\ Scheck et al.\ 2008) and becomes the dominant 
multi-dimensional phenomenon
to initiate shock expansion. This was seen in the
studies by Scheck et al.\ (2008) and is confirmed by the
linear stability analysis of Yamasaki \& Yamada (2007). The SASI
stimulates large-amplitude bipolar shock oscillations, which create
strong entropy variations in the postshock flow and thus trigger
violent secondary convection. This helps pushing the stalled shock
to larger radii and therefore stretches the time accreted matter
is exposed to neutrino heating in the gain layer,
establishing healthy conditions 
for the success of the neutrino-driven mechanism.

Such a decisive role of the SASI was indeed observed in case of
an 11.2$\,M_\odot$ progenitor, whose collapse and explosion were 
simulated with the {\sc Prometheus-Vertex} code (Buras et al.\ 2006b).
Only when the setup of the numerical grid did not prohibit the growth of
the lowest (dipolar and quadrupolar) SASI modes\footnote{In previous
simulations with failed explosions (Buras et al.\ 2003a) the 
computational volume 
was constrained to a lateral wedge of 90 degrees ($\pm45^\circ$) around
the equatorial plane, using periodic boundary conditions. This setup
prevented the occurrence of the dipole and quadrupole SASI modes.},
which have the largest growth rates, could an explosion be obtained.
Convection alone was too weak to yield sufficient support for
the neutrino-heating mechanism. Instead it was the SASI in the
first place which pushed the shock farther out and helped to
bring the ratio of advection timescale to 
neutrino-heating timescale closer to the critical value of
unity, which is necessary for reversing accretion to explosion.

The underlying physical mechanism that is responsible for the
SASI phenomenon is still controversial and currently a matter of 
vivid debate. Foglizzo et al.\ (2007) explain the SASI by an
advective-acoustic cycle (see also Foglizzo 2001, 2002, 2008). 
This interpretation is in agreement with many properties of the
instability observed in 
the hydrodynamic simulations of Scheck et al.\ (2008)
and of Ohnishi, Kotake, \& Yamada (2006). In contrast, 
Blondin \& Mezzacappa (2006) and Blondin \& Shaw (2007) advocate
a purely acoustic cycle as the cause of the low-mode SASI and 
see this hypothesis supported by their numerical studies.

In this paper we will present evidence that the SASI-aided 
neutrino-heating mechanism may initiate an explosion not only
in the case of an 11.2$\,M_\odot$ progenitor as recently seen
by Buras et al.\ (2006), but in a wider
mass range of progenitor stars. We study here a 15$\,M_\odot$
model (s15s7b2 from Woosley \& Weaver 1995) which is frequently used 
in previous and present core-collapse studies because it can be
considered as representative 
of stars in a larger mass interval. Stellar evolution 
calculations by Woosley, Heger, \& Weaver (2002), for example, produced
a very similar core structure also for other supernova progenitors 
up to about 20$\,M_\odot$ (see model s20.0 in Figs.~A.2 and A.3 of
Buras et al.\ 2006b). We discuss 
results of a 2D simulation in which a modest amount of rotation
was assumed and which develops an explosion at 600$\,$ms after
core bounce, and compare it with non-exploding spherically 
symmetric (1D) simulations on the one hand, and 2D 
models without rotation on the other, in which two different 
equations of state (EoSs) for supernova matter are used, namely the 
Lattimer \& Swesty (1991) EoS with a soft nuclear phase and
the Hillebrandt, Wolff, \& Nomoto (1984; see also 
Hillebrandt \& Wolff 1985) EoS with a significantly stiffer
nuclear phase (Hillebrandt 1994). Both are consistent with current
neutron star mass determinations but the former leads to 
significantly more compact proto-neutron stars than the latter.
The rotating pre-collapse iron core
has a spin period of about 12 seconds. This is roughly a factor 
of ten more rapid than predicted by recent stellar evolution models
including magnetic fields (Heger, Woosley, \& Spruit 2005), but 
considerably slower than
necessary for magnetohydrodynamics to be important in triggering
the explosion, which requires pre-collapse spin periods of 
2--3 seconds (Burrows et al.\ 2007a, Thompson et al.\ 2005).

We have also evaluated our simulations for the presence of 
gravity waves in the nascent neutron star. Burrows et al.\
(2006, 2007b) found in their 2D supernova models that the
anisotropic accretion flow (as a consequence of the SASI) excited 
the neutron star to vigorous g-mode oscillations, in particular
also the dipole ($l = 1$) mode. Later than about one second
after bounce, the amplitude of these 
oscillations became very large (about 3$\,$km) at the neutron star 
surface and the neutron star as a whole showed a large-amplitude
periodic displacement along the $z$-axis of the numerical grid 
(Burrows et al.\ 2007c). 
As a consequence, a sizable acoustic energy flux was driven into
the surrounding gas and was transporting a significant
amount of acoustic energy to the stalled supernova shock.
Burrows et al.\ did not find any neutrino-driven explosions
in their models. However, at late times (typically at $t > 1\,$s
after bounce), and therefore much later than the shock is
revived by the neutrino-heating mechanism in the 11.2$\,M_\odot$
simulation of Buras et al.\ (2006b) and in the 15$\,M_\odot$
model discussed here, the acoustic 
energy flux caused by the core g-mode
oscillations became dominant compared to neutrino heating and
was able to cause an explosion. In our simulations the
g-mode oscillations of the newly formed neutron star
have always such small amplitudes that the acoustic energy input
to the developing blast is negligible compared to neutrino energy
deposition. In order to figure out whether our
neutrino-hydrodynamics code is able to deal with this potentially
important physical effect, we have also performed numerical tests.
These demonstrate that our code is well able to 
track large-amplitude g-modes, also of dipole character, if such
modes are excited in the neutron star core. We do not find any
reason why our third-order hydrodynamics scheme with the 
chosen numerical resolution should not
be able to follow the long-term excitation of neutron star 
oscillations by anisotropic accretion and turbulence in the
SASI region between neutron star and shock front. Whether the excitation
could be more efficient at times later than currently followed in our
simulations (at most about 630$\,$ms after bounce) remains unanswered
and must be left for future exploration.

The reason why Burrows et al.\ (2006, 2007b) did not see
any neutrino-driven explosions, which is in contrast to our 
findings, is not clear. However, the numerical approaches
of both groups and the included physics are different in many 
aspects, and therefore it is not astonishing that for example
the rate of neutrino energy deposition behind the supernova
shock differs significantly between the calculations.
While our ``ray-by-ray plus'' treatment of
neutrino transport is highly sophisticated in dealing with the
energy dependence of the problem but describes its dependence 
on the polar angle only approximately, Burrows et al.\
applied a 2D neutrino diffusion scheme, in which the
energy dependence was only incompletely taken
into account, because energy-bin coupling was ignored
and Doppler effects due to the motion of the stellar fluid were
neglected. In addition, the Tucson collaboration
made use of a different numerical method to solve the 
hydrodynamics equations, e.g.\ the gravitational force was 
implemented in the momentum equation in an uncommon, 
momentum-conserving way. In their \textsc{Vulcan} code they
employed a polar grid only in the outer regions of the computational 
volume but cartesian coordinates in the central part. They
performed their simulations with the nuclear EoS of Shen et al.\ 
(1998), which is
significantly stiffer than the Lattimer \& Swesty (1991) EoS, though
fairly similar to the Hillebrandt \& Wolff EoS (Marek et al.,
in preparation; see also Janka et al.\ 2007).
Moreover, their models were
computed with purely Newtonian gravity, whereas a significantly
stronger relativistic gravitational potential according to
Marek et al.\ (2006) is applied in our models.

In fact,
Newtonian 2D simulations in which the confluence of neutrino
heating, the SASI, and nuclear burning produced explosions of
11 and 15$\,M_\odot$ progenitors, have recently been reported
by Bruenn et al.\ (2006) and Mezzacappa et al.\ (2007) and 
seem to support the successful shock revival found in an
11.2$\,M_\odot$ star by Buras et al.\ (2006b). All these results
were obtained with the EoS of Lattimer \& Swesty (1991).
Like Buras et al.\ (2006b), the Oak Ridge group employed for the
neutrino transport in two dimensions the ``ray-by-ray-plus''
approximation to deal with the lateral dimension. Their transport
scheme, however, was based on a flux-limited diffusion description,
in which the energy-dependence of the transport was treated
in full detail. We consider the discrepant results obtained by
the Tucson and Oak Ridge collaborations as additional motivation
for us to contribute to the discussion by presenting our current
simulations in this publication.

The paper is structured as follows: In Sect.~\ref{sec:numerics}
we give a brief summary of the numerical scheme and physics input 
used in the presented simulations, an overview of which follows
in Sect.~\ref{sec:models}. Section~\ref{sec:results} 
contains our results. For reference, we present in 
Sect.~\ref{sec:1dmodels} results of our
1D simulations for the two employed nuclear EoSs,
then describe the dynamical evolution
of our exploding model in Sect.~\ref{sec:dynamics}, compare the 
different computational runs in Sect.~\ref{sec:comparison}, and
discuss our neutrino results in 
Sect.~\ref{sec:neutrinos}. An analysis of neutron star g-modes
including numerical tests is presented in 
Sect.~\ref{sec:gmodes}. In Sect.~\ref{sec:summary} we summarize
our findings and draw conclusions.

\section{Methods and models}
\label{sec:mm}

In this section we briefly summarize the basic properties of
our numerical code and the input physics used in this work. We
then give a compilation of the core-collapse simulations whose
results are described afterwards.

\subsection{Numerical approach and input physics}
\label{sec:numerics}

The core-collapse and post-bounce calculations presented in this
work are performed with the \textsc{Prometheus-Vertex}
neutrino-hydrodynamics code, details of which were published
by Rampp \& Janka (2002) and Buras et al (2006a).
The code module that integrates the nonrelativistic hydrodynamics
equations is a conservative and explicit Eulerian implementation of a
Godunov-type scheme with higher-order spatial and temporal accuracy.
The self-gravity of the stellar gas is treated in an approximation
to general relativity, in which the monopole term
of the Newtonian potential is replaced by an effective 
relativistic scalar potential that was constructed from close 
comparison of the Newtonian and relativistic equations of motion
(for details, see Marek et al.\ 2006). For this effective 
relativistic monopole potential we consider the Cases~A and R
of the Marek et al.\ paper. The higher-order terms of
the gravitational potential are taken as solutions of the Poisson
equation for the two-dimensional gas distribution (see 
M\"uller \& Steinmetz 1995). 

The time-implicit transport routine solves the moment
equations of neutrino number, energy, and momentum in spherical
symmetry, employing a variable Eddington factor for
closing the set of equations (Rampp \& Janka 2002). 
The closure factor is obtained from a 
simplified Boltzmann equation, whose integro-differential 
character is tackled by iterating the coupled system of Boltzmann
and moment equations until convergence is achieved.
Our treatment of the neutrino transport and of neutrino-matter 
interactions is energy dependent and includes the full redistribution
of neutrinos in energy and momentum 
space by scattering reactions with electrons,
positrons, and nucleons. Effects caused by the motion of the
stellar fluid are taken into account as well as general relativistic
redshifting and time dilation. In two spatial dimensions the transport 
is based on the ``ray-by-ray plus'' approximation, in which 
spherically symmetric transport problems are solved for the
conditions present in the radial ``rays'' corresponding to the different
lateral bins of the polar grid. This means that we assume the
local neutrino phase-space distribution to be axially symmetric
around the radial direction and thus to depend only on one
instead of two angles, as a consequence of which the
neutrino pressure tensor remains diagonal and the lateral 
component of the neutrino flux vanishes. This simplifies the 
transport from a five dimensional, time-dependent problem (with
variables being the radius, energy, polar angle, and two
angles that characterize the neutrino momentum direction) to
only four dimensions and reduces the 
complexity of the moment equations significantly (see Appendix~B
in Buras et al.\ 2006a) at the expense of
neutrino shear and non-radial flux of the neutrinos, which
are disregarded in such an approach.
However, the moment equations still contain the terms that account
for the advection of trapped neutrinos with the stellar fluid
motion in the lateral direction. Moreover, the lateral component
of the neutrino pressure gradient is included in the hydrodynamics
equations.

Comparison of our approximative treatment of general relativity
with fully relativistic simulations revealed excellent agreement 
in spherical symmetry and still good agreement in two dimensions 
when the rotation did not become too extreme (see Marek et al.\ 2006,        
Liebend\"orfer et al.\ 2005).

A state-of-the-art treatment of the interactions of neutrinos
($\nu$) and antineutrinos ($\bar\nu$) of all flavors 
(Buras et al 2006a, Marek et al.\ 2005) is applied in
our simulations, including all relevant $\beta$-processes, thermal
processes, and the scattering off electrons, positrons, nucleons,
and nuclei. In neutral-current and charged-current
neutrino-nucleon reactions the effects of nucleon recoil and
thermal motions as well as nucleon correlations in the dense medium
are taken into account (see Buras et al.\ 2006a, and references therein).
Neutrino-nucleon scattering just like neutrino-electron scattering is
therefore a channel for energy exchange between neutrinos and the
stellar medium and fosters neutrino thermalization, thus leading
to a noticeable reduction of the mean energy and a spectral 
pinching of radiated neutrinos, in particular of muon and tau
flavor (Keil, Raffelt, \& Janka 2003). Electron captures on heavy 
nuclei are described by making use of the improved data of
Langanke et al.\ (2003), who constructed a table for
the capture rates of a large ensemble of nuclei in nuclear 
statistical equilibrium. Also interactions of neutrinos of different
flavors, scatterings as well as neutrino-antineutrino pair conversion
(Buras et al.\ 2003b) are included in our treatment.

%
%



We employ two different nuclear equations of state (EoSs) in our
simulations. On the one hand, we use a soft version of the 
compressible liquid drop EoS of Lattimer \& Swesty 
(1991; see also Lattimer et al.\ 1985; ``LS-EoS'') 
with an incompressibility 
modulus of bulk nuclear matter of 180$\,$MeV and a symmetry 
energy parameter of 29.3$\,$MeV. On the other hand we perform
simulations with the considerably stiffer EoS of Hillebrandt \&
Wolff (1985, see also Hillebrandt et al.\ 1984; ``HW-EoS''), 
which is based on a Hartree-Fock calculation,
assuming a Skyrme force for the nucleon-nucleon interaction with
parameters given by K\"ohler (1975). Its incompressibility has
a value of 263$\,$MeV and the symmetry energy was chosen to be 
32.9$\,$MeV. Details of the calculation can be found in 
Hillebrandt et al.\ (1984) and Hillebrandt \& Wolff
(1985). Both EoSs yield different maximum masses of
stable, nonrotating neutron stars. In case of the 
LS-EoS the maximum gravitational mass is 1.84$\,M_\odot$, whereas
it is 2.21$\,M_\odot$ for the HW-EoS, which are both compatible 
with measured neutron star masses (Lattimer \& Prakash 2007).
Both EoSs also lead to a significantly different 
evolution of the radius of the nascent neutron star as a function
time (see Fig.~7 in Janka et al.\ 2007).
While the LS-EoS produces rather compact neutron stars
with a radius of $R_{\mathrm{ns}} = 11.9\,$km for a cold star with 
1.4$\,M_\odot$ and $R_{\mathrm{ns}} = 10.0\,$km for a cold star
near the maximum mass, the corresponding radii are 13.94$\,$km and 
13.5$\,$km, respectively, for the HW-EoS (cf.\ Marek 2007).

We apply the nuclear EoS (either LS or HW) only above
a transition density $\rho_{\mathrm{EoS}}$. In the 
low-density regime, $\rho < \rho_{\mathrm{EoS}}$, the EoS includes
the ideal gas contributions of electrons
and positrons (of arbitrary degrees of relativity and degeneracy),
photons, and a mixture of non-relativistic classical Boltzmann gases
of neutrons, protons, $\alpha$ particles, and 14 kinds of heavy nuclei.
Coulomb lattice corrections of the pressure, energy density, entropy, 
and adiabatic index are taken into account. Above a temperature of
$T_{\mathrm{NSE}} = 0.5\,$MeV the nuclear constituents are assumed to 
obey nuclear statistical equilibrium (NSE), whereas below this 
temperature only nuclear burning of silicon and oxygen (and similarly 
neon and magnesium), and
carbon (implemented according to the ``flashing''
treatment by Rampp \& Janka 2002, Appendix~B.2) 
can change the composition.
The value of $\rho_{\mathrm{EoS}}$ is chosen such that
a smooth transition of pressure, internal energy density, and chemical
potentials as functions of density is guaranteed. During the collapse
phase $\rho_{\mathrm{EoS}}$ is set to $6\times 10^{7}$g$\,$cm$^{-3}$
in case of the LS-EoS and $1.5\times 10^{9}$g$\,$cm$^{-3}$ for the
HW-EoS, while after core bounce, when $\alpha$ particles can reach a 
significant mass fraction and an error in the LS-EoS 
(Fryer, private communication) is potentially relevant
(see, however, Buras et al.\ 2006a), we use 
$\rho_{\mathrm{EoS}} = 10^{11}\,$g$\,$cm$^{-3}$.

%

\subsection{Investigated models}
\label{sec:models}

Except for a simulation with an 11.2$\,M_\odot$ progenitor from
Woosley et al.\ (2002), for which we show results from a 
continuation of the model run published by Buras et al.\ (2006b), 
all core-collapse and supernova calculations in this work are 
based on the 15$\,M_\odot$ progenitor s15s7b2 from
Woosley \& Weaver (1995) and are listed in Table~\ref{tab:models}.
Details of the 11.2$\,M_\odot$ model and a comparison of 1D and
2D results can be found in the Buras et al.\ (2006b) paper.
Two of the 15$\,M_\odot$
calculations are conducted in spherical symmetry, one with the
LS-EoS (indicated by ``LS-1D'' in the model name) and another one 
with the HW-EoS (labeled by ``HW-1D''). These are compared with
four 2D (axially symmetric) simulations, two of which (M15LS-2D
and M15HW-2D) are computed without rotation, using the EoSs also
employed in the 1D models. In the other two simulations (i.e.,
our reference long-time run, M15LS-rot, and Model~M15LS-rot9) 
we impose rotation on the progenitor core with initially a constant
angular frequency of 0.5$\,$rad$\,$s$^{-1}$ (corresponding to a
rotation period of about 12$\,$s) inside the Fe core and an $r^{-3/2}$
decline outside (see Fig.~1 in M\"uller et al.\ 2004 and Sect.~3.4
in Buras et al.\ 2006b). This rotation rate is roughly a factor
ten faster than current predictions for pre-collapse stellar cores
by Heger et al.\ (2005), but too slow for 
the free energy reservoir in the secularly evolving proto-neutron
star to be sufficient to power magnetohydrodynamic explosions
(see Burrows et al.\ 2007a and Thompson et al.\ 2005).

In all of our nonrotating models we have implemented for the
monopole of the gravitational potential the effective relativistic
potential according to Case~A of Marek et al.\ (2006),
which in that paper was identified to yield the best results in comparison 
to spherically symmetric, fully relativistic simulations. For models
with core rotation, however, the best choice of the effective 
gravitational potential is less obvious. While for slow rotation 
Case~A still produces the most satisfactory results, this monopole
term is not able to account for the relativistic gravity of rotationally
deformed cores. In such a situation, Marek et al.\ (2006) found
Case~R to be the better representation, because, e.g., it leads to central
densities closer to those of the relativistic calculation. 
Several 100$\,$ms after core bounce even our slowly rotating pre-collapse
model develops a sizable oblate deformation of the nascent neutron star.
Since late-time effects are the main interest of our present investigations, 
we therefore decided to employ the effective relativistic TOV potential of
Case~R of Marek et al.\ (2006) in our reference long-time 2D
simulation (Model~M15LS-rot) and to compare this model with another 
simulation that includes rotation and uses Case~A for the monopole 
term of the gravitational potential (Model~M15LS-rot9).

Our models are computed with 400--1000 non-equidistant radial zones,
which are chosen such that the resolution $\Delta r/r$ is typically 
better than three percent in the interior of the neutron star and
between less than one percent and 1.5
percent around the neutrinospheres and outside of the neutron star. 
During the simulations the number of radial grid zones $N_r$ is 
increased (Table~\ref{tab:models} gives $N_r$ at the beginning and at
the end of the different runs) and the region
of higher resolution is moved to smaller radii in order to follow
the contraction of the neutron star and to ensure good resolution
at the neutron star surface where the density gradient steepens 
with time. We make sure that at least 20 radial zones are used
per density decade, corresponding to at most 12\% density variation 
between neighboring zones. The representation of hydrostatic
equilibrium as well as the accurate computation of the neutrino
transport in the decoupling layer require at least this numerical
resolution; in case of the transport the luminosity variation 
from zone to zone is a limiting factor. Convergence tests for our
code showed that insufficient radial resolution in particular
in this region may not only lead to bloated surface layers of
the neutron star and to an overestimation of the neutrino
luminosities, but as a consequence also to artificial 
explosions even of spherically symmetric models. In order to
make sure that our exploding 2D simulation, Model~M15LS-rot,
is well resolved, we recomputed a significant part of the
crucial evolution phase of this model with about half the 
radial zone width in the neutron star surface layer.
This high-resolution run, Model~M15LS-rot-hr, is not listed in 
Table~\ref{tab:models}, but some of the plots will also show 
results from this calculation for comparison.

Our 2D simulations are performed
with 128 or 192 equidistant lateral zones from pole to pole of
the spherical coordinate grid (Table~\ref{tab:models}). 
Comparing results for both resolutions we found excellent
quantitative agreement of the neutron star and neutrino properties
(which implies that the lateral resolution is sufficient for 
describing neutron star convection) and very good qualitative
agreement of the shock evolution (because of the chaotic character
of hydrodynamic instabilities in the hot-bubble region, a detailed
quantitative agreement cannot be expected).
At the grid center, however, a small 
region (1.7$\,$km in radius or six radial zones) 
is computed in spherical symmetry to avoid the most severe
limitation of the hydrodynamics timestep due to the
Courant-Friedrich-Lewy condition. 

In the transport module of our code we usually employ 17 geometrically
distributed energy bins for neutrinos between 0 and 380$\,$MeV. One
of our models, M15LS-rot9, is computed only with nine energy bins,
which leads to a reduced steepness of the high-energy tail of 
the neutrino spectrum roughly three orders of magnitude and more 
below the spectral peak. Spherically symmetric test calculations,
however, show that this lower accuracy of the neutrino transport
has hardly any influence on the hydrodynamic evolution of the
supernova core and also causes only small differences (of order 10\%)
in the luminosities and mean energies of the radiated neutrinos. 
Using nine instead of 17 energy bins reduces 
the CPU time requirements of a simulation by nearly a factor
of seven.

The 2D runs without rotation are started at core
bounce after mapping the corresponding 1D models to the
2D grid and imposing random zone-to-zone variations of the
radial velocity component (or density) with an amplitude of one percent. 
The rotating models
are computed in 2D from the onset of the collapse with
random perturbations and the initial rotation law imposed on the 
spherically symmetric progenitor star 
(for details, see Buras et al.\ 2006b).

Running 2D models with a sophisticated energy-dependent treatment
of the neutrino transport for many hundred milliseconds after 
bounce poses a considerable numerical challenge, because the 
rapid variations of the accretion flow in the presence of violent
hydrodynamic instabilities in the postshock region do not allow
the (implicit) neutrino transport timestep to become larger than
about $2\times 10^{-6}\,$s and therefore about 500.000 timesteps
and $3\times 10^{18}$ 
floating point operations were needed for our most advanced 2D model
MS15LS-rot, which was evolved to more than 600$\,$ms after bounce.
For this reason, CPU-time limitations did not permit us to follow
all of our 2D models to such very late post-bounce times. Although the
set of calculations of Table~\ref{tab:models} is not finished in this
sense (and currently we have no way to do that in an acceptable
time frame), it nevertheless allows for some interesting conclusions
concerning the possibility of getting neutrino-driven explosions
late after bounce and the dependence of this result on the 
properties and the neutrino emission of the forming neutron star.

\section{Results}
\label{sec:results}

In this section we discuss our simulations, starting with 
the 1D results for reference, then continuing with the exploding 
long-time 2D simulation, and finally turning to the other 2D
models, which allow us by comparisons to better understand the
factors that are decisive for the explosion in the successful
run.

\subsection{Spherically symmetric 15$\,M_\odot$ simulations}
\label{sec:1dmodels}

\begin{figure*}
\plottwo{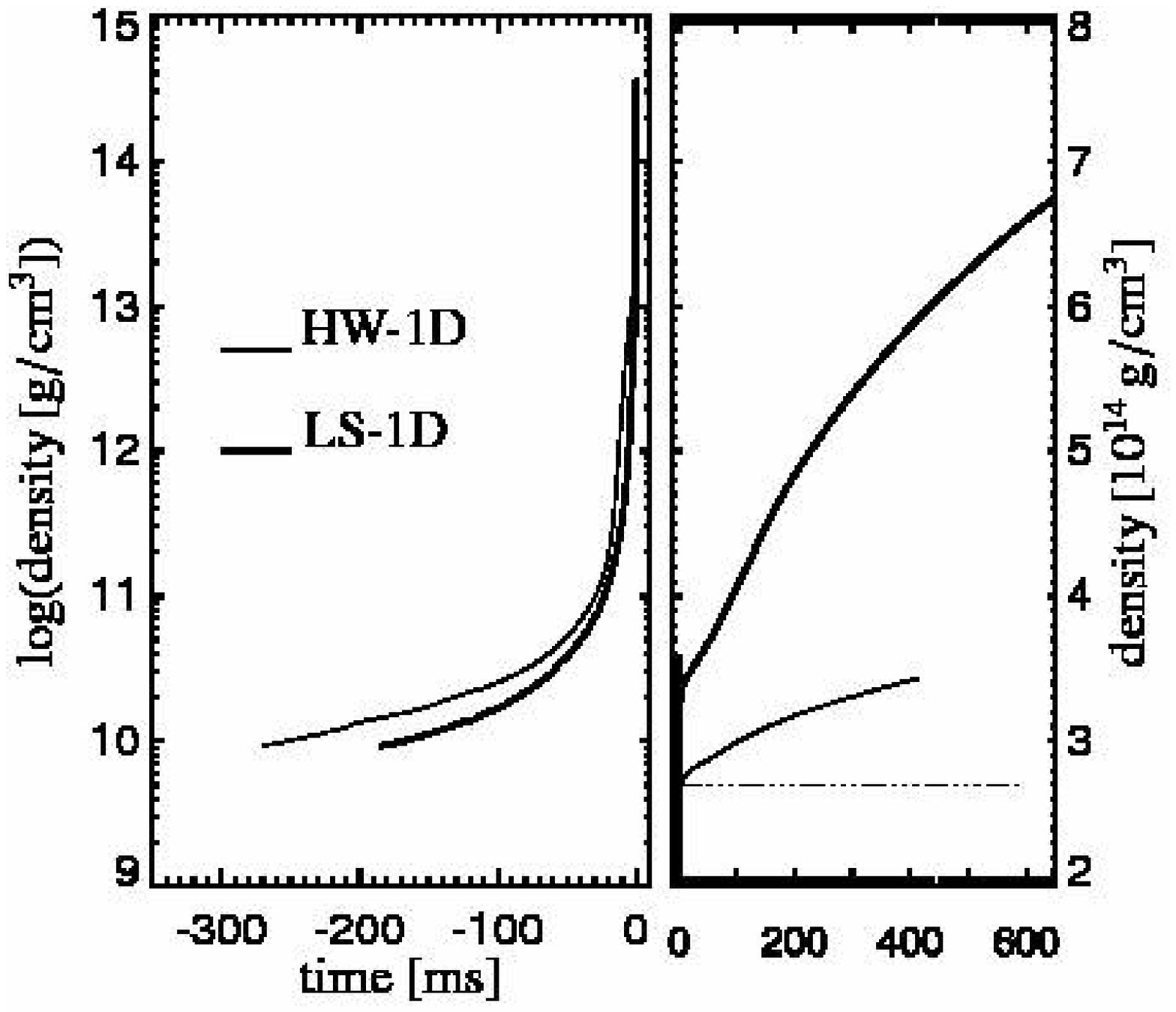}{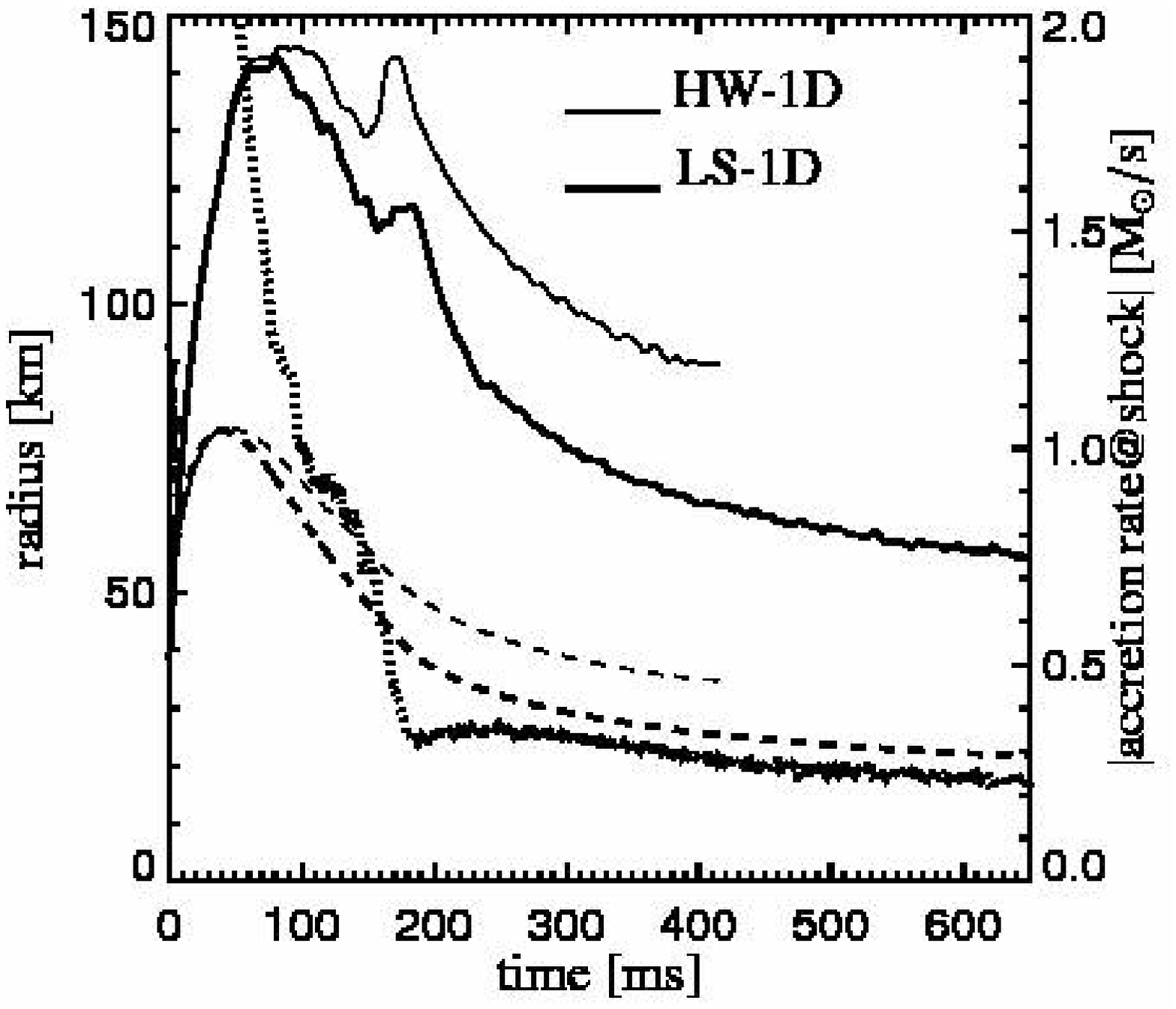}
\plottwo{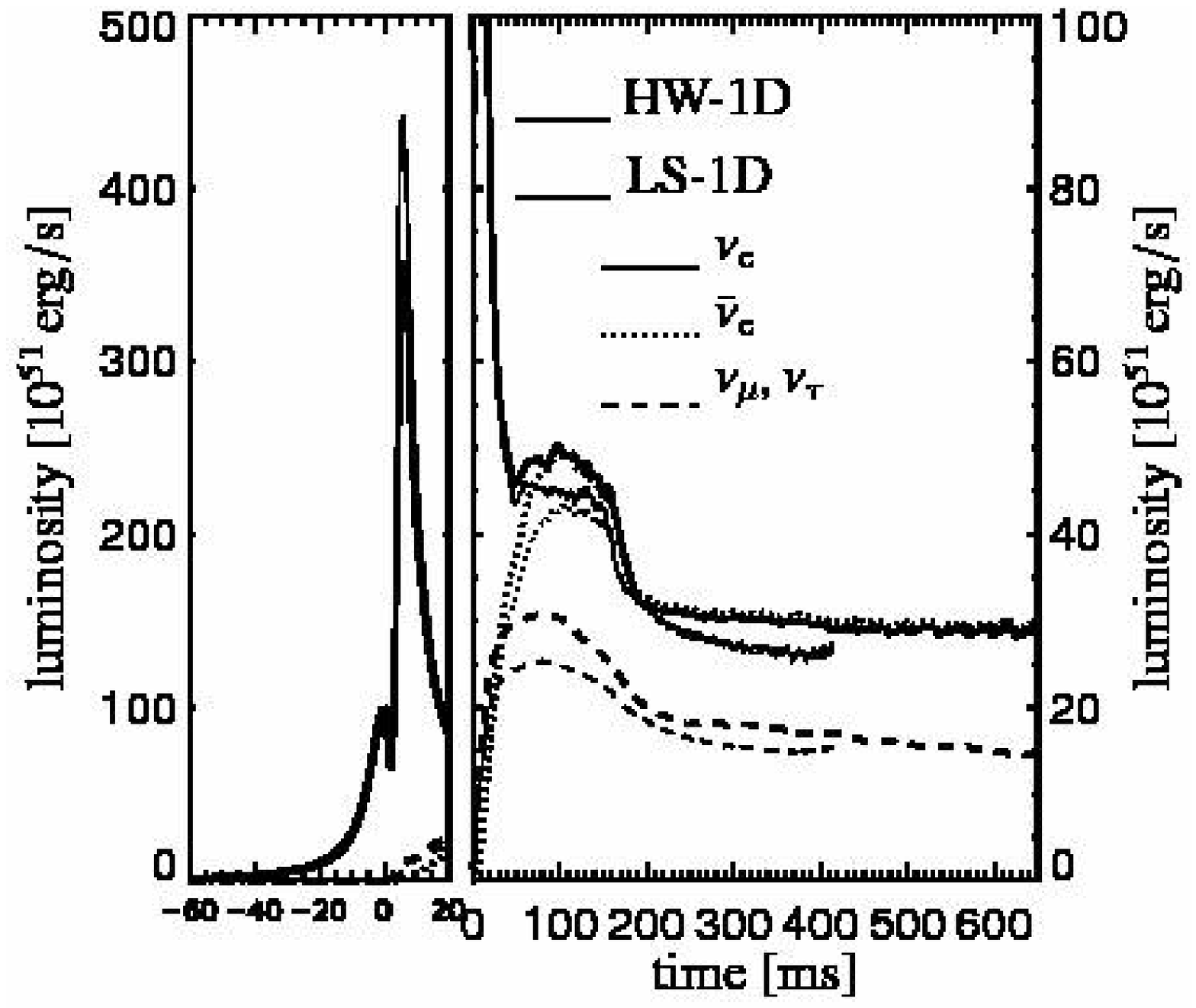}{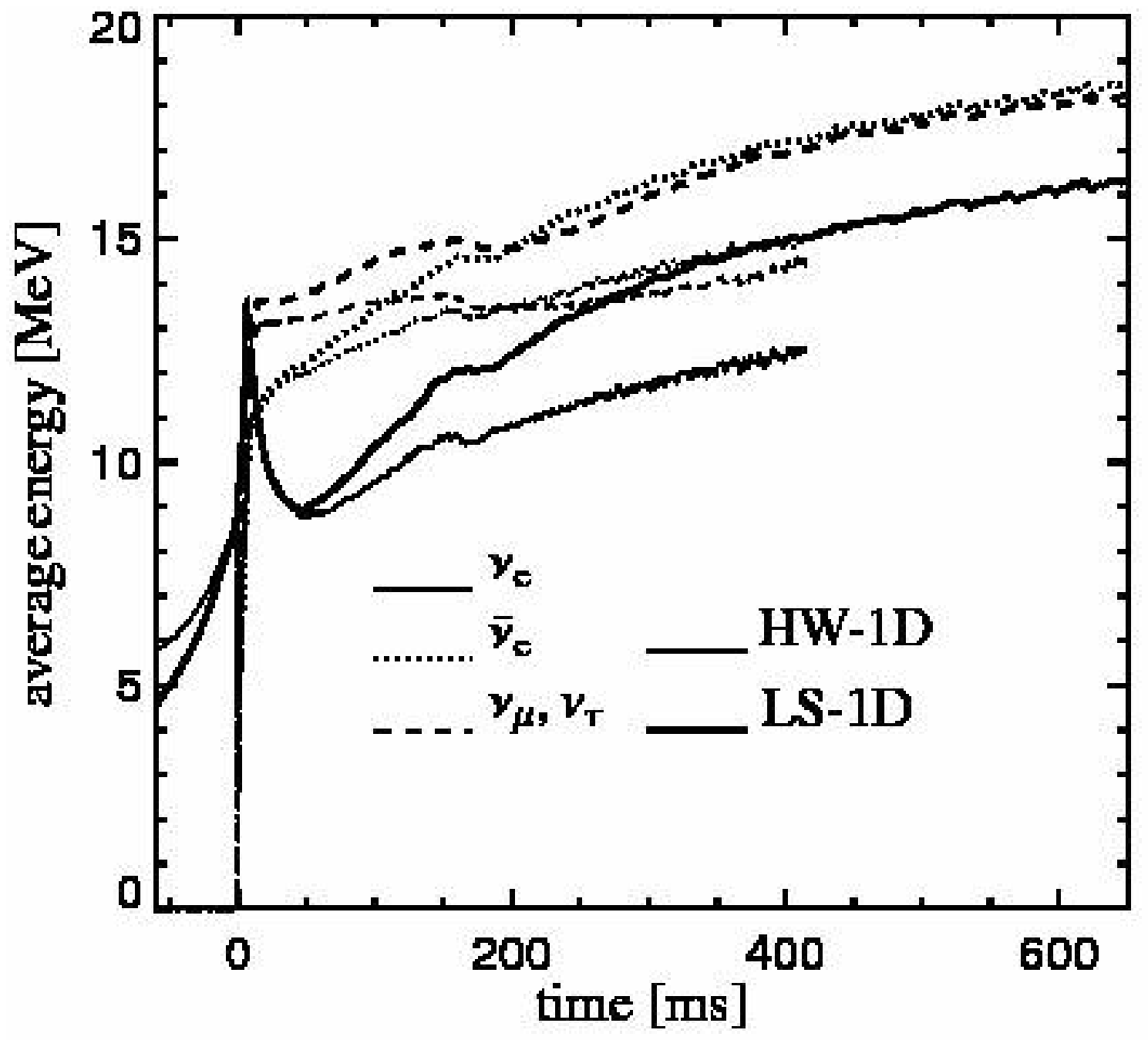}
 \caption{Results of our two 1D models, M15LS-1D (bold lines)
and M15HW-1D. Time is normalized to the moment of core bounce.
{\em Top left:} Central densities as functions of time. The
horizontal dash-triple-dotted line marks the value of the
nuclear saturation density, $\rho_0 \approx
2.7\times 10^{14}\,$g$\,$cm$^{-3}$.
{\em Top right:} Radii of shock (solid lines) and electron
neutrinosphere (dashed lines) as functions of time. The transient
shock expansion around 170$\,$ms after bounce is caused by a
composition interface with sudden decrease of the mass accretion
rate arriving at the shock. The mass accretion rate is displayed
by the bold dotted line. 
{\em Bottom left:} Luminosities of electron neutrinos, electron
antineutrinos, and muon or tau neutrinos (or their antiparticles)
as functions of time.
{\em Bottom right:} Mean energies of radiated neutrinos (computed
as the ratio of energy to number flux) for electron neutrinos,
electron antineutrinos, and heavy-lepton neutrinos and antineutrinos.
The luminosities and mean energies are shown as measured by an 
observer at rest
relative to the stellar center at 400$\,$km (from where the
gravitational redshift to infinity is negligibly small).
The softer LS-EoS leads to a much larger central density, a
significantly more compact neutron star and therefore considerably
higher luminosities and mean energies of the radiated neutrinos.
\label{fig:1dresults}}
\end{figure*}

In Fig.~\ref{fig:1dresults} the results of our two 1D simulations
are shown in overview. The collapse of the stellar core until bounce
is somewhat faster with the softer (i.e.\ here: its adiabatic index
in the subnuclear regime is lower) LS-EoS, and the maximum density
reached at bounce is $3.6\times 10^{14}\,$g$\,$cm$^{-3}$ compared
to $3.0\times 10^{14}\,$g$\,$cm$^{-3}$ in case of the HW-EoS. The
central density is measured at a distance of 1.65$\,$km (i.e., in
the sixth radial grid zone) where numerical fluctuations caused by
the inner grid boundary are absent). During the post-bounce 
evolution the higher stiffness of the nuclear phase in the 
HW-EoS leads to a considerably
lower central density and a clearly less rapid contraction of the 
nascent neutron star, visible from the larger radius of the electron
neutrinosphere in the right upper panel of Fig.~\ref{fig:1dresults}.
Since the postshock layer is nearly in hydrostatic equilibrium
during the quasi-stationary accretion phase of the stalled shock,
the shock position $R_{\mathrm{s}}$ is a sensitive function of the 
neutron star radius, $R_{\mathrm{ns}}$, roughly as
\begin{equation}
R_{\mathrm{s}}\,\propto\,
\left({R_{\mathrm{ns}}^4 T_{\mathrm{ns}}^4\over |\dot M| \,
M_{\mathrm{ns}}^{1/2}} \right)^{\! 2/3}\ ,
\label{eq:rshock}
\end{equation}
which follows from Eqs.~(39), (56), and (63) of Janka (1991).
Here $M_{\mathrm{ns}}$ is the neutron star mass, $T_{\mathrm{ns}}$
the temperature at the neutrinosphere, and $\dot M$ ($< 0$)
the rate of mass accretion by the shock. In case of the
HW-EoS, the stagnation radius of the shock is therefore significantly
larger during the phase of shock retraction ($t \ga 80$--100$\,$ms 
after bounce).

In contrast, the luminosities $L_\nu$ and mean energies $\left\langle
\epsilon_\nu\right\rangle$ of the neutrinos radiated during shock 
accretion are appreciably higher in case of the LS-EoS, because the
larger neutrinospheric temperature of the more compact neutron
star overcompensates for the smaller radius (roughly, the $\nu_e$ 
emission behaves like blackbody radiation and thus
$L_\nu\propto R_{\mathrm{ns}}^2 T_{\mathrm{ns}}^4$ and 
$\left\langle \epsilon_\nu\right\rangle \propto T_{\mathrm{ns}}$).
Interestingly, the prompt $\nu_e$ burst during shock breakout
reveals the opposite dependence on the nuclear EoS: it is more
luminous in case of the HW-EoS because of a stronger deleptonization
in a wider spatial region, which is facilitated by a less steep 
increase of the optical
depth in the deleptonization region and thus an easier escape of
the electron neutrinos.

We also point out that in the simulations with both EoSs the
average energy of the radiated $\bar\nu_e$ gets very close to 
that of the emitted muon and tau neutrinos or becomes
even slightly higher after
about 200$\,$ms of post-bounce accretion (Fig.~\ref{fig:1dresults},
bottom right). This effect is visible in the mean spectral energies,
which are defined as ratio of the energy density to the 
number density of neutrinos,
$\left\langle \epsilon_\nu \right\rangle = \int_0^\infty \mathrm{d}\epsilon\,
J_\nu(\epsilon)/\int_0^\infty \mathrm{d}\epsilon\,\epsilon^{-1}J_\nu(\epsilon)$,
with $J_\nu(\epsilon)$ being the zeroth energy moment of the specific 
intensity\footnote{At sufficiently large radii, where neutrinos in the
bulk of the spectrum propagate nearly radially (i.e., the flux factor
in the laboratory frame is near unity, which is well fulfilled at the
chosen radius of evaluation at 400$\,$km), the local energy and number 
densities are essentially identical with the energy and number flux
densities.}. In contrast, the rms energies of the energy spectrum,
$\left\langle \epsilon_\nu \right\rangle_\mathrm{rms} \equiv 
\left [ \int_0^\infty \mathrm{d}\epsilon\, \epsilon^2
J_\nu(\epsilon)/\int_0^\infty \mathrm{d}\epsilon\,
J_\nu(\epsilon)\right ]^{1/2}$,
still follow the standard order sequence, 
$\left\langle \epsilon_{\nu_e} \right\rangle_\mathrm{rms} <
\left\langle \epsilon_{\bar\nu_e} \right\rangle_\mathrm{rms} <
\left\langle \epsilon_{\nu_x} \right\rangle_\mathrm{rms}$,
although the difference between the last two is considerably smaller than
in older simulations, in which the transport treatment of 
heavy-lepton neutrinos $\nu_x$ did not take into account 
the energy exchange through neutrino-nucleon scatterings and the 
production of $\nu_x\bar\nu_x$ pairs by nucleon-nucleon bremsstrahlung
and by the annihilation of $\nu_e\bar\nu_e$ pairs (for more details,
see Buras et al.\ 2003b, Raffelt 2001, Keil et al.\ 2003). 
We will come back to
a closer discussion of these interesting spectral properties in 
Sect.~\ref{sec:neutrinos}.

\subsection{Two-dimensional 15$\,M_\odot$ model with explosion}
\label{sec:dynamics}

In contrast to the spherically symmetric simulations, 
the two-dimensional Model~M15LS-rot turns out to approach 
an explosive runaway situation after more than 500$\,$ms of
post-bounce accretion (Figs.~\ref{fig:rotmod1} and
\ref{fig:rotmod2}). Some snapshots of
the entropy distribution in the central region (with radii
between $\sim$400$\,$km and $\sim$800$\,$km) for characteristic 
stages of the evolution are displayed in Fig.~\ref{fig:snapshots}.

\subsubsection{Post-bounce evolution}
\label{sec:postbounce}

The results of a 90-degree equivalent of Model~M15LS-rot  
(computed with a pole-to-equator wedge of the spherical
coordinate grid and with reflecting conditions at both boundaries) 
until 300$\,$ms after bounce were described by 
Buras et al.\ (2006b), see Sect.~3.4 there. Postshock
convection is triggered 
by the small-scale perturbations that evolve from a
1\% random initial (pre-collapse) seed during infall. The convective
activity exhibits the most rapid growth in high angular modes and
visible inhomogeneities begin to show up at 50--60$\,$ms after bounce.
In the 180-degree calculation discussed here, the first
large-scale differences between both hemispheres appear at about
100$\,$ms and only shortly later (at $\sim$120$\,$ms, see
also the top left snapshot in Fig.~\ref{fig:snapshots}) low-amplitude
bipolar shock oscillations set in, manifesting the development
of a global, low-mode asymmetry. This can be seen in the top panel of
Fig.~\ref{fig:modes}, where 
large amplitudes of the dipolar ($l=1$) and quadrupolar ($l=2$)
terms of a spherical harmonics decomposition of the shock radius as a 
function of the polar angle $\theta$ occur at $t \ga 100\,$ms post 
bounce. The plot suggests that the dipole mode gains strength 
slightly faster and reaches large amplitudes some ten milliseconds
earlier than the quadrupole mode.

The appearance of strong low-mode, quasi-periodic shock deformation 
is associated with sloshing motions of the whole accretion layer
between shock and neutron star. The process is very similar to
the phenomenon identified as SASI in the idealized accretion 
setup studied by Blondin et al.\ (2003) and seen
also in more sophisticated post-bounce accretion simulations by
Scheck et al.\ (2008). In those calculations
convective activity was initially absent and set in only later,
and therefore the oscillatory
growth of the SASI in the linear regime could be clearly
identified. In contrast, in the model discussed here, sufficiently
strong neutrino heating and sufficiently large seed perturbations
cause convective activity in the gain
layer to develop readily on a short timescale after bounce.
The early presence of convection masks the characteristic
growth behavior of the SASI. We suspect that the
acoustic waves and vorticity created by convection even
accelerate the evolution of the SASI and the conditions for
a linear description never exactly apply.

The growth of the SASI modes in the linear regime was analysed
in detail by Blondin \& Mezzacappa (2006). Their 2D axisymmetric
hydrodynamic simulations with a simplified representation of the
conditions in supernova cores revealed that the $l = 1$ mode is
always unstable,
grows faster than the $l = 2$ mode, and is the only mode that
makes the transition to the nonlinear stage. Our simulations
disagree with their results concerning the faster initial
growth of the quadrupole mode (see the evolution during the first
$\sim$50$\,$ms after bounce in the upper and middle panels of
Fig.~\ref{fig:modes}), which however was also seen in some 
model runs performed by Scheck et al.\ (2008). In the highly
nonlinear phase we observe that the dipole and quadrupole modes
are present with similar strengths. During episodes in 
which the $l=2$ component dominates the low-mode power (e.g.\
at $t\sim$150, 290, 380, 480$\,$ms, see middle panel of
Fig.~\ref{fig:modes}), the spectrum is indeed enhanced 
at frequencies well above 100$\,$Hz
(Fig.~\ref{fig:modes}, lower panel), consistent with 
the expectation that the $l=2$ mode should be distinguished from
the $l=1$ mode by a higher frequency (Blondin \& Mezzacappa
2006, Fig.~4). However, we cannot exclude that some of the 
activity we find associated with quadrupolar deformations is not
a true $l=2$ but a ``ghost'' $l=1$ mode with the frequency of
the latter, because the $l=2$ spectra show a very broad distribution
with significant power also below 100$\,$Hz. 
Perturbations associated with prompt postshock convection in the
first tens of milliseconds after bounce (see Marek et al.\ 2008)
and violent neutrino-driven convection at $t \ga 80\,$ms, which are
present in all of our simulations --- different from the idealized
accretion models considered by Blondin and Mezzacappa 2006, where
the setup was chosen to be convectively stable ---, but
possibly also numerical errors  
must be expected to give rise to the excitation of $l=2$ motions
and thus might prevent the presence of a purer and more 
clearly dominant $l=1$ mode.

In Fig.~\ref{fig:rotmod2} the 
the radii of the shock near the north pole and south pole are
plotted as functions of time. One can well see the alternating
phases of shock expansion in the northern and southern 
hemispheres. The snapshots in Fig.~\ref{fig:snapshots} 
at 454$\,$ms and 524$\,$ms after bounce capture two such
moments in which the shock is inflated in opposite directions.
There is an inverse relation between the typical period of these
SASI oscillations and the mean shock radius. The period is
about 10--12$\,$ms when the shock is near 150$\,$km (around
300--400$\,$ms after bounce) and more like 15--17$\,$ms when
the average shock radius reaches 180--200$\,$km (around 500$\,$ms), 
corresponding to power maxima at typical frequencies beween 
50$\,$Hz and 100$\,$Hz. Towards the end of the simulation, as the
shock expands farther, the oscillation period becomes even longer.
This can be seen in the
time-dependent frequency spectrum of the low-mode SASI 
power (obtained from a fourier analysis of the quantity given
in Eq.~(2)) in the bottom panel of Fig.~\ref{fig:modes}. The
power spectrum has a very broad peak in the mentioned frequency
interval and decays steeply towards higher frequencies.

\begin{figure*}
\plotone{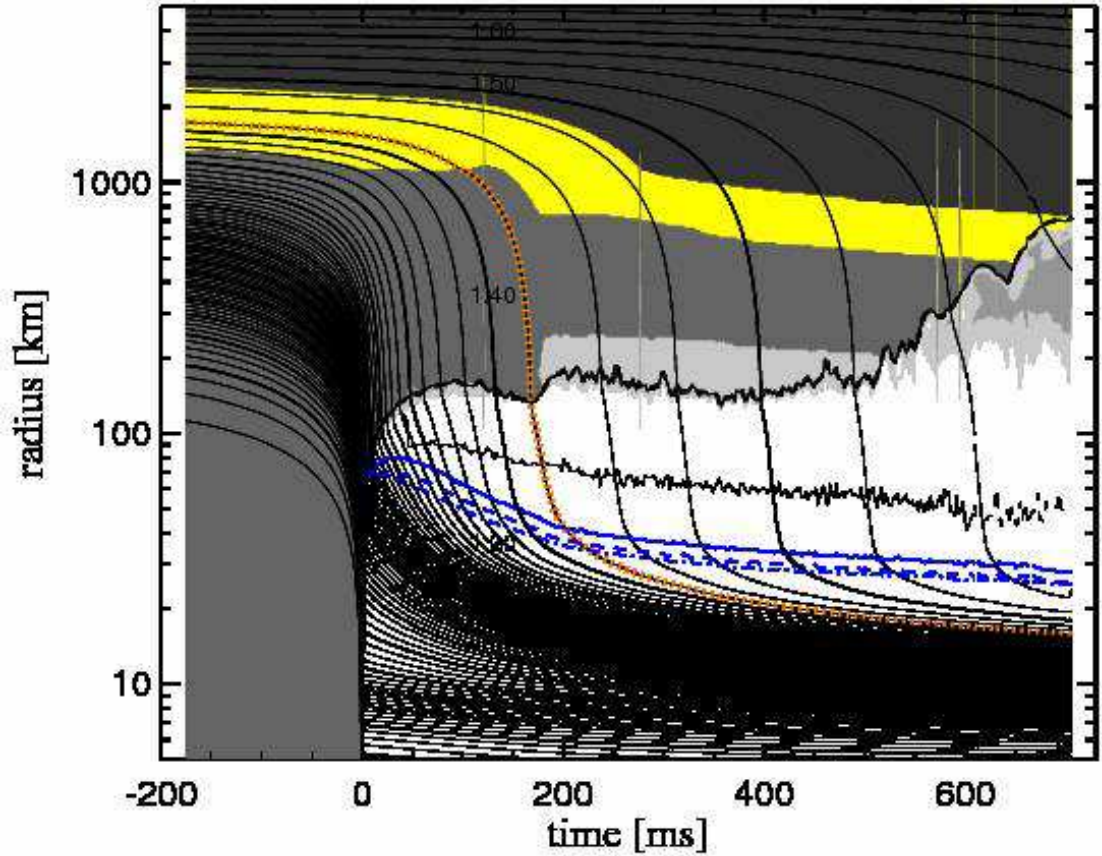}
 \caption{Time evolution of Model~M15LS-rot visualized
by mass shell trajectories. In this 2D
simulation with rotation, the mass-shell lines mark the radii of
spheres that contain certain values of the rest mass (the plot
is based on an evaluation of the mass-weighted lateral average
of the 2D data set).
They are spaced in steps of 0.025$\,M_\odot$ with bold lines
every 0.1$\,M_\odot$. The thick solid line starting at $t = 0$
denotes the mass-averaged
shock position, the blue lines represent the mean neutrinospheres
of $\nu_e$ (solid), $\bar\nu_e$ (dashed), and heavy-lepton
neutrinos (dash-dotted), the black dashed curve shows the
mean gain radius, and the location of the
composition interface between the silicon
shell and the oxygen-enriched Si-layer of the progenitor star
at 1.42$\,M_\odot$ is highlighted by a red dashed line. Different
shadings indicate regions with different chemical composition.
Dark grey marks the layer where the mass fraction of oxygen is 
larger than 10\% (which corresponds to the inner boundary of the
layers that contain significant amounts of oxygen), 
medium grey the region  where the mass fraction 
of heavy nuclei with mass numbers $A \ge 56$ exceeds 70\%, the
yellow band in between is the layer where both abundance constraints
are not fulfilled (in this region silicon and sulfur are abundant),
light grey indicates those regions
where more than 30\% of the mass is in $\alpha$-particles, and
the white areas enclosed by the shock front contain mostly free
nucleons and only a small
mass fraction (less than 30\%) of $\alpha$-particles.
At times $t \ga 600\,$ms post bounce, slightly darker grey patches 
in the light-grey postshock regions contain a mass fraction
of more than 60\% helium nuclei. This signals that the nucleon 
recombination becomes more complete and/or that the dissociation of
alpha particles to free nucleons is less complete in the matter 
expanding behind the outgoing shock because of 
low postshock temperatures when the shock reaches larger radii.
Note that compressional heating triggers nuclear burning 
(described in our simulations by a ``flashing treatment'',
see Sect.~\ref{sec:numerics}) and 
leads to changes of the chemical composition in the infalling 
stellar layers. 
\label{fig:rotmod1}}
\end{figure*}

\begin{figure*}
\plotone{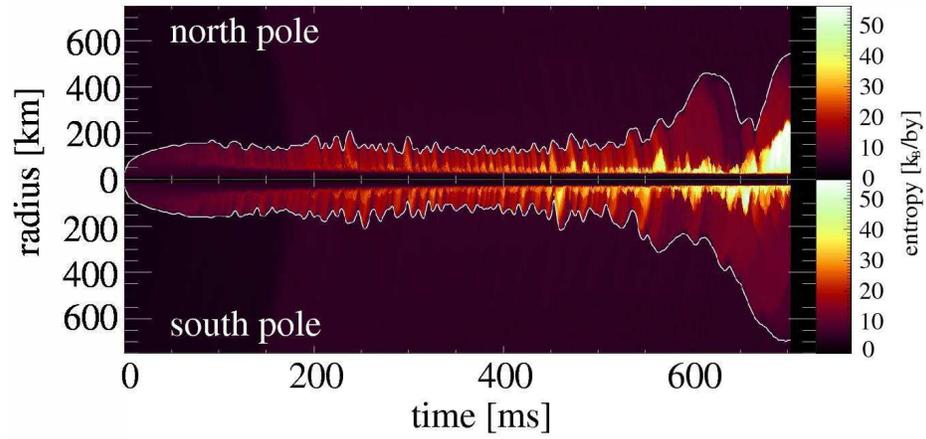}
 \caption{Radial positions of the shock near the north
and south poles of Model~~M15LS-rot as functions of 
post-bounce time (white lines).
The color coding represents the entropy per nucleon of the
stellar gas. The quasi-periodic, bi-polar shock expansion and
contraction due to the SASI can be clearly seen.
\label{fig:rotmod2}}
\end{figure*}

\begin{figure*}
\plottwo{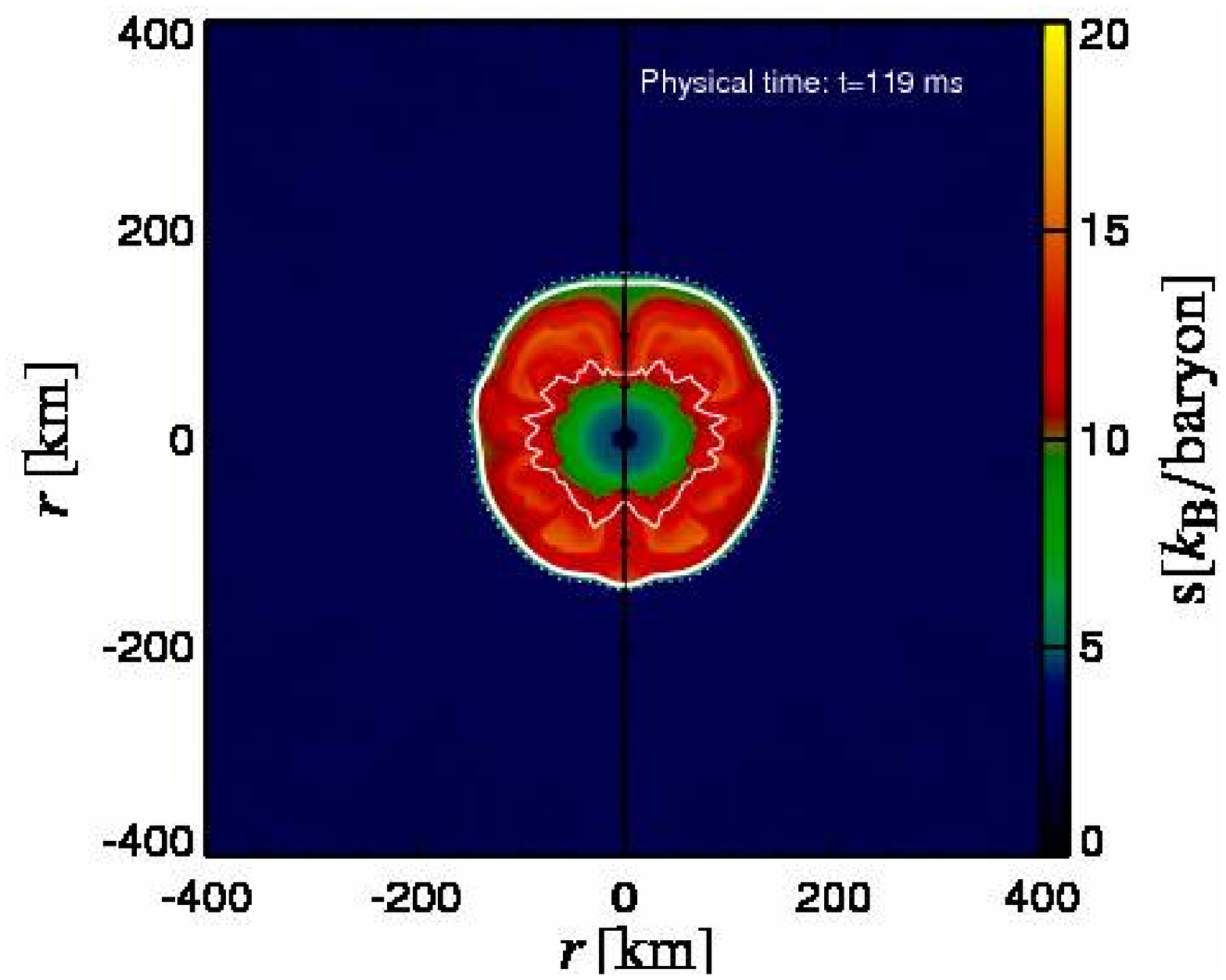}{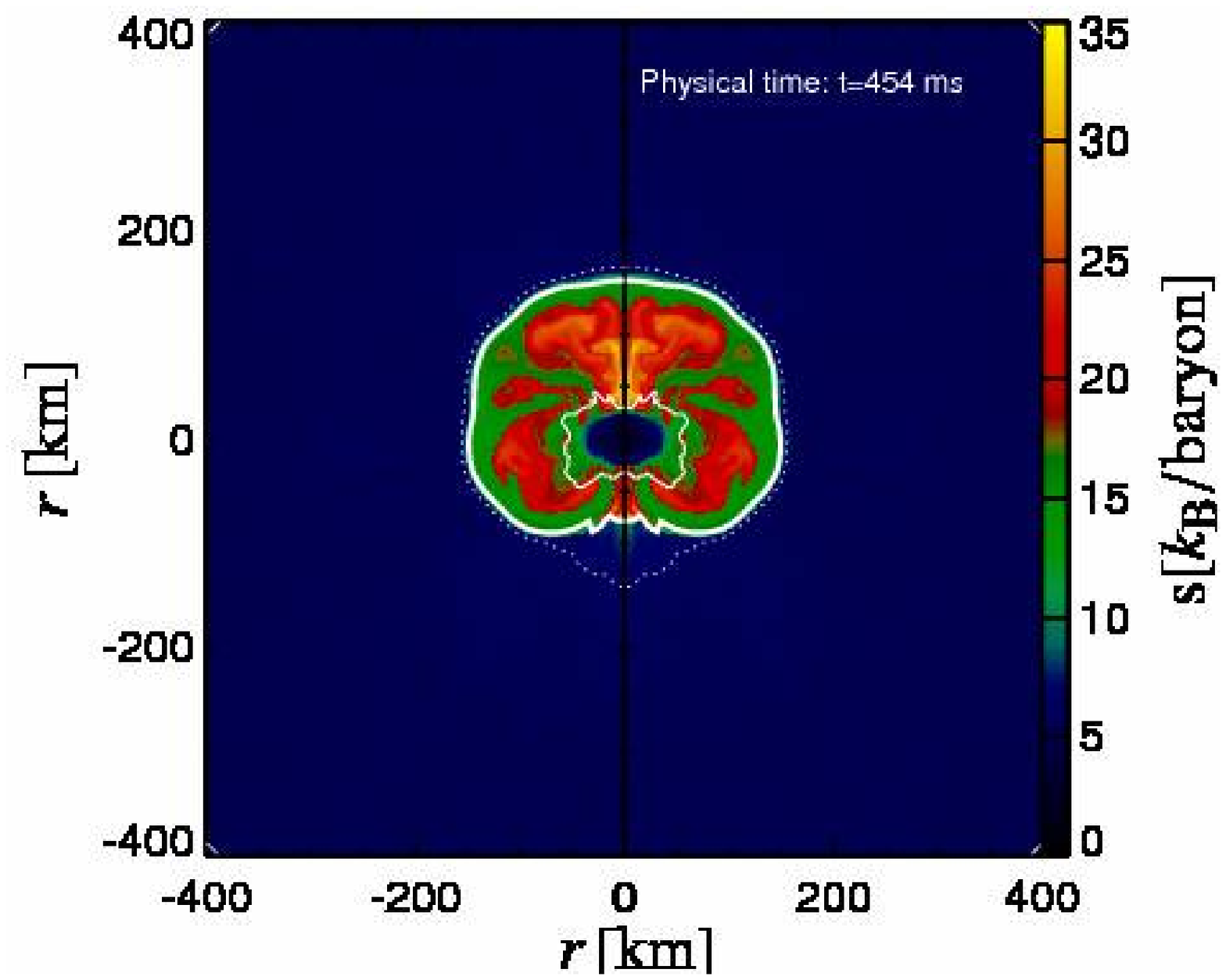}
\plottwo{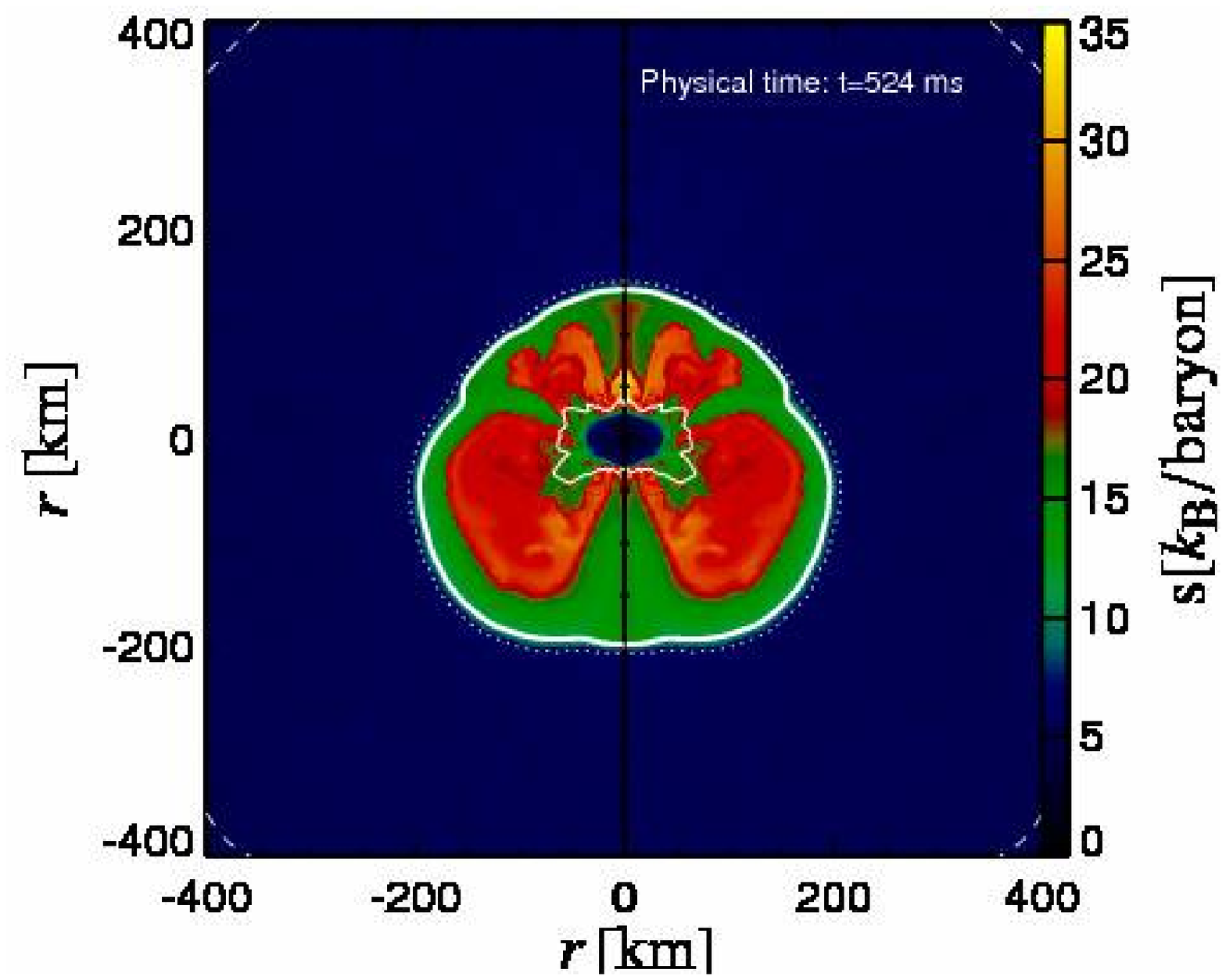}{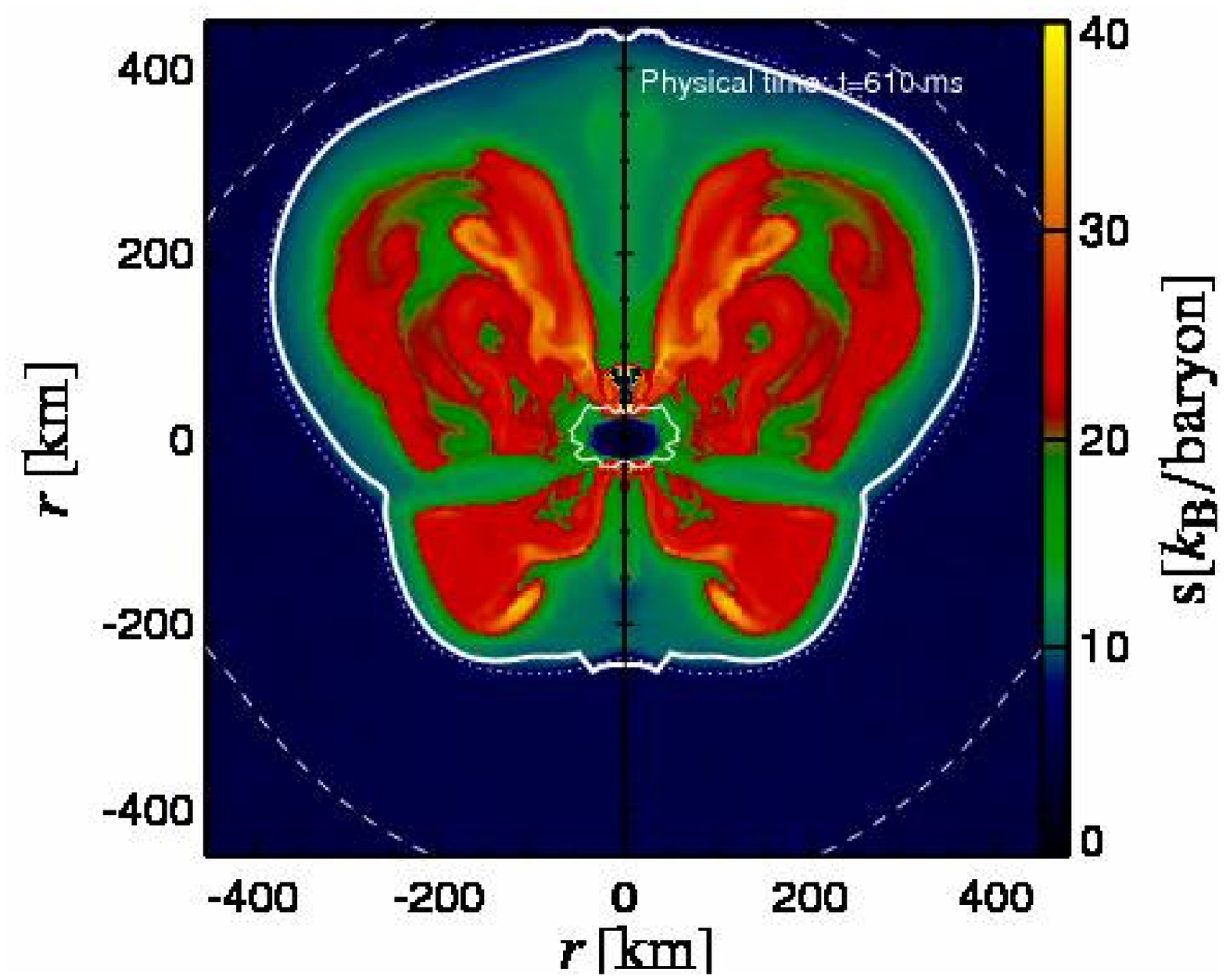}
\plottwo{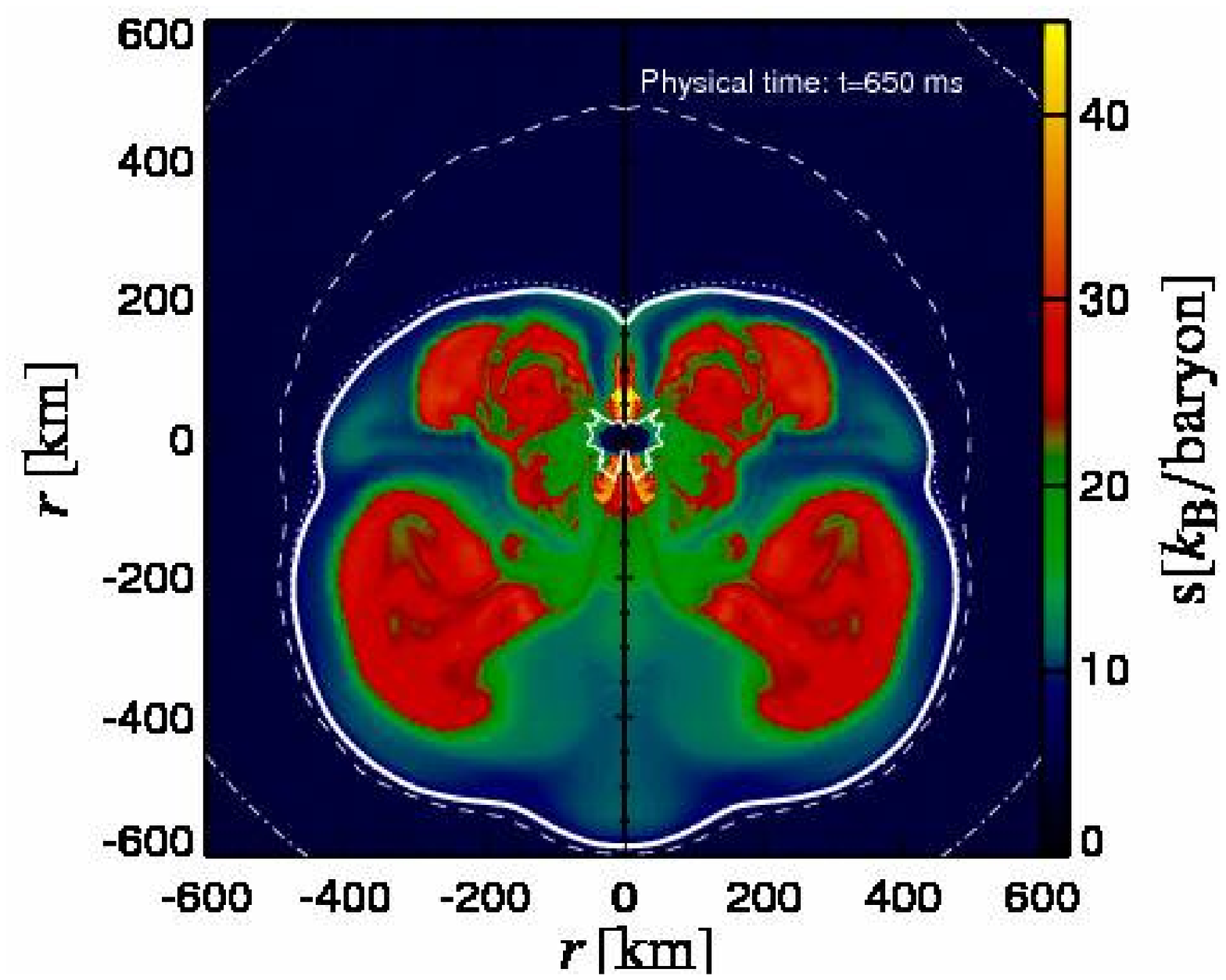}{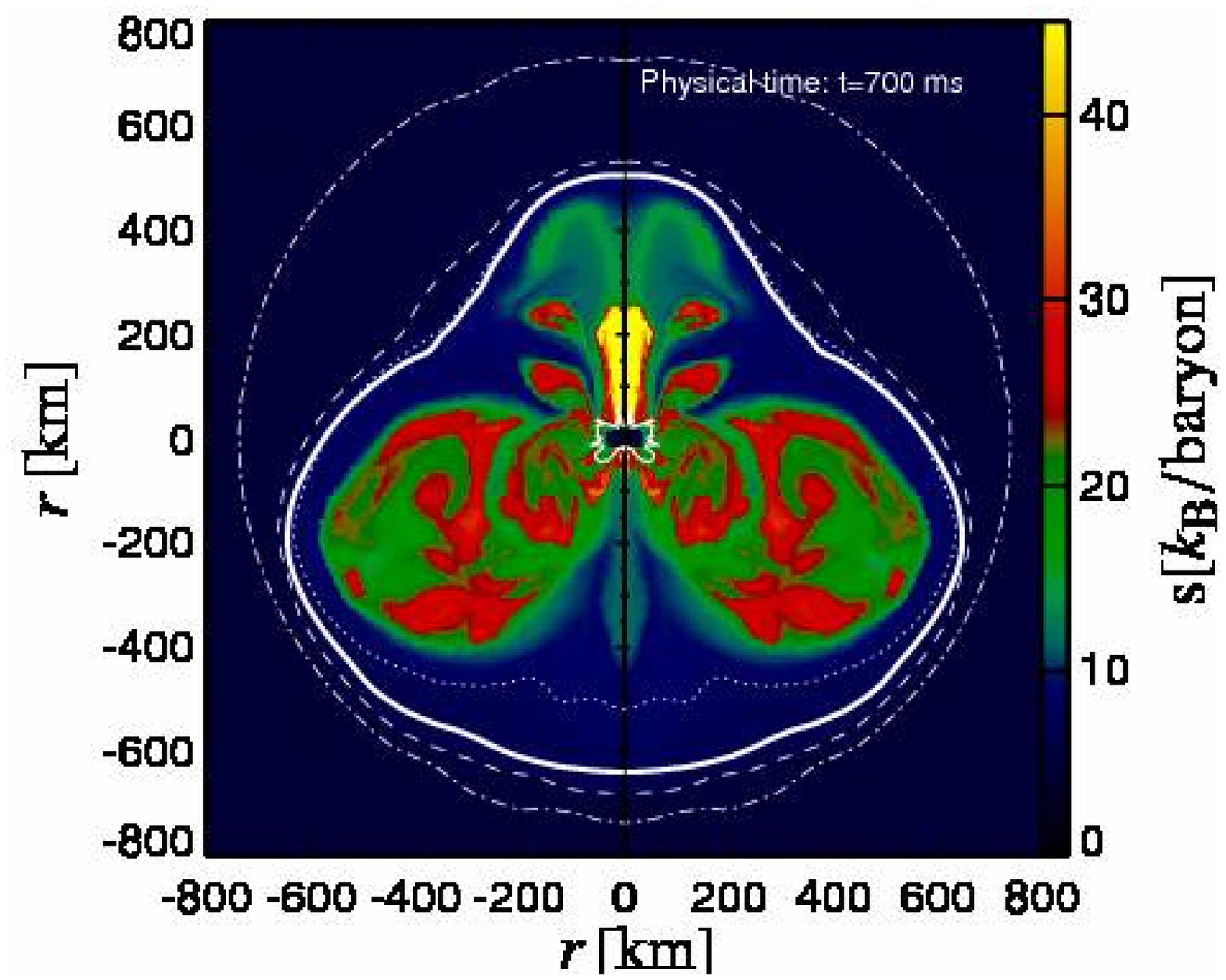}
\caption{Six snapshots from the post-bounce
evolution of Model~M15LS-rot. The color coding represents the
entropy of the stellar gas. The shock is visible as deformed
sharp discontinuity between low-entropy, infalling matter 
in the upstream region and high-entropy, boiling matter 
behind the shock; its position is highlighted by a bold, 
solid white contour. The top left plot
shows the entropy distribution at $t= 119\,$ms after bounce,
about 40$\,$ms after the postshock convection has reached the
nonlinear regime and the shock develops first small nonsphericities. 
The top right and middle left plots ($t = 454\,$ms and
and 524$\,$ms after bounce, respectively) demonstrate the presence 
of very strong bipolar oscillations due to the SASI, the middle
right plot ($t = 610\,$ms p.b.) displays the beginning of a rapid
outward expansion, and the lower two plots (for $t = 650\,$ms 
and 700$\,$ms post bounce) show the onset of the explosion with a
largely aspherical shock that possesses a dominant $l=1$ deformation 
mode. Note that the radial scale was adjusted in the last three
snapshots and that the contracting nascent neutron star
exhibits a growing prolate deformation because of the rotation
considered in this simulation. The thin, solid white line in each panel
marks the direction-dependent location of the gain radius, and the
thin dotted, dashed, and dash-dotted white lines indicate the 
inner boundaries of the regions where iron-group elements, 
silicon, or oxygen, respectively, dominate the composition 
(the contours are defined by mass fractions of 30\% iron-group
elements, 30\% silicon, and 10\% oxygen, respectively). 
In some of the panels not all these composition interfaces
are located within the plotted area, and the iron-dissociation
line or the iron-silicon interface can (at least partly) overlap
with the shock contour. We point out that the rotation of the
model is so slow that the composition interfaces in the preshock
region exhibit no visible centrifugal deformation.
\label{fig:snapshots}}
\end{figure*}

\begin{figure}
\epsscale{0.45}
\plotone{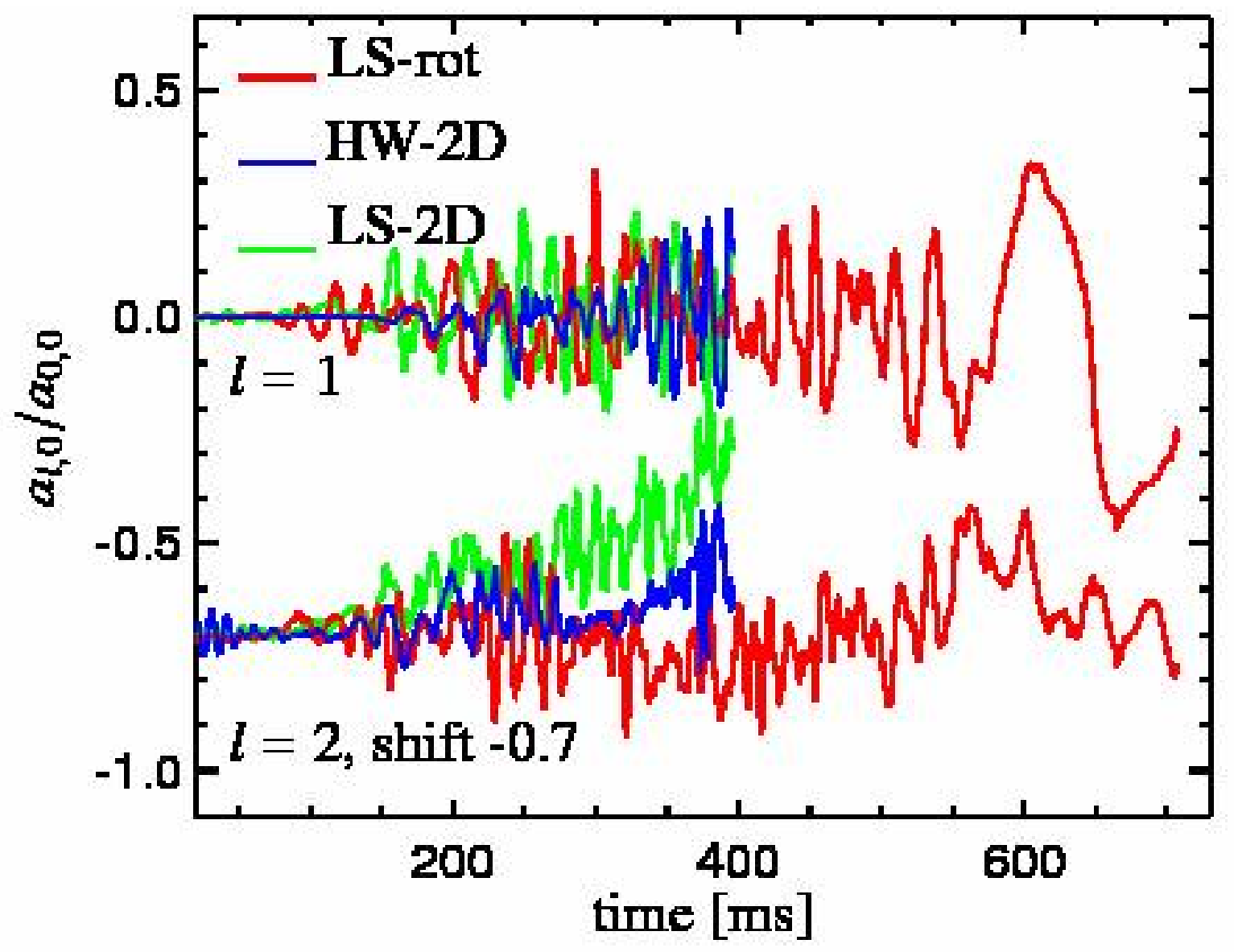}
\plotone{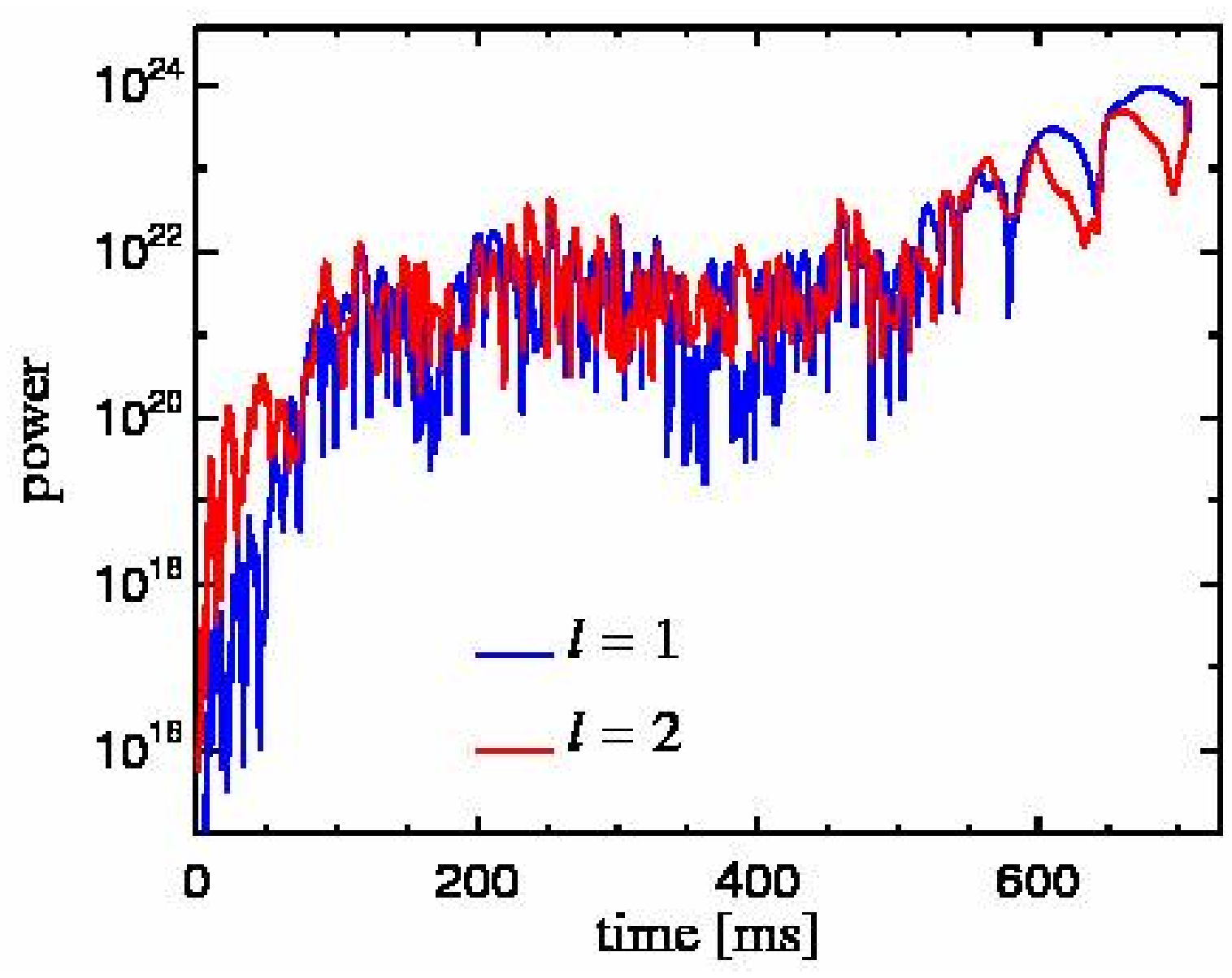}
\plotone{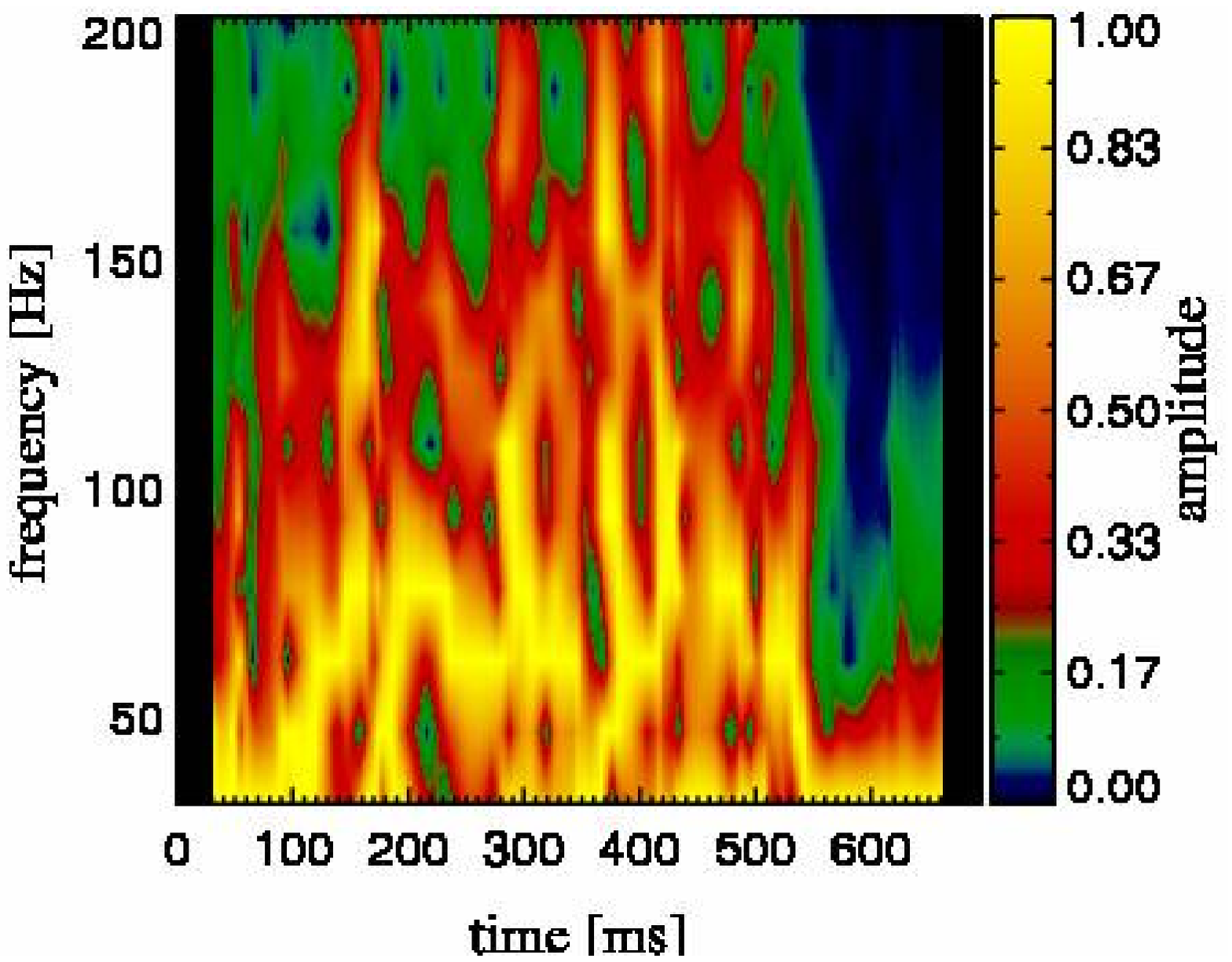}
 \caption{SASI and convective activity in Model~M15LS-rot 
versus post-bounce time.
{\em Top:} Coefficients of the dipole ($l=1$) and 
quadrupole ($l=2$) modes of the spherical harmonics expansion
of the angle-dependent shock position, normalized
to the amplitude of the $l=0$ mode. For better visibility the
curve of the $l=2$ mode is shifted downward by 0.7 units.
The results for Models~M15LS-2D and M15HW-2D are also shown
for comparison.
{\em Middle:} Power of the $l=1$ and $l=2$ modes of the 
spherical harmonics expansion of the fractional pressure 
variations $[P(r,\theta)-\left\langle P(r,\theta)
\right\rangle_\theta]/\left\langle P \right\rangle_\theta$, 
integrated over the volume between average electron 
neutrinosphere and average shock radius.
{\em Bottom:} Frequency spectrum versus time of the combined
power in $l=1$ and $l=2$ modes shown in the middle panel.
The fourier analysis was performed every two milliseconds,
sampling information in time windows of 64$\,$ms width
(which leads to discrete frequencies of 15.6$\,$Hz and
multiples), and the spectra were normalized on each time slice. 
\label{fig:modes}}
\end{figure}

The middle panel of Fig.~\ref{fig:modes} displays the corresponding
time evolution of the total power in the $l=1$ and $l=2$ modes,
computed as the volume integral of the squared dipole and 
quadrupole amplitudes of the fractional pressure variations 
between average neutrinosphere radius $R_\nu$ and average 
shock radius $R_{\mathrm{s}}$ (cf.\ Blondin \& Mezzacappa 2006),
\begin{equation}
{\mathrm{Power}}(l;t)
\,\equiv\,2\pi \int_{R_\nu}^{R_{\mathrm{s}}} {\mathrm{d}}r\,r^2 \,
a_{l,0}^2(r,t) \ ,
\label{eq:power}
\end{equation}
where $a_{l,0}(r,t)$ are the amplitudes of the spherical harmonics 
expansion of the normalized pressure fluctuations according to
\begin{equation}
{P(r,\theta,t)-\left\langle
P(r,\theta,t)\right\rangle_\theta \over \left\langle 
P(r,\theta,t)\right\rangle_\theta}\, = \, 
\sum_{l=0}^\infty a_{l,0}(r,t) {\cal P}_l^0(\cos\theta) 
\label{eq:expansion}
\end{equation}
with ${\cal P}_l^0(\cos\theta)$ being the Legendre polynomials 
(the index for the azimuthal modes is $m=0$ because of the axial 
symmetry of our 2D models). The combined power of the two
lowest modes turns out to reflect rather sensitively the dynamical
activity in the accretion layer. One should note that during phases
of relative quiescence of the dipole mode the quadrupole mode is 
dominant and vice versa. A high level of activity is reached shortly
after convection has become strong and the SASI deformation of the
shock has set in ($t \ga 100\,$ms after bounce). The following slight
reduction of the power is a consequence of the shock retraction between
100 and 150$\,$ms post bounce. When the jump in the entropy, density, 
and mass accretion rate associated with the 
composition interface between the Si-layer and the oxygen enriched 
Si-shell of the progenitor reaches the shock at
$\sim$170$\,$ms after bounce, transient shock inflation is triggered
(cf.\ Figs.~\ref{fig:1dresults}, \ref{fig:rotmod1}, \ref{fig:rotmod2}, 
and \ref{fig:shockpositions}). As a consequence, the SASI power increases
again before it decays once more during another period of shock 
contraction. At $t \ga 400\,$ms a phase of basically continuous, 
slow expansion of the average
shock radius begins and the low-mode power grows. After 500$\,$ms
until the end of our simulation at $\sim$700$\,$ms, the low SASI
modes gain another factor of 100 in power (middle panel of 
Fig.~\ref{fig:modes}), which is accompanied by an accelerating 
mean shock inflation. Finally, after
550$\,$ms, the expansion of the average shock radius speeds up strongly,
and Model~M15LS-rot enters a runaway situation
that initiates an explosion 
(Figs.~\ref{fig:rotmod1}, \ref{fig:rotmod2}, \ref{fig:snapshots}, 
and \ref{fig:shockpositions}). The shock 
acceleration is supported by the onset of 
nucleon recombination to $\alpha$-particles in the hot-bubble 
medium or/and by only partial dissociation of $\alpha$-particles
in swept-up matter as the shock reaches radii beyond 200$\,$km and 
temperatures of less than $\sim$1$\,$MeV are present behind the 
shock (cf.\ the light-grey region
behind the shock in Fig.~\ref{fig:rotmod1}). A nuclear binding energy 
of about 7$\,$MeV per nucleon is either released by the formation of
$\alpha$-particles or saved when helium is not dissociated. This
increases the thermal energy of  
the stellar gas and thus adds to the energy deposited by neutrino 
heating in bringing the gas in the gain layer to a 
gravitationally unbound state (for a further discussion of this
point, see Sect.~\ref{sec:expenergy}).

During the outward acceleration of the average shock radius
at $t\ga 550\,$ms, the two-dimensional shock contour
in the axisymmetric Model~M15LS-rot develops a 
pronounced butterfly-like shape 
(middle right and bottom panels of Fig.~\ref{fig:snapshots}) with 
alternatingly stronger expansion in either the northern or the southern
hemisphere (Fig.~\ref{fig:rotmod2}) and a significantly more inflated 
southern lobe at the end of the simulated evolution. The nearly 
unipolar onset of the explosion indicates the dominance of the $l=1$
SASI mode, consistent with the fact that the amplitude of the
$l=1$ component is clearly dominant in the upper and middle
panels of Fig.~\ref{fig:modes} at $t \ga 600\,$ms. 
The swelling high-entropy plumes of neutrino-heated gas in
both hemispheres push the shock outward
and create a kink in the shock surface near the equator. The
kink feature causes accreted gas to hit the shock at an oblique
angle and therefore to be deflected at passing the shock. The
deflected gas streams from the northern and southern hemispheres
collide and form a long-lived and waving
equatorial downdraft. This accretion funnel collects the major
fraction of the equatorially infalling gas and channels it towards
the neutron star with supersonic velocities. The existence and 
stability of this feature has important consequences for the
long-lasting energy input by neutrino heating into the developing 
explosion (see Sect.~\ref{sec:expenergy}). Gas with low angular
momentum, however, collapses closer to the rotation axis and 
creates very stable
polar downflows in both hemispheres. These polar downdrafts are
a common feature of all 2D simulations with rotation 
and their durability and
universality can be understood by the Solberg condition, which
says that a situation in which the specific angular momentum
increases with distance from the rotation axis is stable 
(see the results and discussion in Scheck et al.\ 2006).

Rotation, although ``modest'' in the sense discussed in 
Sect.~\ref{sec:models}, has an important influence in 
Model~M15LS-rot, but it is not crucial for the evolution
after bounce and the final outcome obtained in this simulation.
This will be further discussed in Sect.~\ref{sec:comparison} by
comparing the different models of our set of runs. Until the
end of the computation (at $\sim$615$\,$ms), the neutron star 
in Model~M15LS-rot
has contracted to an average radius of 26$\,$km (compared
to 21$\,$km for the 1D Model~M15LS-1D) with a pole-to-equator
radius ratio of 0.65 (eccentricity of 0.76). It rotates strongly
differentially with an average period of 3$\,$ms. In the gain
layer the rotation period is typically longer than 10$\,$ms.

\subsubsection{Runaway conditions and explosion criteria}
\label{sec:runaway}

At $t \ga 350\,$ms Model~M15LS-rot evolves gradually and 
steadily towards a situation that is more and more favorable for
an explosion. This is clearly visible in a number of quantities.
Not only the continuous expansion of the shock front 
(Figs.~\ref{fig:rotmod1}, \ref{fig:rotmod2}, and \ref{fig:shockpositions})
but also the mass, energy, and neutrino heating in the gain layer
reveal the trend that the explosion conditions become better
at later times.

Two timescales are crucial for the behavior of the gain layer:
the timescale of accreted matter to be advected from the shock
to the gain radius, 
\begin{equation}
\tau_\mathrm{adv}\, \equiv\, {R_\mathrm{s} - R_\mathrm{g}
\over |\left\langle v_r \right\rangle|} \ ,
\label{eq:tauadv}
\end{equation}
and the neutrino heating timescale of the gas in this region,
\begin{equation}
\tau_\mathrm{heat}\,\equiv\,
{|E_\mathrm{bind}|[R_\mathrm{g},\,R_\mathrm{s}] \over
                      Q_\mathrm{heat}} 
\label{eq:tauheat}
\end{equation}
Here $R_\mathrm{g}$ denotes the average gain radius, 
$\left\langle v_r \right\rangle$ is the average
postshock velocity, $E_{\mathrm{bind}}$ the total energy of the
matter in the gain layer as sum of internal, kinetic, and gravitational
energies, and $Q_\mathrm{heat}$ is the integrated net heating rate 
by neutrinos.
Therefore $\tau_\mathrm{heat}$ measures the time it takes neutrinos to
deposit an energy equal to the binding energy
$E_\mathrm{bind}[R_\mathrm{gain},\,R_\mathrm{shock}]$ of the matter
in the gain layer, whereas
$\tau_\mathrm{adv}$ can be considered as a measure of the 
duration the gas is exposed to neutrino energy deposition.
In multi-dimensional simulations when
non-radial hydrodynamic instabilities play an important role,
Eq.~(\ref{eq:tauadv}) for evaluating the advection timescale should
be replaced by Eq.~(8) of Buras et al.\ (2006b). A ratio of
$\tau_\mathrm{adv}/\tau_\mathrm{heat} > 1$ can be considered as
indicative of favorable conditions for an explosion 
(Janka et al. 2001, Thompson et al.\ 2005, Buras et al.\ 2006b).
Of course, the definitions of both timescales contain a variety
of ambiguities and uncertainties, e.g. whether the total energy or
just the internal energy is the more suitable energy scale in 
the nominator of Eq.~(\ref{eq:tauheat}) or where and how exactly
the mean advection velocity in the denominator of Eq.~(\ref{eq:tauadv})
should be measured. Therefore the timescale ratio is only a 
rough diagnostic number that can be indicative of the presence or
absence of a generally favorable trend, but certainly
$\tau_\mathrm{adv}/\tau_\mathrm{heat} > 1$ cannot be 
taken as a rigorous quantitative criterion for the onset of the
successful blast. Murphy \& Burrows (2008) also stress this 
fact and instead recommend to monitor the residence time distribution 
function of tracer particles, which is better able to capture the 
complex flow dynamics in multi-dimensional simulations.
While this is unquestionably correct, the numerical effort of 
tracking tracer particle paths is considerably bigger than performing
an analysis of the simulations on the basis of Eqs.~(\ref{eq:tauadv})
(or its multi-dimensional analog as given by Eq.~(8) of Buras et al.\
2006b) and Eq.~(\ref{eq:tauheat}).

Figure~\ref{fig:timescales} shows that along with the growing strength of
the low SASI modes (cf.\ Fig.~\ref{fig:modes}) the advection timescale 
in Model~M15LS-rot
rises continuously after 350$\,$ms post bounce, leading to a steady
growth of the timescale ratio $\tau_{\mathrm{adv}}/\tau_{\mathrm{heat}}$
from about 0.6 to values larger that unity. The critical value of unity
is exceeded at $\sim$530$\,$ms after bounce and the ratio remains larger
than unity until the end of the simulation, i.e. for a period of time much
longer than the heating timescale $\tau_{\mathrm{heat}}\approx 20\,$ms.
This is a necessary condition for an explosion (see Buras et al.\ 2006b). 
Note that even after the shock starts a rapid acceleration at 550$\,$ms,
$\tau_{\mathrm{adv}}$ can be evaluated because there is net accretion
still going on. At the end of our simulation for Model~M15LS-rot, 
about 50\% of the matter swept up by the shock is still accreted onto
the neutron star (the fraction is larger at earlier times), 
while the rest stays in the gain layer and causes a
growth of the mass there at a rate of more than 0.1$\,M_\odot\,$s$^{-1}$
(see the top panel in Fig.~\ref{fig:heating}). The dominance of mass
inflow towards the neutron star compared to mass outflow through the 
gain radius until $t\ga 650\,$ms after bounce cannot only be
concluded from the fact that the advection timescale is still defined
at such a late time. It is also suggested by the mass shell 
trajectories in Fig.~\ref{fig:rotmod1} at this time, which indicate
that a significant fraction of the matter in the gain layer is still 
crossing the gain radius with negative velocities.

The neutrino-heating timescale of the gain layer in Model~M15LS-rot
levels off near 20$\,$ms
after $t \sim 350\,$ms (Fig.~\ref{fig:timescales}, middle panel).
This corresponds to a net rate of neutrino energy deposition 
(the ``net rate'' is defined as the rate of energy input by neutrinos
minus the rate of energy loss by neutrino reemission) of about
$4\times 10^{51}\,$erg$\,$s$^{-1}$ (Fig.~\ref{fig:heating}, middle
panel) with a mild increase with time. This increase is a consequence 
of the slowly growing mass and optical depth of the gain layer and 
of the higher mean energies of the neutrinos radiated at later times
(see Fig.~\ref{fig:1dresults} and \ref{fig:rotneutrinos1}).
The corresponding neutrino heating efficiency, defined as ratio of
the total net energy deposition to the sum of the radiated $\nu_e$
and $\bar\nu_e$ luminosities, rises from about 6\% at 350$\,$ms to
peak values of more than 8\% at times later than 500$\,$ms after
bounce.

With the accumulation of mass in the gain layer, also the internal
energy in this region grows (Fig.~\ref{fig:energies}, upper right
panel). The same is
true for the rotation energy of the accreted matter, which rises
because the gas that is accreted at later times comes from larger 
distances and has a higher specific angular momentum. The increase
of both energies, however, is rather moderate (some 10\%) compared 
to the growth of the kinetic energy of fluid motions 
in the gain layer in non-azimuthal directions. Both the energy
associated with gas flows in latitudinal direction and with gas
motions in radial plus lateral directions gain a factor of
roughly four between 350$\,$ms and 600$\,$ms (Fig.~\ref{fig:energies},
upper left plot). This, of course, reflects the increasingly more
violent SASI and convective activity in the shocked flow. These
energies are around
$10^{49}\,$erg, corresponding to $\sim$0.7$\,$MeV per nucleon,
a value that is dwarfed by the neutrino heating rate
(300$\,$MeV$\,$s$^{-1}$ per nucleon), applied for a typical
timescale $\tau_\mathrm{adv}$, and by the recombination energy
of nucleons to $\alpha$-particles. Accordingly, the total energy
per nucleon in the gain layer changes much more strongly
from a mean value of about $-7\,$MeV
at 350$\,$ms to $-3.5\,$MeV around 600$\,$ms and to only 
$-2\,$MeV at 700$\,$ms. 

Towards the end of the computed evolution of Model~M15LS-rot
a sizable and steeply rising fraction of
the mass in the gain layer has obtained a specific energy that
is either positive or only marginally negative 
(Fig.~\ref{fig:energies}, lower two panels). The energy of all
mass in the gain layer with positive total (internal plus
kinetic plus gravitational) specific energy and
positive radial velocity at that time is roughly
$2.5\times 10^{49}\,$erg and also steeply increasing
(Fig.~\ref{fig:energies}), correlated with a very steep rise of
the kinetic energies near the termination point of the simulation.

A variety of energy parameters therefore signals that 
an explosion begins to develop in Model~M15LS-rot. This can 
also be concluded from the behavior of the advection timescale
and the timescale ratio at the end of our simulation. Both show
a steep increase 
because more and more matter in the gain region attains positive 
velocities and starts expanding (Fig.~\ref{fig:timescales}). 
Since a growing fraction of the
matter that is accreted by the shock stays in the gain layer, the
mass in this layer grows and the 
gas infall rate onto the neutron star is reduced. Correspondingly,
less neutrinos are radiated from the accretion layer near the neutron
star surface. This leads to a visible reduction of the 
luminosities of $\nu_e$ and $\bar\nu_e$ and of the mean energies
of all kinds of neutrinos (Fig.~\ref{fig:rotneutrinos1}).

\subsubsection{Role of the SASI}

In the last section we argued that 
the combination of shock expansion, critical timescale
ratio, and growing mass and energy in the gain layer indicates
that Model~M15LS-rot approaches a runaway situation. The described
behavior of these quantities is very similar to the findings of
Scheck et al.\ (2008) for their exploding 
model W00F (cf.\ Figs.~19 and 20 in their paper). In the 
latter model the growth of buoyancy instabilities was initially 
damped due to the choice of small seed perturbations and a low
neutrino-heating rate in the gain layer. In contrast, the larger
perturbations that have developed in the stellar core of Model~M15LS-rot
during the collapse allow convection to begin in this simulation
as soon as neutrino heating achieves to create a convectively 
unstable situation in the gain layer. Therefore the growth of 
the SASI and of convection cannot be studied separately in the
present simulation. 

Nevertheless, the large-amplitude low-mode
shock oscillations seen in Model~M15LS-rot only $\sim$100$\,$ms after
bounce (see Figs.~\ref{fig:rotmod2} and \ref{fig:modes}) 
clearly indicate the presence
of the SASI, whose growth appears to be accelerated by the sonic and
vorticity perturbations associated with the convective activity in
the postshock region. Systematic studies by Scheck et al.\ (2008)
showed that the SASI can play an important supportive
role for neutrino-driven explosions. Scheck et al.\ (2008)
demonstrated that 
SASI shock oscillations can develop even in conditions which per se
are disfavorable for the growth of convection. Thus the SASI can assist
the delayed revival of the stalled shock in at least two ways:
(i) Large-amplitude expansion and contraction phases of the
shock lead to supersonic lateral velocities in the postshock flow
and the formation of sheets with very steep unstable entropy 
gradients. This fosters the development of secondary convection
or strengthens the ongoing convective activity. 
(ii) SASI sloshing motions as well as strong postshock convection 
do not only produce large non-radial velocity components in the
postshock flow but also push the
accretion shock to larger radii. This again reduces the average
infall velocity in the postshock layer, because the velocity in
the free-fall region ahead of the shock drops like 
$R_{\mathrm{s}}^{-1/2}$. A larger shock radius thus leads to a
significantly longer advection timescale, roughly $\tau_{\mathrm{adv}}
\propto R_\mathrm{s}^{3/2}$. 

These effects help to keep accreted matter in the gain layer for
a longer period of time (a consequence of which is the increasing
mass in this layer). This supports the energy deposition by
neutrinos and thus facilitates the explosion. So, although
the kinetic energy associated with the SASI remains negligible for
the explosion energetics, once the SASI and convection have pushed
the shock sufficiently far out, prolonged neutrino heating and 
ultimately also the additional energy release by nucleon recombination 
(or, similarly, energy savings by only partial dissociation of nuclei
in the shock-accreted gas) set up the conditions for runaway.
 
The contributing effects in the postshock accretion flow, the 
SASI, convection, and neutrino heating, however, are strongly
linked and interdependent in a complex way. In such a highly 
nonlinear situation it is therefore extremely difficult to exactly
determine the influence of each aspect individually. It is, for example,
unclear which conditions are primarily responsible 
for the growth of the SASI power 
at late post-bounce times ($t\ga 300\,$ms), and which role neutrino 
heating may play in this context. By comparing 
in Sect.~\ref{sec:comparison} Model~M15LS-rot with the
other simulations of our sample --- although it was not yet possible
to carry these other runs to equally late times as our reference 
model --- we will 
strive for a better understanding of the influence of at least some 
potentially important aspects like rotation and the contraction and
growing compactness of the forming neutron star.

\subsection{Two-dimensional 11.2$\,M_\odot$ model with explosion}
\label{sec:model11.2}

As discussed in the previous sections, the
strong amplification of the SASI and convective activity
lead to runaway conditions and the initiation of an explosion
in Model~M15LS-rot for a 15$\,M_\odot$ star.
Buras et al.\ (2006b) observed a similar situation in their
180-degree simulation of the collapse and post-bounce 
evolution of a (non-rotating) 11.2$\,M_\odot$ progenitor from 
Woosley et al.\ (2002). Different from our successful 15$\,M_\odot$ 
run, the neutrino-powered explosion was found to set
in much earlier, namely already at about 200$\,$ms after bounce.

The importance of the presence of low-mode SASI oscillations
for the development of the blast in this case was concluded 
from the comparison with a simulation that was computed with 
the same input
physics but with the lateral grid constrained to a 90-degree 
wedge around the equatorial plane of the polar coordinate system
and periodic boundary conditions. This simulation
did not show an explosion until the same post-bounce
time\footnote{Of course, this does not permit the conclusion
that an explosion could not occur later. Such a possibility, 
however, seems
disfavored because the sensible parameters for a runaway situation
(advection timescale, neutrino heating rate, mass in the gain
layer, etc.) exhibit a much more pessimistic trend than in the 
successful 11.2$\,M_\odot$ run with the 180-degree grid.}. 
In the latter setup the 
choice of the grid did not allow for the occurrence of 
$l=1$ and $l=2$ SASI modes. Interestingly, Marek (2007)
obtained an explosion for the 11.2$\,M_\odot$ progenitor when
he reran the model with a 90-degree grid extending from the
pole to the equator with reflecting boundaries, thus assuming 
equatorial symmetry and 
excluding the $l=1$ mode but giving room for the mirror symmetric
$l=2$ mode. This result was recently confirmed by Murphy \& Burrows
(2008) and shows that obviously the support of strong 
quadrupolar SASI motions makes already a significant difference
compared to a case where both lowest-order spherical harmonics
are suppressed.

Here we present some results for a continuation of the 
11.2$\,M_\odot$ explosion simulation of Buras et al.\ (2006b). 
Due to the larger timesteps that were possible in this case,
we could evolve the model further into the explosion. The
results are therefore of interest in the present context 
because they provide insight into the way the explosion 
strengthens at evolution stages that we were unable to 
reach in the 15$\,M_\odot$ run.

Figure~\ref{fig:massshells11} displays the mass-shell evolution
as deduced from the (mass-weighted) angle-averaged 2D data of
the simulation. Grey or yellow shading of different regions 
gives rough information about the nuclear composition in 
the corresponding layers. One can see that the onset of the shock
expansion coincides with only partial dissociaton of helium
nuclei to free nucleons in the postshock region. Whether this
is causal for the shock expansion or just a consequence of it,
is not finally clear. In any case, the larger abundance of 
$\alpha$-particles
reduces the consumption of thermal energy for breaking up 
strongly bound nuclei. Figure~\ref{fig:massshells11} also 
shows that the shock reaches the infall layers containing a 
mass fraction of more than 10\% oxygen at $t\approx
270\,$ms after bounce only after it
has already propagated to more than 700$\,$km. 

At that time the shock is already half
way through the supersonically collapsing inner shells
of the progenitor star and on average rushes outward with a radial
velocity of 10,000$\,$km$\,$s$^{-1}$ (Fig.~\ref{fig:expener11},
left panel).
The shock surface has a highly deformed prolate shape with an
$z:x$-axis ratio of more than 2:1 (Figs.~\ref{fig:expener11} and
\ref{fig:snapshots11}). Large lobes filled with neutrino-heated,
high-entropy gas expand into the northern and southern 
hemispheres and are continuously fed with fresh material
that is channelled in a long-lived, very stable equatorial accretion
funnel from behind the shock to the close vicinity
of the gain radius to absorb energy there from neutrinos.
The situation is very similar to what we observed in the case 
of our 15$\,M_\odot$
explosion model (see Sect.~\ref{sec:postbounce}). Until the end of 
the simulated evolution, the outward moving, hot
gas in the two big lobes (the northern one being slightly
larger) has accumulated
a positive total (internal plus kinetic plus gravitational)
energy of nearly $2.5\times 10^{49}\,$erg with a tendency of 
steep rise (Fig.~\ref{fig:expener11}, right panel). 

The pronounced asphericity has the consequence that the shock
reaches the silicon-oxygen interface near the equatorial plane 
later than in the polar directions (see the panels for 
$t = 250\,$ms and 275$\,$ms
after bounce in Fig.~\ref{fig:snapshots11}).
Therefore a wedge-like region around the equator remains
for some time, where silicon and sulfur are still present with 
higher abundances between the shock and the oxygen layer, while 
the matter swept up by the shock consists mostly of iron-group
nuclei and $\alpha$-particles. The mass-shell plot of 
Fig.~\ref{fig:massshells11}, which is constructed from the 
laterally averaged 2D data at each radius, is misleading
by the fact that this preshock
material appears to be located behind the angle-averaged shock 
radius (at post-bounce times $270\,\mathrm{ms}\la t \la 300\,$ms).
We note that the penetration into the oxygen-rich infalling shells,
beginning at $t\sim 250\,$ms p.b.,
does not have any obvious supportive or strengthening effect on
the outgoing shock.
 
In Fig.~\ref{fig:heating11} we provide information about the 
conditions and neutrino-energy deposition in the gain layer of
the 11.2$\,M_\odot$ model. As in the 15$\,M_\odot$ case the 
mass in the gain layer increases when the shock begins its 
outward expansion. At the same time the infall (advection)
timescale of matter between the shock and the gain radius 
increases, but continues to be well defined. Again, as in the
15$\,M_\odot$ explosion model, this suggests
the presence of ongoing accretion of gas through the gain layer
to the neutron star (which can also be concluded from the 
continued contraction of mass shells in this region
in Fig.~\ref{fig:massshells11}). Shortly after the 
(net) neutrino-heating rate has reached a pronounced 
peak of about $7.5\times 10^{51}\,$erg$\,$s$^{-1}$ at 
$t\approx 70\,$ms, it makes a rapid drop to around
$3\times 10^{51}\,$erg$\,$s$^{-1}$. This decline is a 
consequence of the decay of the neutrino luminosities at 
the time when the mass infall rate onto the 
shock and the neutron star decreases. The decrease occurs when
the steep negative density gradient (and positive entropy step)
near the composition interface between
the silicon layer and the oxygen-enriched Si-layer of the 
progenitor star (near 1.3$\,M_\odot$) arrives at the shock
(at $t\approx 100\,$ms after bounce). Nevertheless, the heating
timescale shrinks essentially monotonically, which points to an
evolution of the matter in the gain layer towards an unbound  
state, i.e., the absolute value of the total gas energy in the 
numerator of Eq.~(\ref{eq:tauheat}) goes to zero.

\subsection{Explosion energy}
\label{sec:expenergy}
 
In both our 11.2$\,M_\odot$ and 15$\,M_\odot$ explosions, 
the energy of the matter in the gain layer with positive
radial velocities (``explosion energy'') 
reaches $\sim$2.5$\times 10^{49}\,$erg at the
end of the computed evolutions and rises with a very steep 
gradient (Figs.~\ref{fig:energies} and \ref{fig:expener11}). 
Therefore reliable estimates of the final explosion energy
cannot be given at this time.
For that to be possible, the simulations would have
to be continued for many hundred milliseconds
more (which is numerically a challenging task and 
currently impossible for us with the sophisticated and
computationally expensive neutrino
transport and chosen resolution). This
is obvious from the neutrino-driven explosion models 
investigated by Scheck et al.\ (2008; Figs.~9 and 10 there) 
and Scheck et al.\ (2006; Appendix~C there) and has
several reasons. (1) When rapid shock acceleration sets in,
only a smaller fraction of the mass in the gain layer starts to
attain positive total specific energy (see Fig.~\ref{fig:energies}
for our 15$\,M_\odot$ explosion model and Fig.~24 in Buras et al.\ 
2006b for the 11.2$\,M_\odot$ simulation),
and it takes several ten milliseconds until a major part of the
mass between shock and gain radius follows the outward motion
of the shock and the explosion energy reaches even only
some 10$^{49}\,$erg (cf.\ Fig.~10 in Scheck et al.\ 2008,
Fig.~\ref{fig:expener11} and Fig.~24 in Buras et al.\ 2006b
for the 11.2$\,M_\odot$ explosion, and Fig.~\ref{fig:energies}
for the exploding 15$\,M_\odot$ model).
(2) Also then the energy of the explosion grows
only gradually and initially roughly linearly with a rate of 
some 10$^{50}\,$erg per 100$\,$ms. During this phase ongoing 
accretion transports fresh gas towards the gain radius, where
a part of the gas absorbs energy from neutrinos and begins
to rise again, while the rest of the accreted gas is advected into
the cooling layer below the gain radius and is added 
into the forming neutron star. (3) After some time the downdrafts
of accreted matter may be quenched and the inflow of gas
towards regions near the gain radius may stop. Nevertheless,
more neutrino-heated gas is ejected in the neutrino-driven
wind that sheds off matter from the surface layer of the nascent
neutron star. During this phase the increase of the explosion
energy levels off, but still a significant fraction of the 
final energy may be added (though with a much lower rate) over 
timescales of several more seconds. The corresponding wind power 
is a very sensitive function of the neutrino emission properties 
(i.e., of the time-dependent luminosities and mean energies of 
the radiated neutrinos) and of the neutron star mass and radius.
Some $10^{50}\,$erg, in optimistic estimates even more than 
$10^{51}\,$erg, of energy might be pumped into the supernova
this way after the onset of the blast (see Burrows \& Goshy 1993;
Qian \& Woosley 1996; Thompson, Burrows, \& Meyer 2001).
(4) The total energy injected at the 
explosion center has to be corrected for the negative total
energy of the gravitationally bound outer stellar shells, 
which are going to be swept out in the blast instead of 
falling back to the neutron star.

Point~(2) implies that 
simultaneous accretion and outflow of neutrino-heated gas as
a generically multi-dimensional phenomenon is an essential feature 
of the supernova explosion mechanism. The rate at which accreted,
cool gas is channelled through the neutrino-heating region,
gains energy from neutrinos, and finally rises outward again 
(the process described in point~(2) above)
decides about the power that is accumulated in ejected material.
Let the corresponding outflow rate of neutrino-heated gas
be a fraction $\zeta$ of the mass 
accretion rate $\dot M_{\mathrm{acc}}$ through the shock,
i.e., $\dot M_{\mathrm{out}} = \zeta \dot M_{\mathrm{acc}}$.
The rate at which energy is pumped into the ejecta can then
be estimated very roughly as
\begin{eqnarray}
\dot E_{\nu} &\,\sim\,& \zeta \dot M_{\mathrm{acc}}
                     \dot q_\nu \tau_{\mathrm{adv}}  \nonumber \\
   &\,\sim\,& 2\times 10^{51}\,\mathrm{{erg\over s}}\ 
    \left({\zeta\over 0.5}\right)
    \left({\dot M_{\mathrm{acc}}\over 0.2\,M_\odot/\mathrm{s}}\right)
    \times  \nonumber \\
   &\phantom{\sim}&\phantom{2\ 10^{51}\,\mathrm{{erg\over s}}}\times
    \left({\dot q_\nu m_\mathrm{B} \over 300\,\mathrm{MeV/s}}\right)
    \left({\tau_{\mathrm{adv}}\over 30\,\mathrm{ms}}\right) \,.
\label{eq:exprate}
\end{eqnarray}
Here $m_\mathrm{B}$ is the baryon mass and $\dot q_\nu$ the 
average net neutrino energy deposition rate
per unit of mass in the gain layer, which can be estimated by the 
ratio of the net neutrino heating rate in the gain layer,
$\dot E_{\nu,\mathrm{gain}}$ (middle panel of Fig.~\ref{fig:heating}),
to the mass $M_{\mathrm{gain}}$ in the gain layer (upper panel of 
Fig.~\ref{fig:heating}):
\begin{equation}
\dot q_\nu \,\approx\, {\dot E_{\nu,\mathrm{gain}} \over M_{\mathrm{gain}}}
\,.
\label{eq:qdot}
\end{equation}
The advection timescale 
$\tau_{\mathrm{adv}}$ through the gain region in Eq.~(\ref{eq:exprate}) 
is a measure of
the mean exposure time of the gas to neutrino heating\footnote{The
product $\dot M_\mathrm{acc}\tau_\mathrm{adv}$ turns out to be a fairly
accurate representation of the mass in the gain layer at all times after 
bounce. This fact confirms that the advection time according
to the definition in Eq.~(8) of Buras et al.\ (2006b), which is used
to evaluate $\tau_\mathrm{adv}$ in our models, is reasonable and
meaningful.}. The numerical
values used in Eq.~(\ref{eq:exprate}) are guided by the conditions
at $t\ga 600\,$ms in run M15LS-rot (see Figs.~\ref{fig:shockpositions},
\ref{fig:timescales}, \ref{fig:heating}, and Sect.~\ref{sec:runaway}),
for which we therefore estimate 
$\dot E_{\nu} \sim (1...2)\times 10^{51}\,$erg$\,$s$^{-1}$.
In the case of our explosion simulation of the 11.2$\,M_\odot$ star,
Eq.~(\ref{eq:exprate}) with the numbers taken from the latest 
evolution stage shown in  
Fig.~\ref{fig:heating11} yields a neutrino energy input rate to
the ejecta of 
$\dot E_{\nu}\sim (3...6)\times 10^{50}\,$erg$\,$s$^{-1}$, which is 
consistent with the growth of the positive energy in the gain layer
at the end of the model run (compare Fig.~\ref{fig:expener11}, right
panel).
The energy deposition rate according to Eq.~(\ref{eq:exprate}) is of the
order of the net neutrino heating rate of the gain layer, 
$\dot E_{\nu,\mathrm{gain}}$ (see Figs.~\ref{fig:heating} and
\ref{fig:heating11}).
In contrast to the latter, however, it accounts for the fact that not
all of the matter that absorbs energy from neutrinos in the gain layer
will finally be added to the outflow (see also Sect.~\ref{sec:runaway}). 
We note that the power of the outflow of neutrino-heated matter is
seriously underestimated in simulations with Newtonian 
instead of relativistic gravity, because in
the Newtonian case the luminosities and energies radiated during 
the accretion phase of the
less compact and cooler neutron star are considerably lower and
$\dot q_\nu$ is correspondingly smaller.

With $\dot E_\nu$ from Eq.~(\ref{eq:exprate}) and the 
duration $\tau_{\mathrm{acc}}$
of the accretion phase, the explosion energy, in a crude way,
can be written as
\begin{equation}
E_\mathrm{exp}\,\approx\, \dot E_\nu\tau_{\mathrm{acc}}\,+\,
E_\mathrm{wind}\,+\,E_\mathrm{burn}\,-\,E_\mathrm{bind} \ .
\label{eq:expenergy}
\end{equation}
Here the first term on the rhs side accounts for the total energy 
that is transferred by neutrinos to the ejecta during the phase
of simultaneous accretion and outflow. The second term measures
the integrated power of the neutrino-driven wind that starts 
after accretion has ceased (the phase mentioned in point (3) above).
Wind matter in contrast to accreted gas must be lifted all the
way out of the deep gravitational potential at the neutron star
surface. The
third term takes into account additional energy release by
nuclear burning in shock-heated matter when the shock wave
reaches layers of still unburned nuclear fuel, and the fourth term
corrects for the gravitational binding energy of the overlying
stellar material (as addressed in point (4) above). For stars
around 15--20$\,M_\odot$ the last two terms can be estimated to
approximately compensate each other (see Janka et al.\ 2001). 
In the neutrino-driven wind phase (when accretion of infalling 
gas onto the neutron star has ended) the gain radius retreats to a
location closely outside of the neutrinospheres (this phase is not
yet reached in both the 11.2$\,M_\odot$ and 15$\,M_\odot$ explosions 
discussed in this paper, see Figs.~\ref{fig:rotmod1} and 
\ref{fig:massshells11}).

The release of nuclear recombination energy $E_\mathrm{rec}$
when free nucleons in the neutrino-driven wind ejecta assemble to 
$\alpha$-particles and heavier nuclei, must be included as a positive
contribution to the $E_\mathrm{wind}$ term so that 
$\dot E_\mathrm{wind} = \dot E_{\nu,\mathrm{gain}} + \dot E_\mathrm{rec}$
during the wind phase. For matter
that is accreted through the shock and gets heated by neutrinos
(an effect that is accounted for in the first
term on the rhs of Eq.~(\ref{eq:expenergy})), however, such nucleon
recombination does {\em not} necessarily mean a gain in the
energy budget. The gas swept up 
by the shock and falling inward to the neutron star 
in downflows is initially composed of nuclei (iron, silicon, or 
oxygen, depending on the location of the shock, see Fig.~\ref{fig:rotmod1}).
Compressional heating within the shock and in the downflows (or 
ultimately energy deposition by neutrinos) produces 
such high temperatures that the nuclei get disintegrated, which
consumes a fair amount of dissociation energy. This energy
can subsequently be recovered in the fraction of the matter
that reexpands and gets blown out again. Different 
from the energy input by neutrinos, the recombination of
nucleons, however, yields a net energy gain only when the
nucleon recombination leads to a state with higher nuclear binding
energy per baryon. This means that the net gain of energy by nuclear 
processes in the accreted material can at most be equal to the difference 
in the binding energy of oxygen or silicon (initially present in the
accretion flow) compared to iron-group elements (the most strongly bound
nuclei), which is much smaller than the complete recombination energy
of free nucleons to nuclei.
 
For both components of the ejecta, however, for the neutrino-driven 
wind material as well as accreted and reexpanding matter, nuclear 
processes are unlikely to be the main source of the energy excess 
(and thus of the explosion energy)
that the matter carries outward in the supernova blast. On the one
hand, gas residing
in the gain layer before the explosion sets in as well as matter 
ablated from the neutron star surface in the later neutrino-wind phase
is initially at rest in the strong gravitational field
very close to the neutron star and has a largely negative total energy
(${\cal{O}}$($-10\,$MeV/nucleon) or more). 
The energy liberated by the recombination of free nucleons to nuclei 
in this matter, though large ($\sim$7--8.8$\,$MeV per nuclen), 
can typically only assist the revival of the
stalled shock but cannot account for all the energy needed to make
this matter gravitationally unbound. Neutrino energy deposition is
also needed to overcome the gravitational binding and to produce the
positive energy of the ejecta at infinity. On the other hand, gas
that is accreted from large distances, channelled through the 
neutrino-heating region, and then reexpanding outward right away is 
initially gravitationally only weakly bound (its total specific 
energy is typically only of the order of $-(1...2)\,$MeV per nucleon). 
A change of the nuclear composition in this matter by assembling
heavier nuclei from lighter ones can release only a relatively small
amount of energy (also of the order of 1--2$\,$MeV per nucleon) and 
can bring
the gas at most to a gravitationally marginally unbound energy level.
So again the ultimate energy excess that drives the supernova explosion 
has to come from another source. Neutrino energy deposition plays this
dominant role in our simulations.

In this context it is crucial to answer the question  
how long accretion can continue and how long efficient energy input 
by neutrinos according to Eq.~(\ref{eq:exprate})
can therefore apply. The time when accretion ends can be deduced in a
crude way from the requirement that at this moment the outgoing shock
(with velocity $v_\mathrm{s}$ and postshock velocity $v_2$ in the
observer frame)
must accelerate the postshock matter to escape velocity. The
corresponding radius is therefore determined by
\begin{equation}
R_\mathrm{esc}\,\approx\,{2\,G M\over v_2^2}
\,\sim\,5400\,\mathrm{km}\ M_{1.5}\, v_{\mathrm{s},9}^{-2}\ ,
\label{eq:rescape}
\end{equation}
where $M_{1.5}$ denotes the enclosed mass normalized to
1.5$\,M_\odot$, $v_{\mathrm{s},9}$ the shock velocity in 
$10^9\,$cm$\,$s$^{-1}$, and the postshock velocity $v_2$ was 
determined from the 
preshock velocity $v_1$ and the shock velocity $v_\mathrm{s}$ by
the first shock-jump condition as
\begin{equation}
v_2\,=\, (1-\beta^{-1})v_\mathrm{s}
+ \beta^{-1}v_1\,\approx\, {6\over 7}\,v_\mathrm{s} \ ,
\label{eq:vpostshock}
\end{equation}
with $\beta = \rho_2/\rho_1\approx 7$ and $v_1 \sim 0$. 
For the duration of the accretion phase after the onset of outward
shock expansion we therefore get:
\begin{equation}
\tau_{\mathrm{acc}}\,\approx\,{R_\mathrm{esc}\over v_\mathrm{s}}\,\sim\,
0.5\,\mathrm{s}\ M_{1.5}\, v_{\mathrm{s},9}^{-3}\ .
\label{eq:taccretion}
\end{equation}
This means that accretion can easily continue for half a second
or even longer, if the shock expands slowly in the beginning. 
This is confirmed by the supernova simulations of Scheck et al.\ (2006),
in some of which accretion was observed to still go on significantly
longer than one second after the explosion had started.
The explosion energy is therefore not determined by the conditions
just at the moment when the outward shock acceleration begins
or when the shock is in its early stage of expansion. For this
reason it is impossible to conclude from our simulations that the
11.2$\,M_\odot$ and 15$\,M_\odot$ explosions will finally be weak or
strong. In the multi-dimensional environment of the supernova core
with accretion and shock expansion being simultaneously present,
it is essential to take into account the 
long-lasting period of mass downflow to the nascent neutron 
star\footnote{A longer accretion phase of
the nascent neutron star after the onset of the supernova explosion
is also crucial for setting the explosion energy in case of the
acoustic mechanism as discussed by Burrows et al.\ (2007c).}.
Even though the neutrino-driven explosion sets in rather late after
bounce in our simulation, Eqs.~(\ref{eq:exprate}) and (\ref{eq:taccretion})
suggest that the nonspherical nature of the gas flow in the
supernova core, due to which the neutrino-heating region 
around the neutron star is replenished with fresh gas over a 
long period of time, may well allow the explosion energy 
$E_\mathrm{exp}\sim \dot E_\nu\tau_\mathrm{acc}$ to reach several
$10^{50}\,$erg for the 11.2$\,M_\odot$ progenitor and even a value 
around the canonical number of $10^{51}\,$erg in the case of 
our 15$\,M_\odot$ explosion model.

\subsection{Comparison of 15$\,M_\odot$ models}
\label{sec:comparison}

Comparing Model~M15LS-rot with the other 2D simulations of our
sample allows one to gain insight into various aspects that may
have important influence on the post-bounce evolution. 

\subsubsection{High-resolution Model M15LS-rot-hr}

\begin{figure}
\plotone{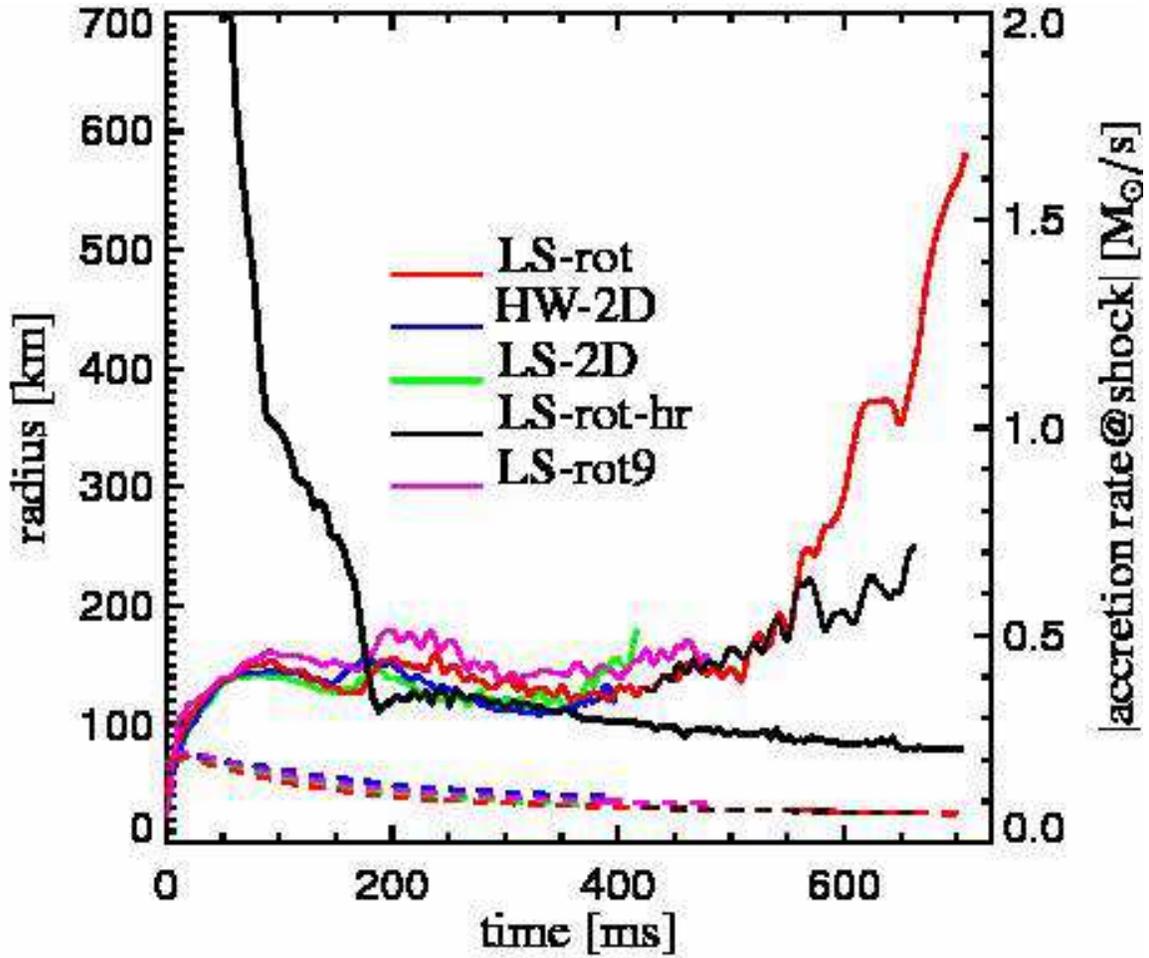}
 \caption{Average shock radii (solid) and average neutron
star radii (dashed) for the set of computed 2D models 
with the 15$\,M_\odot$ progenitor
as functions of post-bounce time (left
vertical scale of the figure). The shock position is 
defined as the arithmetical average over all directions, 
and the mean radius of the (in case of rotation, deformed)
neutron star is determined as the arithmetical mean of all
radial positions where
the density equals $10^{11}\,$g$\,$cm$^{-3}$. The bold
black curve shows the time-dependent mass accretion rate 
just ahead of the shock in Model~M15LS-rot (scale
on the right vertical axis of the plot).
The additional black line between 420 and about
670$\,$ms corresponds to Model~M15LS-rot-hr, which is a
test calculation of Model~M15LS-rot with significantly
increased radial resolution in the neutron star surface
and neutrino-heating layer.
\label{fig:shockpositions}}
\end{figure}

\begin{figure}
\epsscale{0.45}
\plotone{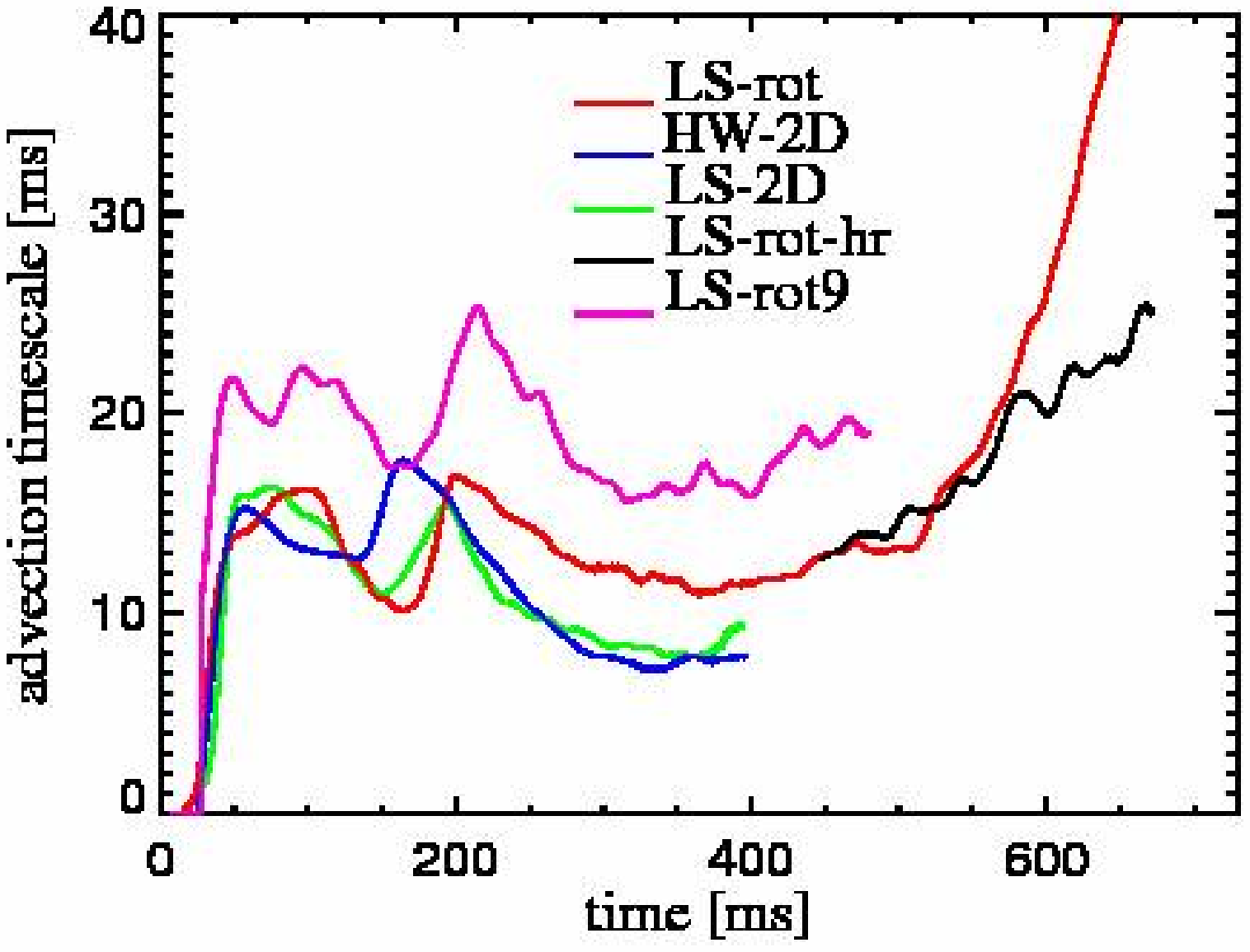}
\plotone{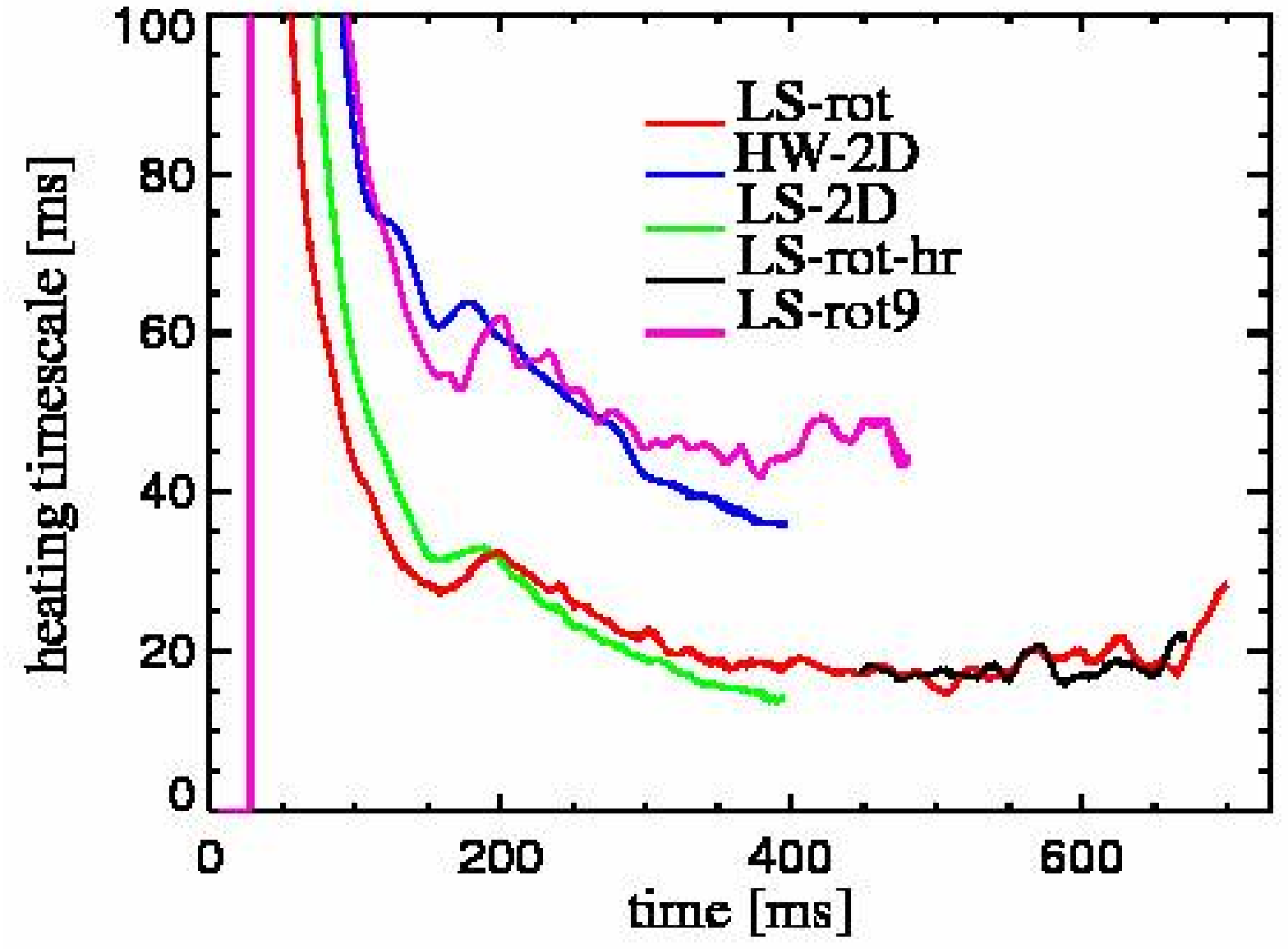}
\plotone{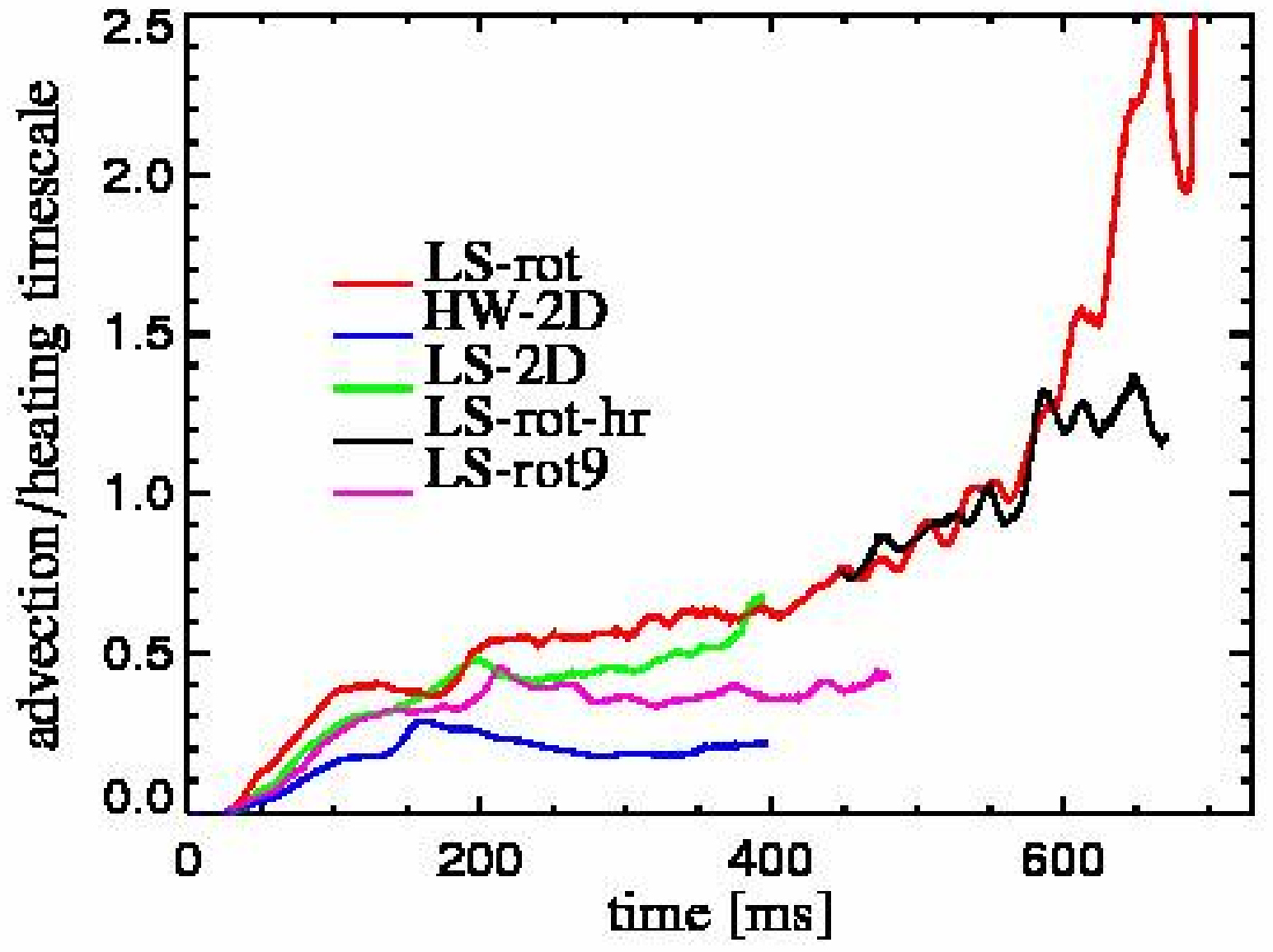}
 \caption{Dynamically important timescales for the set of
computed 2D models with the 15$\,M_\odot$ progenitor  
as functions of post-bounce time.
{\em Top:} Mean timescale for advection of accreted matter
from the shock to the gain radius.
{\em Middle:} Neutrino-heating timescale of matter in the
gain layer.
{\em Bottom:} Ratio of advection to heating timescale. 
The additional black line between 420 and about
670$\,$ms corresponds to Model~M15LS-rot-hr, which is a
test calculation of Model~M15LS-rot with significantly
increased radial resolution in the neutron star surface
and neutrino-heating layer. 
Note that towards the end of our explosion simulation 
(Model~M15LS-rot) the advection timescale and the timescale ratio
increase steeply because
a dominant fraction of the mass in the gain layer begins
to expand outward behind the shock instead of falling towards the
gain radius (this is also reflected by a steep growth of the mass
in the gain layer, see Fig.~\ref{fig:heating}).
\label{fig:timescales}}
\end{figure}

\begin{figure}
\epsscale{0.45}
\plotone{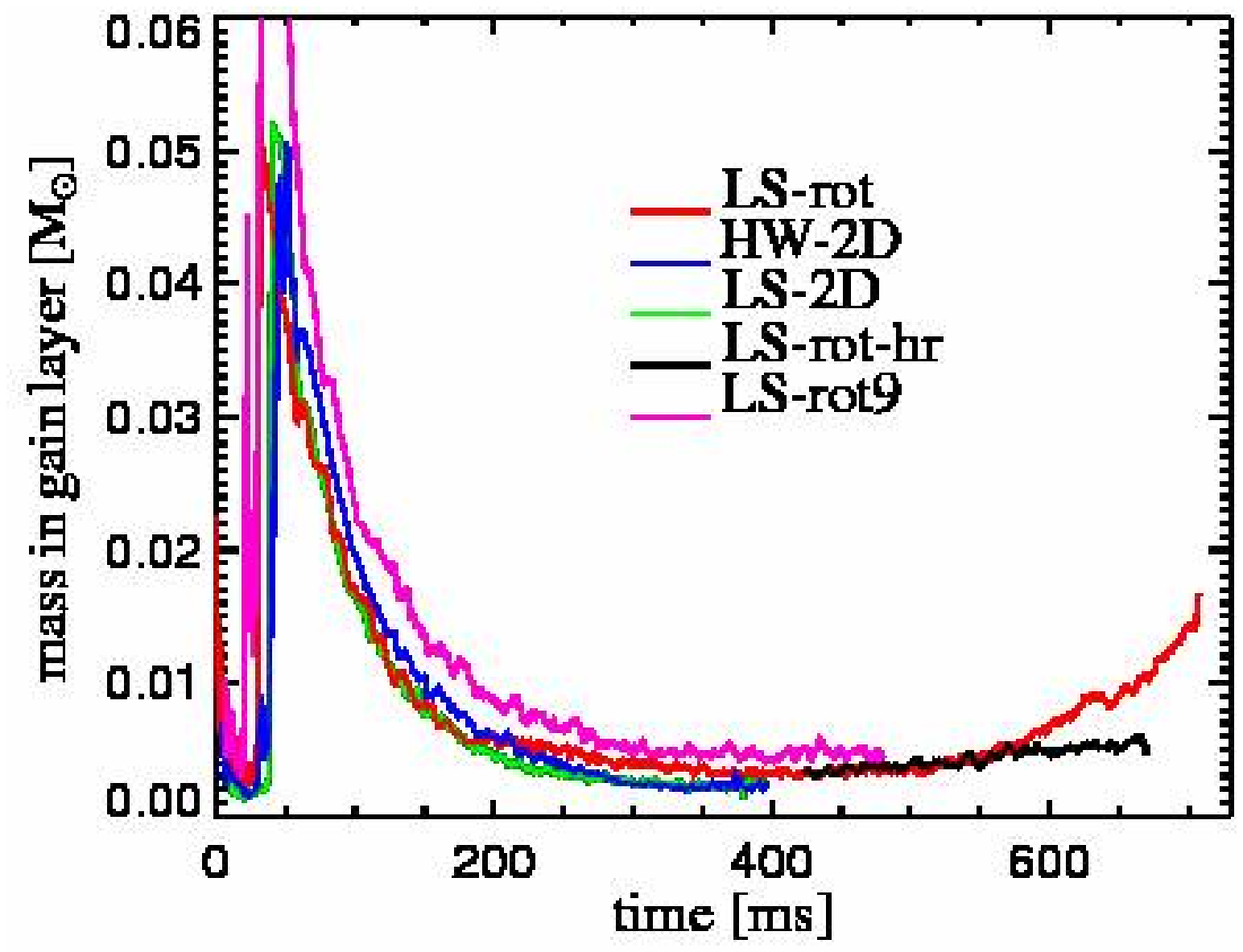}
\plotone{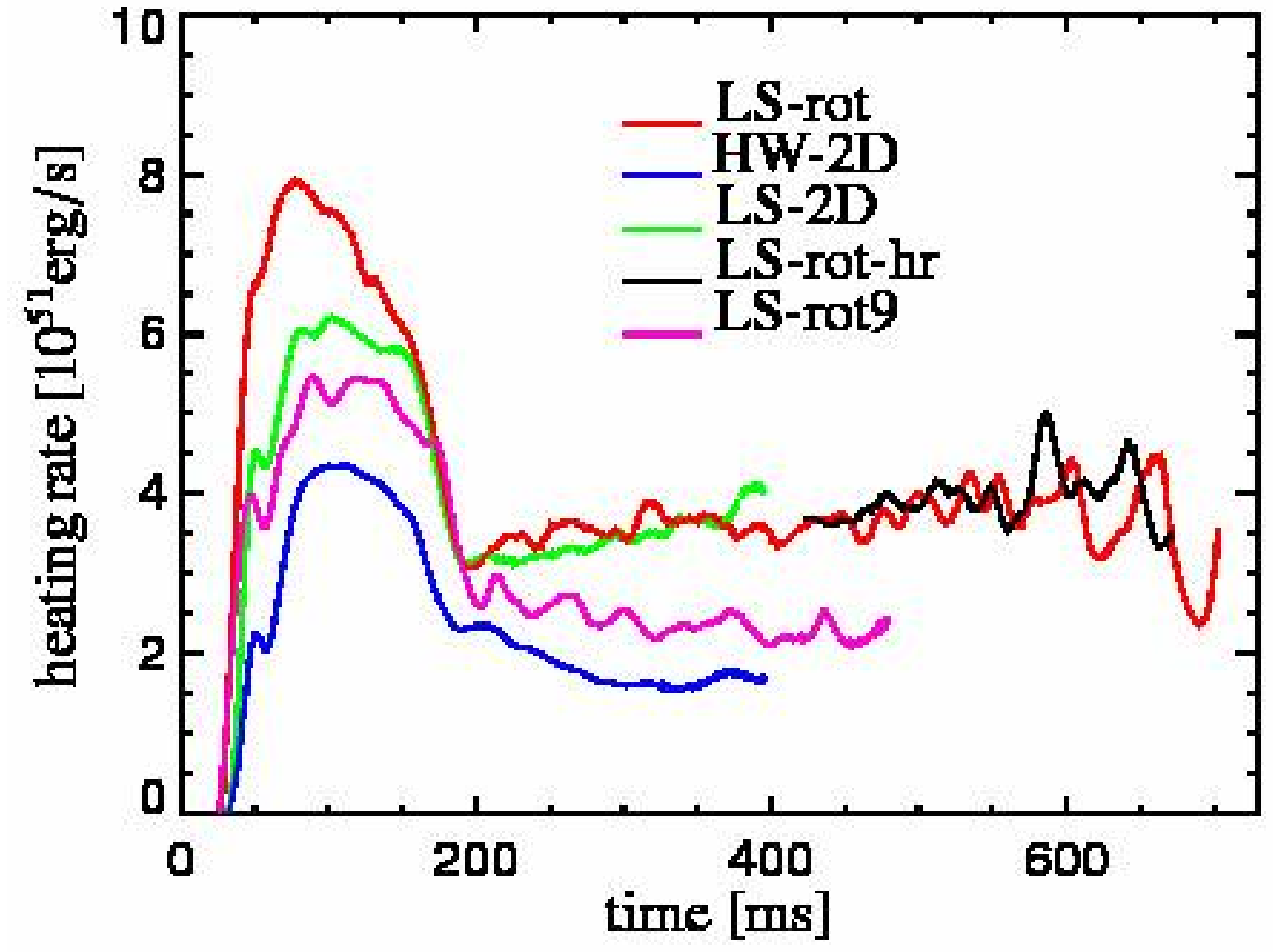}
\plotone{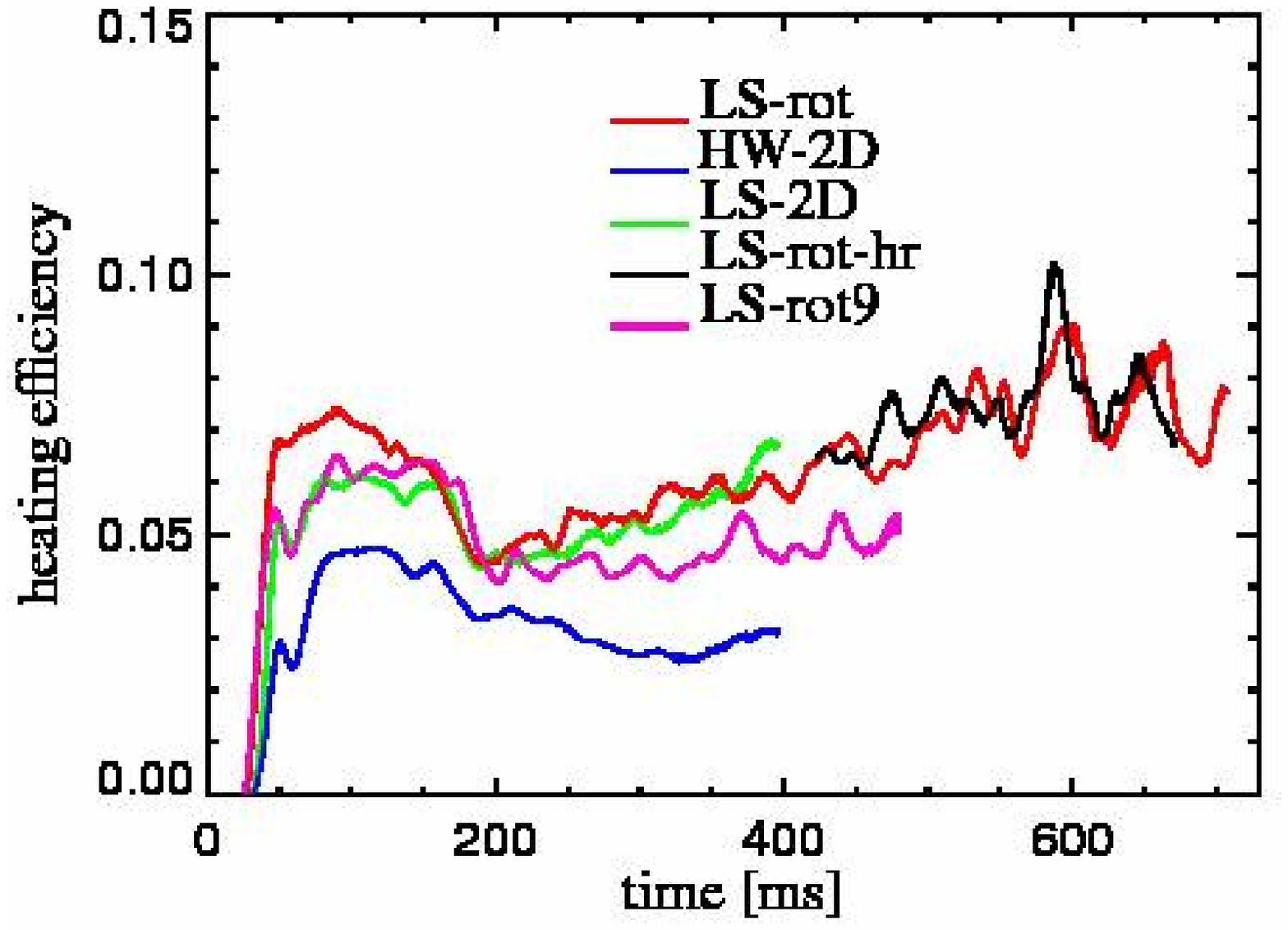}
 \caption{Neutrino heating conditions in the gain layer
for the set of computed 2D models with the 15$\,M_\odot$ 
progenitor. Mass ({\em top}),
total net rate of neutrino heating ({\em middle}), and
neutrino heating efficiency ({\em bottom})
as functions of post-bounce time. The net rate of neutrino
heating is defined as the difference of the neutrino energy deposition
rate in the gain layer and the energy loss rate by the reemission of
neutrinos. The heating efficiency
is computed as the ratio of the total net heating rate to the sum
of $\nu_e$ and $\bar\nu_e$ luminosities as seen in the 
observer frame. The additional black line between 420 and about
670$\,$ms corresponds to Model~M15LS-rot-hr, which is a
test calculation of Model~M15LS-rot with significantly
increased radial resolution in the neutron star surface
and neutrino-heating layer.
\label{fig:heating}}
\end{figure}

\begin{figure*}
\epsscale{1.}
\plottwo{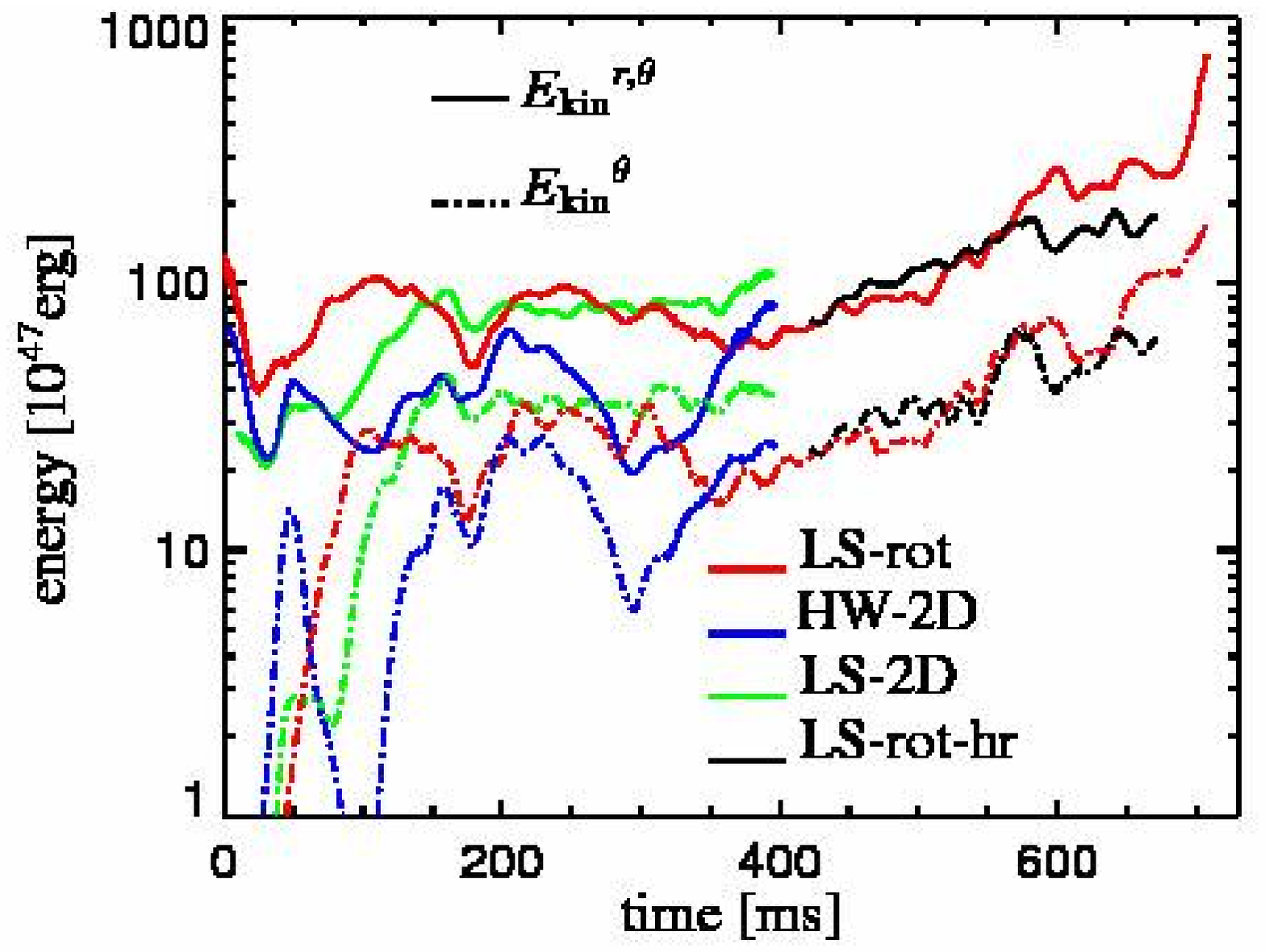}{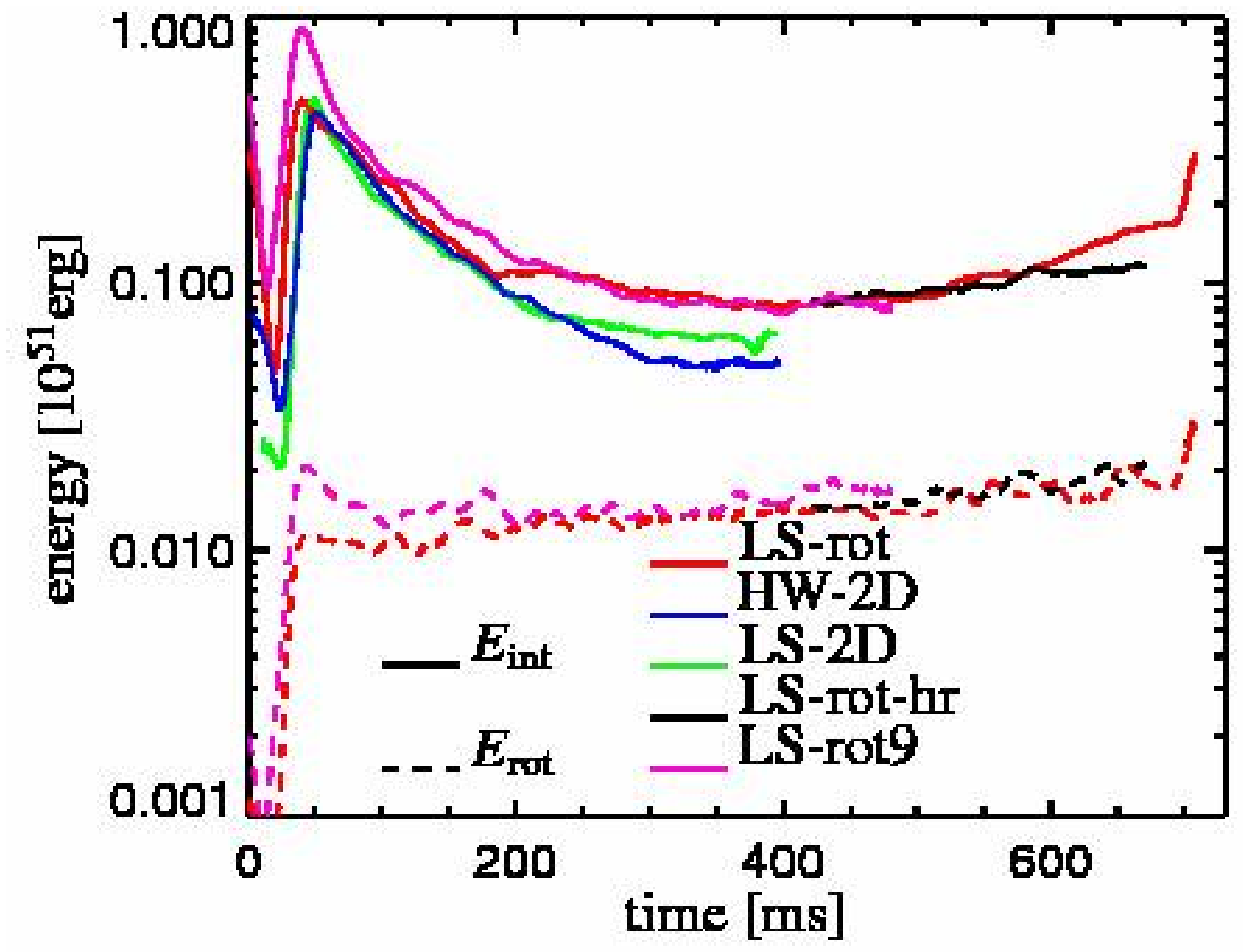}
\plottwo{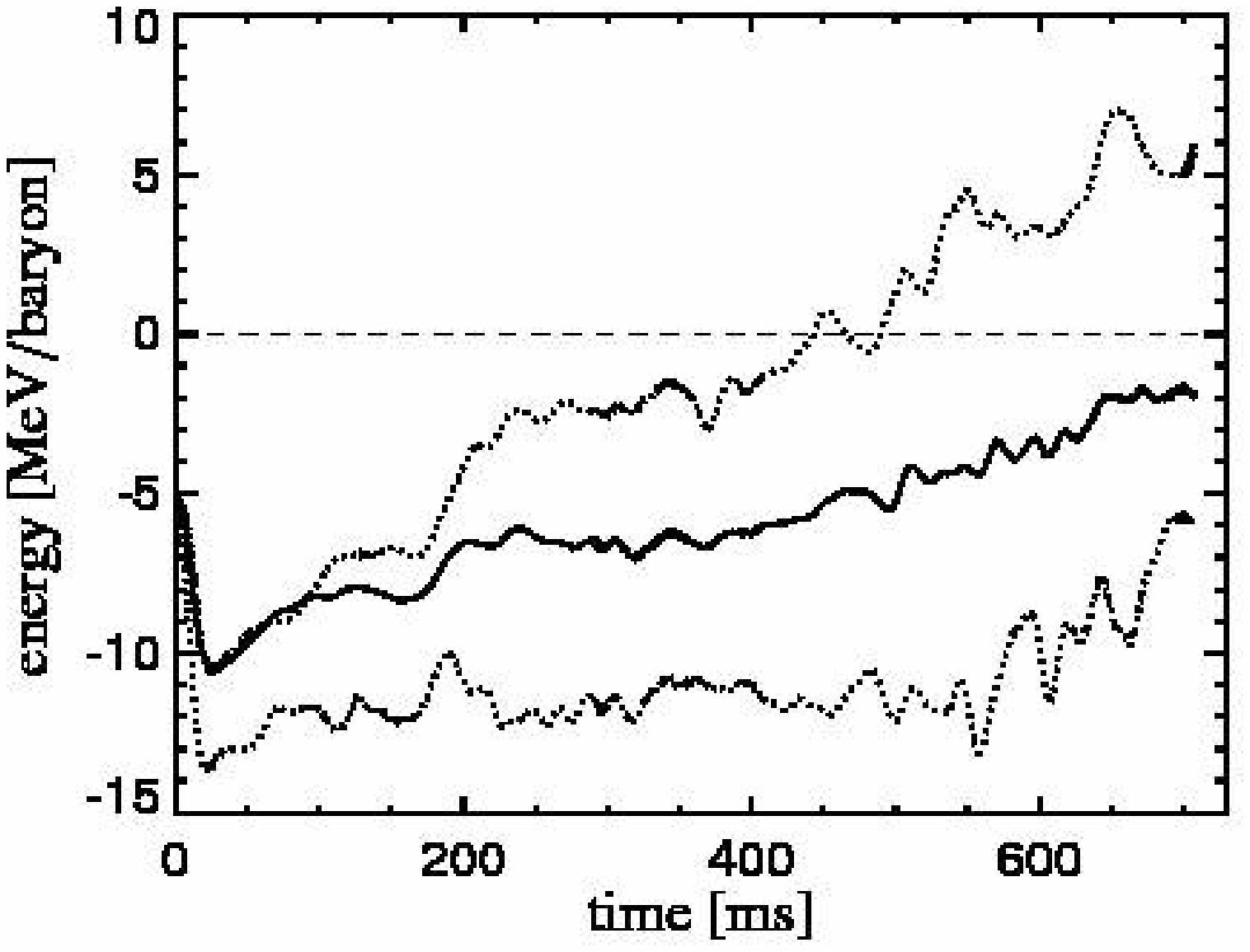}{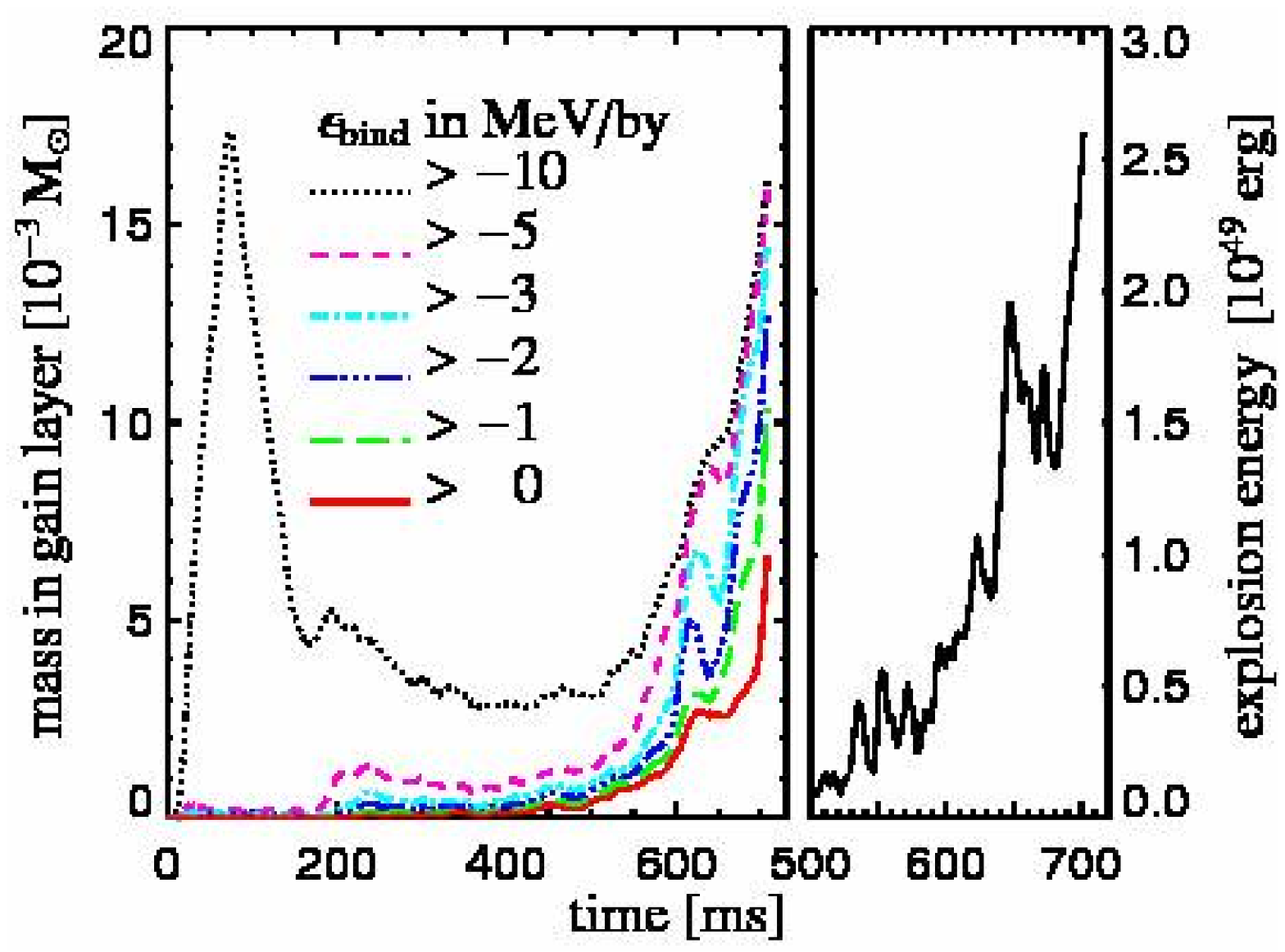}
 \caption{Evolution of energies in the 15$\,M_\odot$ models. 
The upper plots display integrated energies in the 
gain layer as functions of post-bounce time for several models 
of our set of 2D simulations, the lower plots show energies 
per nucleon and corresponding masses in the gain layer as 
functions of time for Model~M15LS-rot.
{\em Top left}: Kinetic energies $E_{\mathrm{kin}}^{r}$ and
$E_{\mathrm{kin}}^{r,\theta}$ associated with the gas motion in radial 
and in radial plus lateral direction, respectively.
{\em Top right:} Internal energy and rotational energy.
{\em Bottom left:} Mean total (kinetic, including rotational, plus 
internal plus gravitational) energy per baryon (thick line) and
energy range that contains 90\% of the mass in the gain layer of
Model~M15LS-rot (dotted).
{\em Bottom right:} Masses in the gain layer of Model~M15LS-rot
with total energies per nucleon above certain values (left panel)
and increase of the ``explosion energy'' of the model, defined 
as total energy of all matter in the gain layer with positive radial
velocity (right panel).
In the upper two plots, the additional black line between 420 and
about 670$\,$ms corresponds to Model~M15LS-rot-hr, which is a
test calculation of Model~M15LS-rot with significantly
increased radial resolution in the neutron star surface
and neutrino-heating layer. The rise of the curves for 
Model~M15LS-rot towards the end of the simulation is a signal of 
the beginning explosion, which is also accompanied by a steep 
increase of the mass in the gain layer (see Fig.~\ref{fig:heating}).
\label{fig:energies}}
\end{figure*}

\begin{figure*}
\plotone{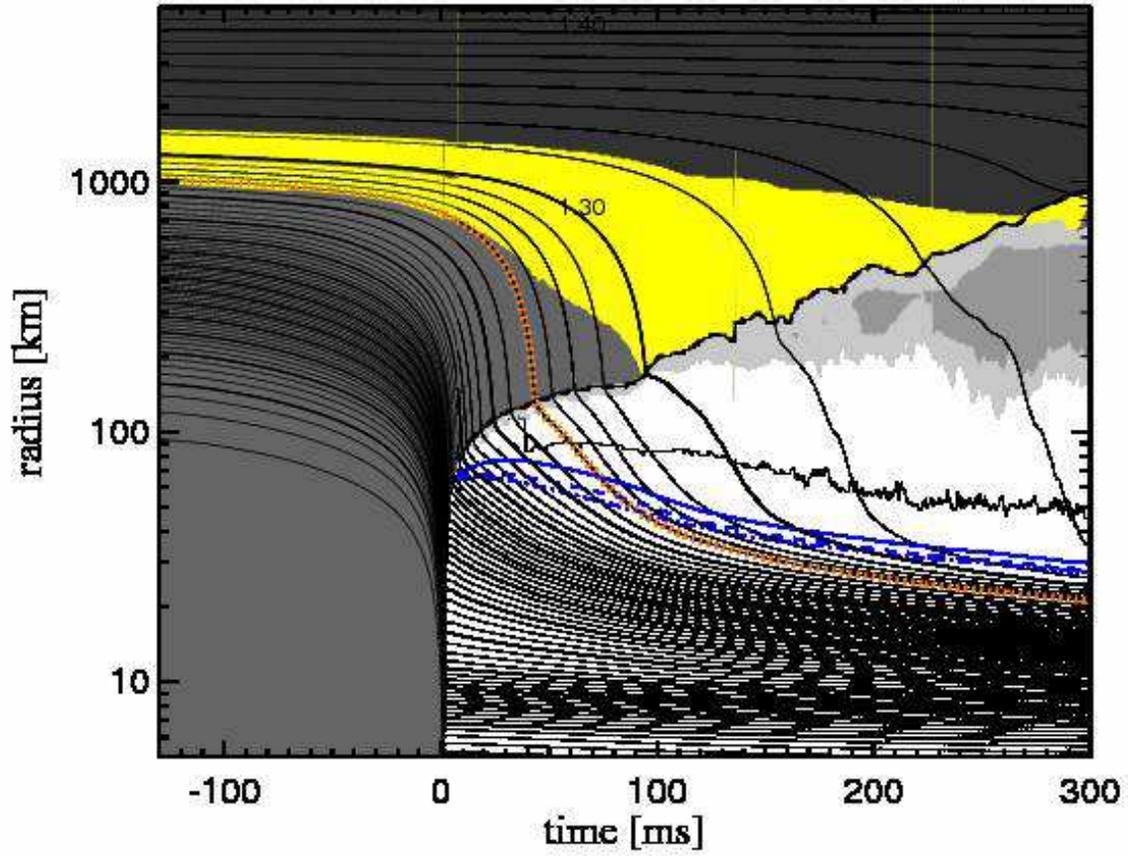}
 \caption{Same as Fig.~\ref{fig:rotmod1} but for our 2D explosion
simulation of an 11.2$\,M_\odot$ progenitor star.
Note that the mass-shell spacing outside of the red dashed
line at an enclosed mass of 1.25$\,M_\odot$ (marking the composition
interface between the silicon layer and the oxygen-enriched Si-shell)
is reduced to steps of $0.0125\,M_\odot$ instead of $0.025\,M_\odot$.
\label{fig:massshells11}}
\end{figure*}

\begin{figure}
\plotone{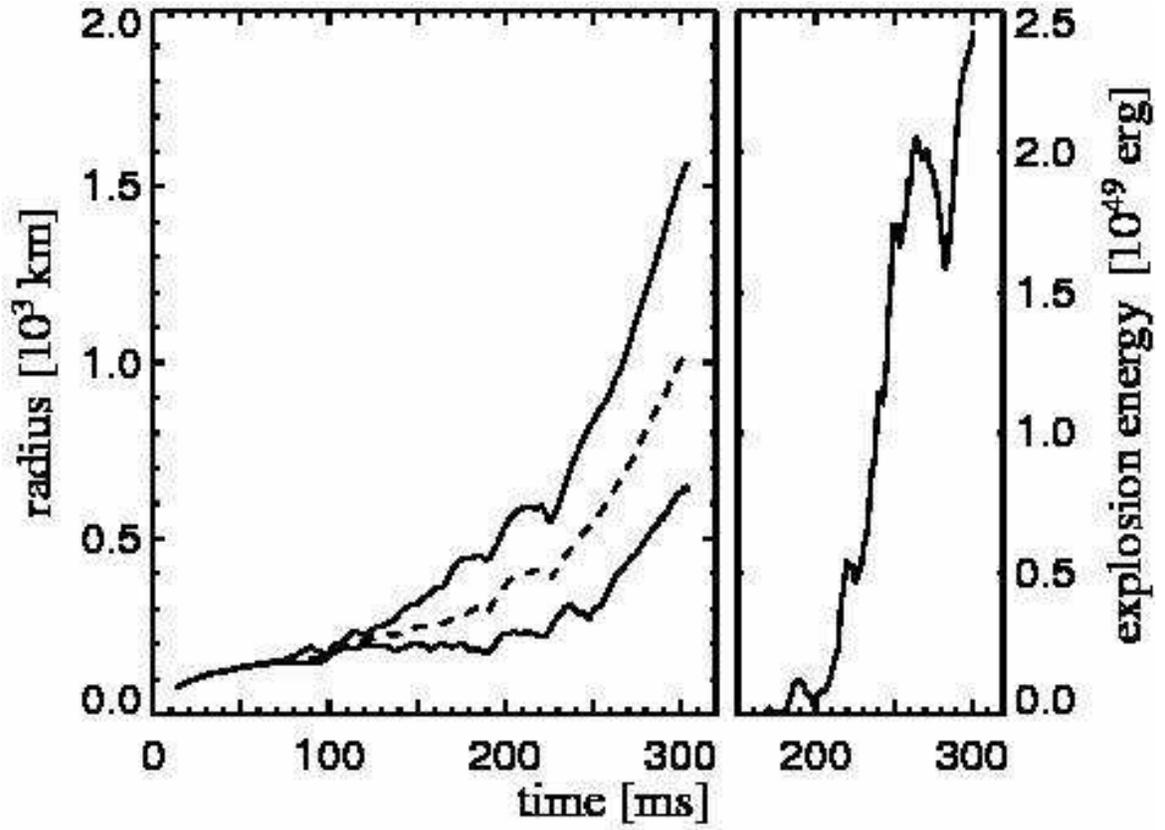}
 \caption{{\em Left panel:} 
Mean shock radius (arithmetical average over all lateral
directions, dashed line) and maximum and minimum shock positions 
as functions of post-bounce time for
our 2D explosion simulation of an 11.2$\,M_\odot$ progenitor.
{\em Right panel:} 
``Explosion energy'' of the 11.2$\,M_\odot$ star, defined as
the total energy (internal plus kinetic plus gravitational) of 
all mass in the gain layer with positive radial velocity,
as function of post-bounce time.
\label{fig:expener11}}
\end{figure}

\begin{figure*}
\plottwo{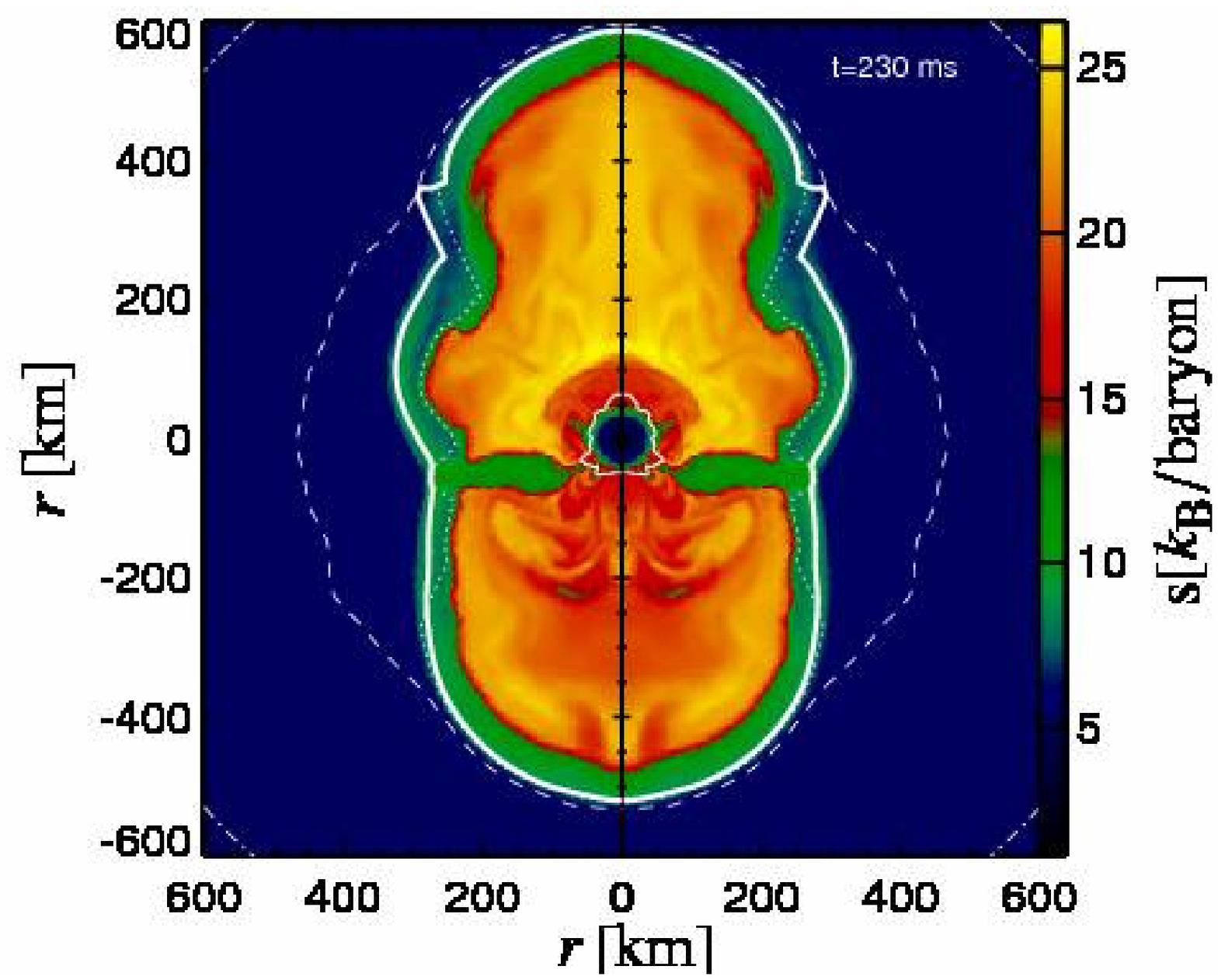}{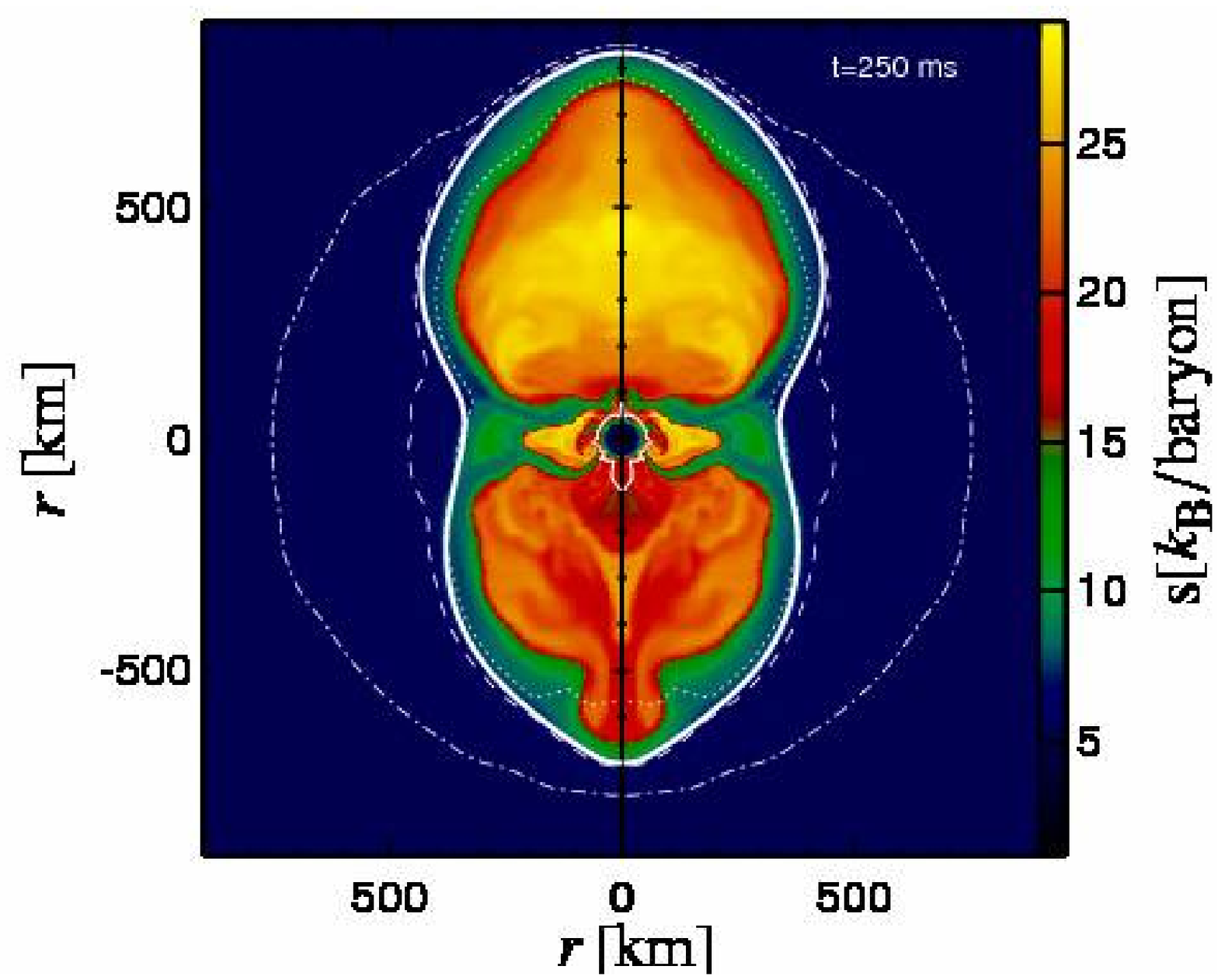}
\plottwo{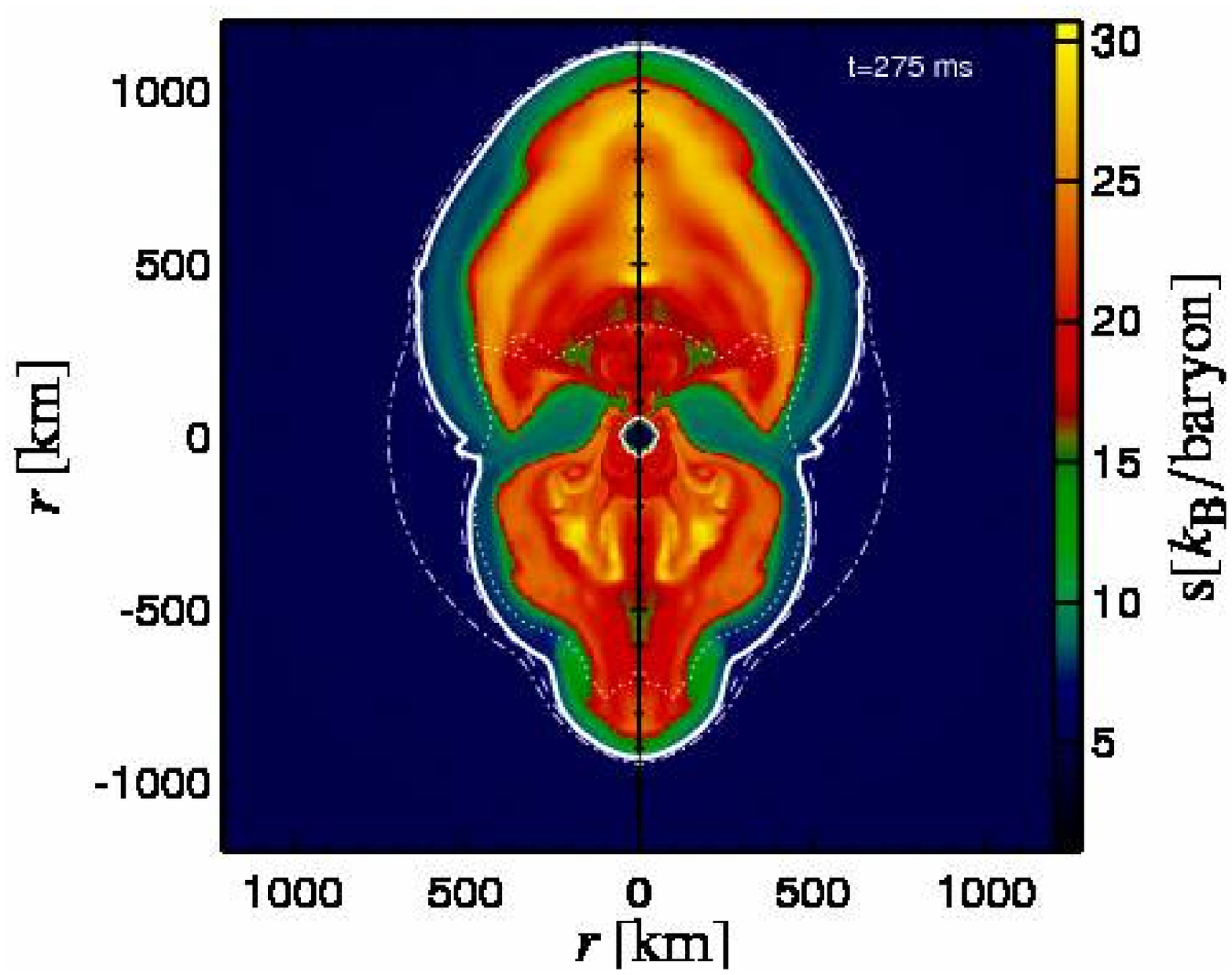}{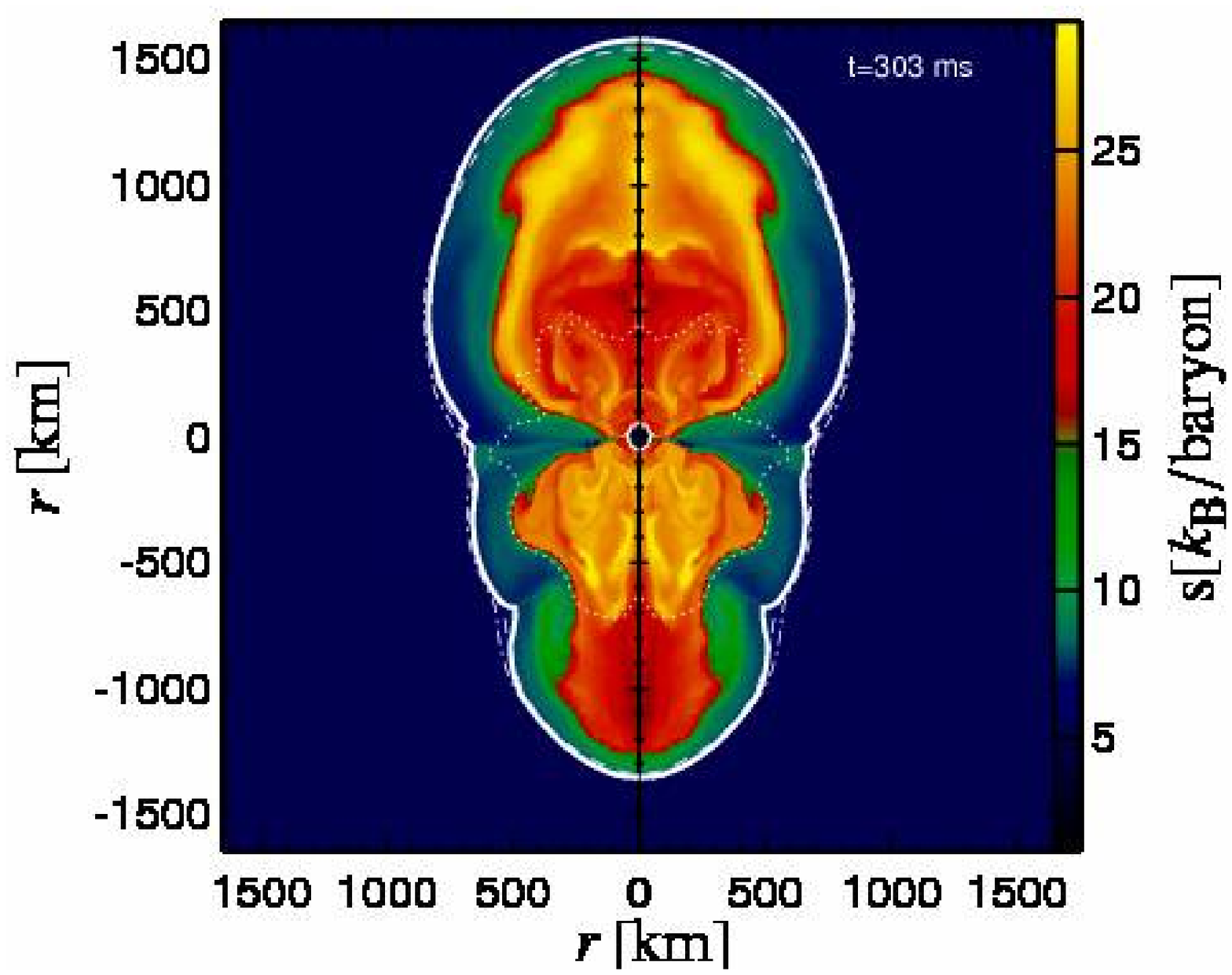}
\caption{Four snapshots from the 
evolution of our 11.2$\,M_\odot$ explosion model at 
times $t = 230\,$ms, 250$\,$ms, 275$\,$ms, and 303$\,$ms 
after core bounce. The figures contain the same features as
shown in Fig.~\ref{fig:snapshots}.
\label{fig:snapshots11}}
\end{figure*}

\begin{figure*}
\plottwo{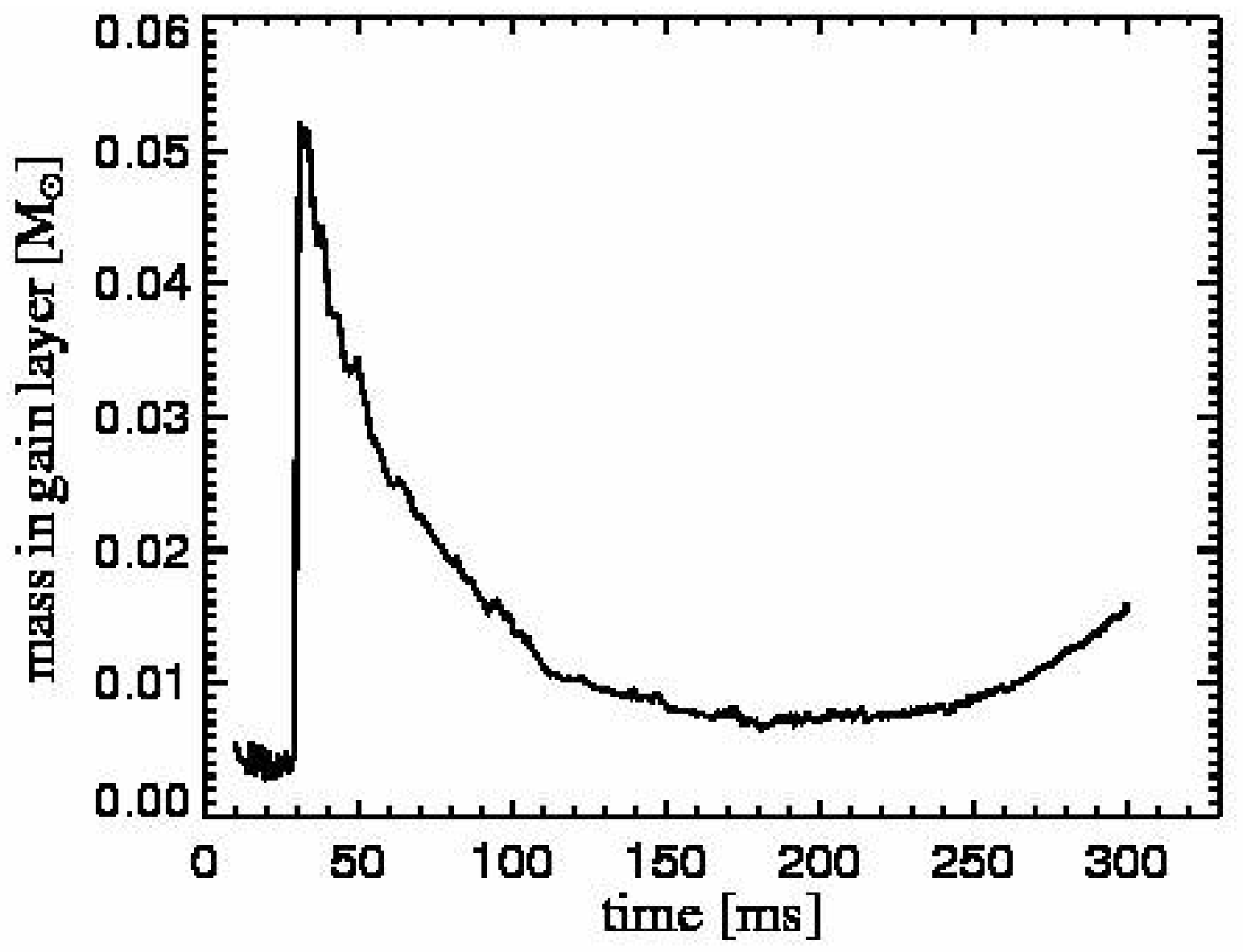}{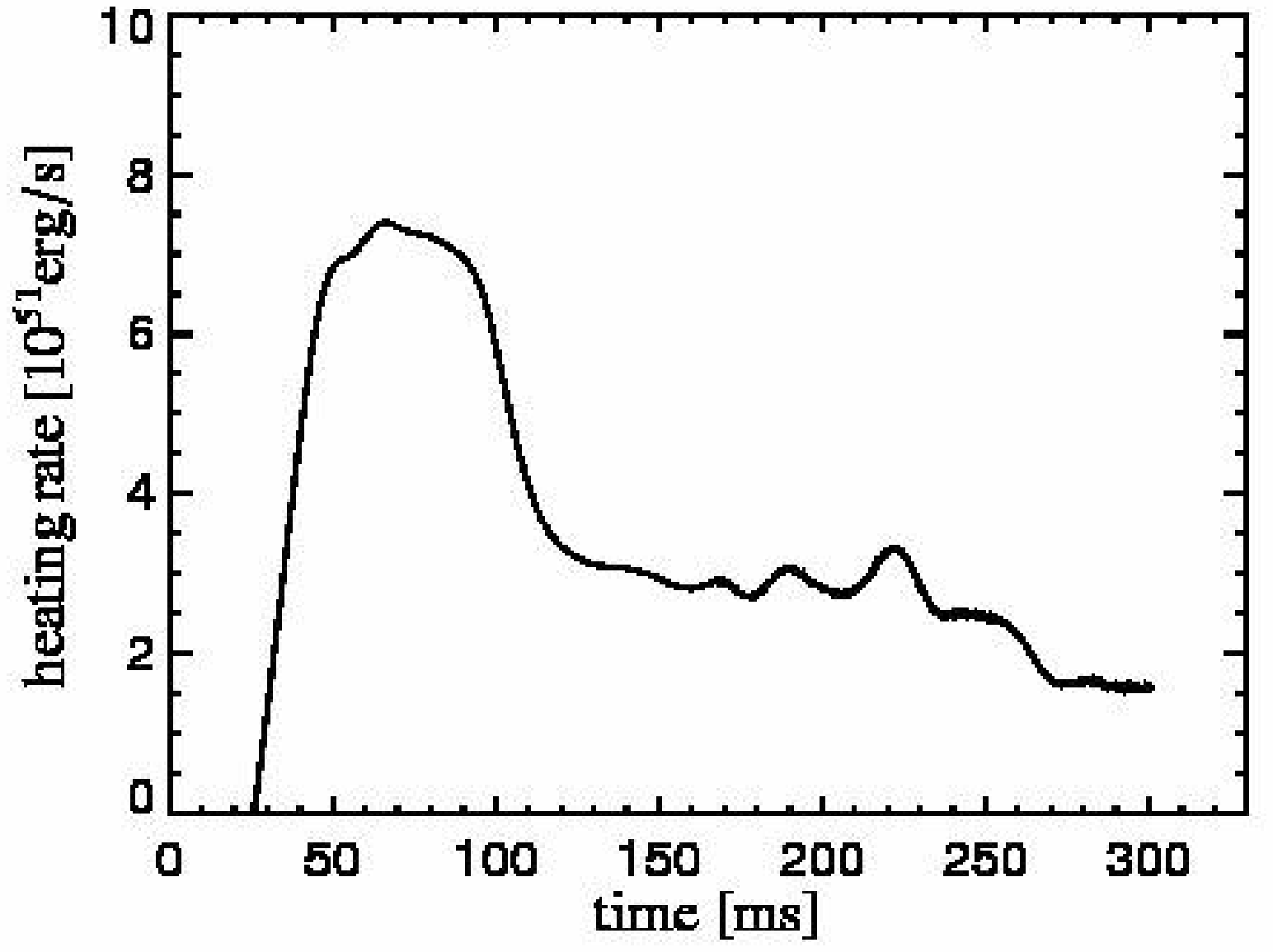}
\plottwo{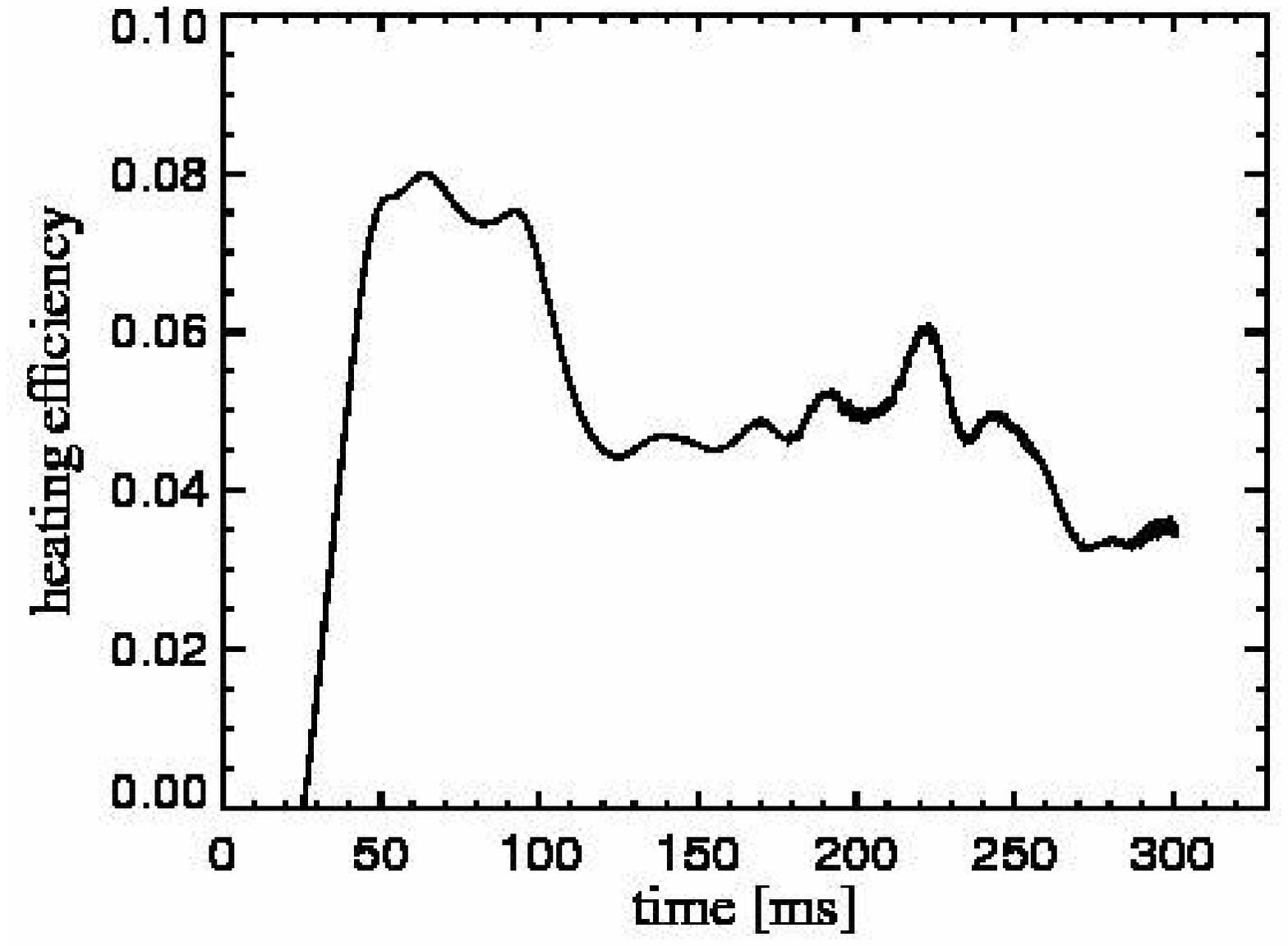}{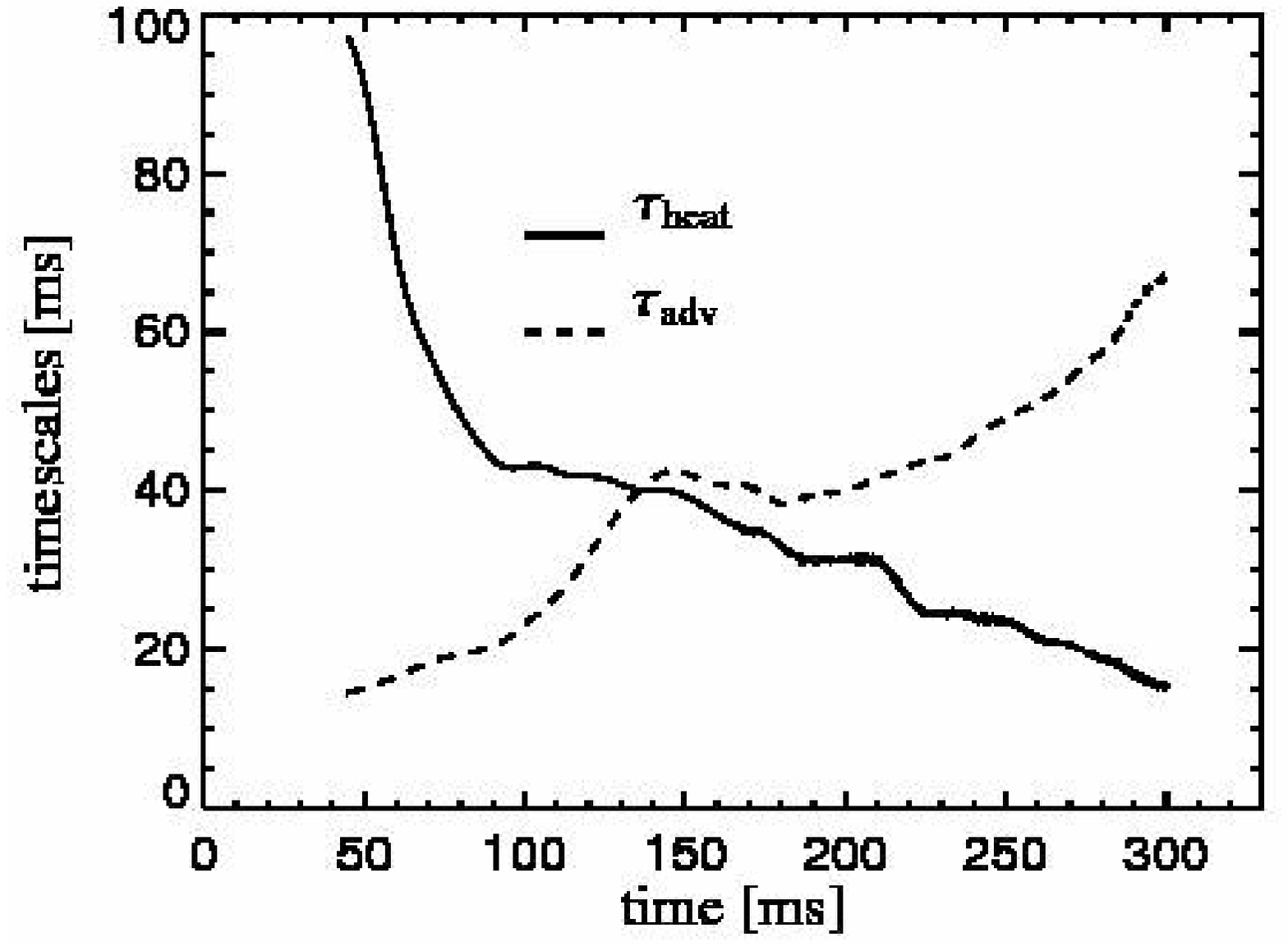}
\caption{Mass ({\em top left}), neutrino-heating rate
({\em top right}), heating efficiency ({\em bottom left}),
and heating and advection timescales ({\em bottom right}) 
in the gain layer as functions of time for our
11.2$\,M_\odot$ explosion model.
\label{fig:heating11}}
\end{figure*}

Repeating the computation of Model~M15LS-rot during the crucial
phase of growing SASI power between $\sim$420$\,$ms and
$\sim$670$\,$ms after bounce with finer radial zoning (Model~M15LS-rot-hr)
does not reveal any significant differences due to the resolution
until 550$\,$ms (cf.\ Figs.~\ref{fig:shockpositions}--\ref{fig:energies}).
Only the transient phase of damped SASI activity around 500$\,$ms
in our reference run of Model~M15LS-rot (see the middle panel of
Fig.~\ref{fig:modes}), which leads to a short pause in the expansion
of the average shock radius (Fig.~\ref{fig:shockpositions}) and a 
step-like interruption of the rise of the advection timescale
(Fig.~\ref{fig:timescales}),
is absent in Model~M15LS-rot-hr. This suggests that the feature
is just a consequence of the chaotic and stochastic behavior of
the non-radial hydrodynamic instabilities that stir up the flow
between shock front and forming neutron star. 

The conditions that can be considered as
favorable for an explosion evolve in a promising way
also in Model~M15LS-rot-hr. The average shock radius 
and the mass in the gain layer exhibit a monotonic trend of 
increase (Fig.~\ref{fig:shockpositions} and top panel of 
Fig.~\ref{fig:heating}, respectively).
The advection timescale shows the same behavior.
Because at the same time the neutrino-heating timescale, 
heating rate, and heating
efficiency settle to stable levels (Figs.~\ref{fig:timescales}
and \ref{fig:heating}), the timescale ratio climbs to a
value of $\sim$1.3, around which it fluctuates at $t \ga 580\,$ms.
Nevertheless, after $t\sim 570\,$ms the evolution of 
Model~M15LS-rot-hr departs from that of
Model~M15LS-rot. The latter develops an explosion after 
$t\sim 570\,$ms, whereas the former reveals SASI
activity with a growing amplitude but no onset of strong outward
shock acceleration until the end of our simulation.

An inspection of the different energies that characterize the
energetic conditions in the gain layer reveals
that the kinetic energy, in particular the part
that is associated with the radial velocity component, 
begins to deviate between the two models around the time when
their evolution starts to diverge (upper left panel of 
Fig.~\ref{fig:energies}). Also the internal energy of the gain layer
begins to grow faster in Model~M15LS-rot than 
in its better resolved counterpart, but the difference
appears slightly later (at $\sim$590$\,$ms instead of $\sim$570$\,$ms)
and grows more gradually than for the kinetic energy. Other
energies as well as the neutrino-heating rate do not show any 
peculiar differences. While the divergence of the internal energies
is likely to be a consequence of the stronger shock expansion 
in Model~M15LS-rot, which leads to growth of the mass in the
gain layer, the difference of the radial kinetic energy could be
causal for the discrepant later evolution of both simulations.
Unfortunately, because of the complex interdependence
of different effects (hydrodynamics, gravity, neutrino physics)
we are not able to unambiguously track down the origin of this
energy difference and to clarify whether it
is really the reason or also 
just the result of the evolution that separates at about 570$\,$ms.

In any case, however, it shows that in the given situation the 
later development of Models~M15LS-rot and M15LS-rot-hr seems to be
extremely sensitive to relatively small effects. An excess of
order $10^{49}\,$erg in the kinetic energy of the radial motions
(and also -- but a factor of $\sim$4 lower -- in the lateral
kinetic energy, see upper left panel of Fig.~\ref{fig:energies})
and a correspondingly stronger SASI and convective activity,
which could be caused by chaotic fluctuations, appears to be 
sufficient to initiate the outward acceleration of the shock and
to cause the bifurcation of the subsequent evolution of both models.
This by itself is a very interesting result,
which, of course, raises questions about the importance of 
stochastic variations or about the robustness of the
explosion found in Model~M15LS-rot, for example with 
respect to differences in the spatial resolution used in the
simulations. We repeat and stress here, however, that a variety
of diagnostic parameters like the growing
shock radius and SASI activity as well as the large value of the
ratio of advection to heating timescale, exhibit an overall 
trend that appears very favorable for an upcoming explosion also
in the case of Model~M15LS-rot-hr. Unfortunately,
without being able to run this model (and any of the other 
15$\,M_\odot$ simulations, in particular with high resolution)
for a sufficiently long post-bounce time --- which is presently 
prevented by the limited computer resources available to us ---
we have no convincing support for this reasoning.

\subsubsection{Rotating Model M15LS-rot9}
\label{sec:rotmod9}

Model~M15LS-rot9 was computed with a somewhat weaker effective
relativistic gravitational potential than our standard 
Model~M15LS-rot (Case~A of Marek et al.\ 2006 instead of Case~R; the
former yields the more accurate treatment for nonrotating stellar
core collapse according to comparisons with fully relativistic
simulations). A comparison of these two simulations with the
nonrotating Model~M15LS-2D, which shares the description of 
the effective relativistic potential with Model~M15LS-rot9 
(cf.\ Table~\ref{tab:models}), allows us to draw conclusions
on the importance of rotation and of the strength of the 
gravitational potential. 

Model~M15LS-2D without centrifugal effects but with weaker gravity
is astonishingly similar to Model~M15LS-rot in many of the
evaluated aspects. This is true, e.g., for the neutrino luminosities
at $t\ga 200\,$ms after bounce 
(bottom panel in Fig.~\ref{fig:rotneutrinos1}) and 
for the neutrino-heating rate in the gain layer and the
heating efficiency (Fig.~\ref{fig:heating}) as well. But also the
average shock radii of both models are fairly similar and therefore
the advection timescales do not differ much 
(Figs.~\ref{fig:shockpositions}, \ref{fig:timescales}). The
ratio of the advection timescale to the heating timescale shows the 
same trend in both cases, with a slight time lag of Model~M15LS-2D
(lower panel of Fig.~\ref{fig:timescales}). This suggests that 
rotation does not play an essential role for the evolution of 
Model~M15LS-rot towards runaway. In fact, with respect to the
overall long-time behavior, rotation seems to just compensate much
of the influence of the stronger gravitational potential used in
Model~M15LS-rot.

The weaker gravity in Model~M15LS-rot9 relative to Model~M15LS-rot
allows the shock to move farther out by $\sim$20$\,$km on average
during the first 300$\,$ms after bounce (Fig.~\ref{fig:shockpositions}).
Since also the infall velocity in the preshock region is lower
due to the reduced gravitational acceleration, the
advection timescale is significantly longer in Model~M15LS-rot9.
At the same time, the accretion luminosities of neutrinos in
Model~M15LS-rot9, in which 
the neutron star is less compact and less hot, are significantly
smaller (Fig.~\ref{fig:rotneutrinos1}, bottom panel) and therefore
the overall postshock heating is weaker 
(up to roughly a factor of two; Fig.~\ref{fig:heating}) and
the heating timescale is considerably longer (typically a factor of 
two; Fig.~\ref{fig:timescales}) than in
Model~M15LS-rot. For this reason, Model~M15LS-rot9 appears to be
less favorable for generating a SASI-aided neutrino-driven
explosion than the nonrotating case with the same description of 
gravity, despite the fact that centrifugal support allows for a
larger average shock radius and correspondingly higher mass in the
gain layer (Fig.~\ref{fig:heating}).
One might therefore conclude that rotation as considered
in our Models~M15LS-rot and M15LS-rot9 does not improve
the conditions for an explosive runaway compared to nonrotating
models in which low SASI modes are able to develop (this comparison
turns out to be reversed, if such modes are suppressed in
the nonrotating case by a constraining choice of the grid setup;
see Fig.~15 in Buras et al.\ 2006b). 

A reliable assessment of the influence of rotation on the 
development of an explosion, however, requires 
a continuation of simulations like Model~M15LS-rot9 to later times.
Centrifugal effects were predicted
to lead to a reduction of the critical neutrino luminosity for the
revival of the stalled shock near the polar axis 
(see the linear analysis by Yamasaki \& Yamada 2005). The effect
grows strongly with higher angular momentum of the accreted matter.
In combination with the fact that rotationally deformed neutron
stars produce enhanced luminosities and/or higher mean energies for
neutrinos radiated in the polar direction (Janka \& M\"onchmeyer 
1989a,b; Kotake et al.\ 2003), this might imply more 
optimistic conditions at very late postbounce times when material
with higher angular momentum from larger initial radii arrives 
at the accretion shock.

\subsubsection{Nonrotating Models M15LS-2D and M15HW-2D}
\label{sec:nonrot}

The comparison of Models~M15LS-2D and M15HW-2D yields information
about the influence of the neutron star EoS, which determines
many aspects of the post-bounce evolution, most importantly
the compactness of the nascent neutron star and the 
properties of the radiated neutrinos. 

The HW-EoS used in Model~M15HW-2D is clearly stiffer and the 
neutron star consequently less
compact (see Figs.~\ref{fig:1dresults} and \ref{fig:shockpositions}).
The neutrino luminosities and mean energies are therefore lower
(Fig.~\ref{fig:1dresults} and the bottom panel of 
Fig.~\ref{fig:rotneutrinos1}), which leads to a correspondingly
reduced heating rate and heating efficiency in the gain layer
(Fig.~\ref{fig:heating}), a much longer heating timescale,
and thus also to a smaller timescale ratio (Fig.~\ref{fig:timescales}).

While in the 1D simulation with the HW-EoS the 
shock position follows the behavior of the neutron star radius 
and expands to a larger distance from the center than in the 
1D model with the LS-EoS (see 
Fig.~\ref{fig:1dresults} and Eq.~(\ref{eq:rshock})), the situation
is different for the 2D runs. Here the sloshing of the shock in 
Model~M15HW-2D reaches a clearly smaller amplitude than in
Model~M15LS-2D during most
phases of the computed $\sim$400$\,$ms of post-bounce evolution.
This is particularly obvious between 200 and 300$\,$ms after bounce
and points to a significantly lower strength of the SASI activity
(for more details, see Marek, Janka, \& M\"uller 2008). 
An inspection of the kinetic energies in Fig.~\ref{fig:energies}
(upper left panel) supports this conjecture, and a quantitative
evaluation reveals that the $l=1$ and $l=2$ SASI modes reach only
20--30\% of the power they have in Model~M15LS-2D. In fact, the
SASI activity in Model~M15HW-2D grows considerably more slowly
and the low-mode power is orders of magnitude smaller than in
Model~M15LS-2D until nearly 200$\,$ms after bounce (see also
Marek et al.\ 2008).

This suggests that the compactness of the forming neutron star
has a very strong influence on the growth and
strength of the SASI and of convective activity in the gain layer,
either directly by the conversion of gravitational binding energy
to kinetic and internal energy in the accretion flow, or 
indirectly via the effects of larger luminosities and mean 
energies of the neutrinos radiated during the period of 
mass accretion and thus stronger neutrino heating.
A more rapidly contracting and more compact neutron star is
obviously more favorable in this context. Such a behavior could
have different reasons, for example a softening of the EoS in
the interior of the nascent neutron star as a consequence of
a microphysical phase transition that is triggered by
evolutionary changes (gravitational settling, neutronization,
heating due to compression and conversion of electron degeneracy
energy to thermal energy) in the star. We therefore emphasize 
that more favorable conditions for an explosion during the 
post-bounce accretion phase of the forming neutron star neither 
require a lower value of the incompressibility modulus of the EoS 
for nearly symmetric matter at saturation density, nor are they
directly linked to the EoS properties around bounce, which affect the 
shock formation and prompt shock propagation but not necessarily
the long-time evolution of the accretion shock. 

The gradual growth of relevant parameters (heating rate, timescale 
ratio, kinetic energy in the gain layer, etc.) towards the end of the
computed evolution appears promising for the possibility of an 
explosion in Model~M15LS-2D at later times: many properties,
of this model near the termination of the run, for example
the neutrino luminosities and neutrino-heating quantities, are very
similar to those of our explosion model M15LS-rot at the same time.
Initially Model~M15HW-2D looks less optimistic than
Model~M15LS-2D, but there are indications that some quantities in
Model~M15HW-2D might reverse their trends in the long run.
For example the shock radius (Fig.~\ref{fig:shockpositions}),
the kinetic energy in the gain layer (Fig.~\ref{fig:energies}), 
the SASI power, and the low-mode shock deformation
(Fig.~\ref{fig:modes}) exhibit a strong rise after $\sim$300$\,$ms,
which leads to an increase of the advection timescale 
(Fig.~\ref{fig:timescales}) and also of the heating in the gain layer
(Fig.~\ref{fig:heating}). This improvement could be a consequence
of the growing compactness of the nascent neutron star.
Both simulations, Model~M15HW-2D and Model~M15LS-2D, however, could
not yet be carried on for sufficiently long evolution periods
after bounce to see which
differences the two employed nuclear EoSs might ultimately could make
for an explosion at even later times.

\subsection{Neutrino emission}
\label{sec:neutrinos}

\begin{figure}
\epsscale{0.45}
\plotone{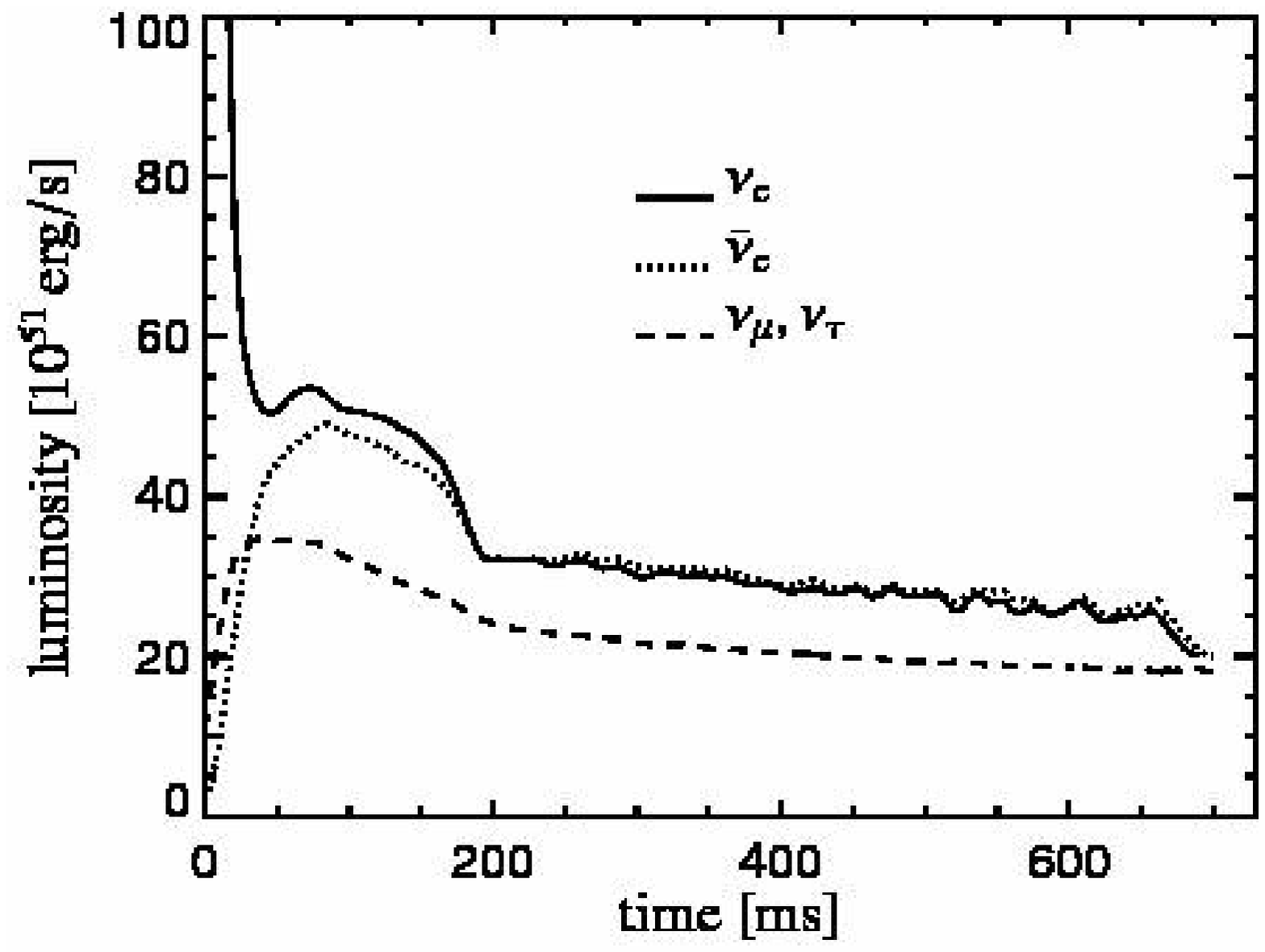}
\plotone{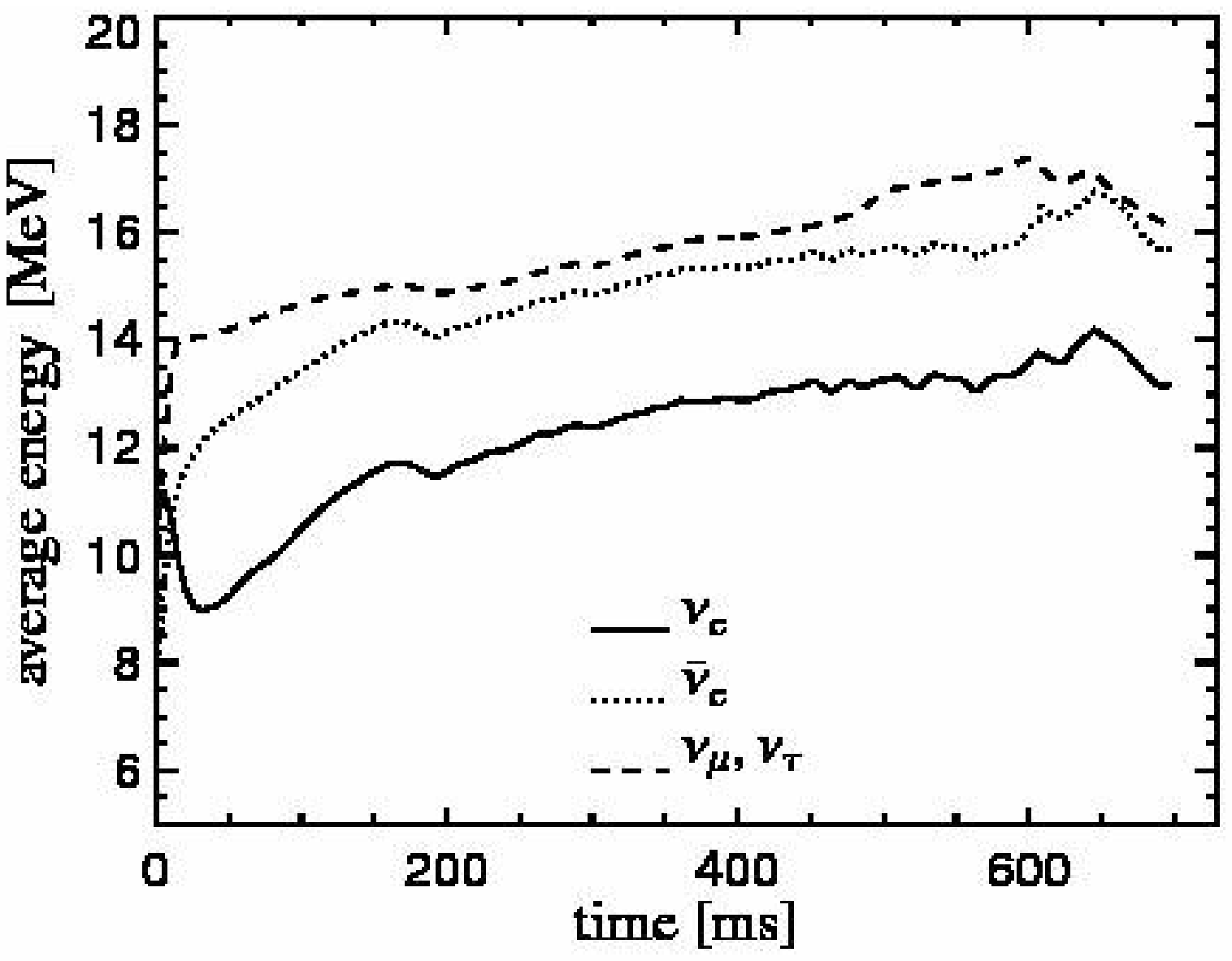}
\plotone{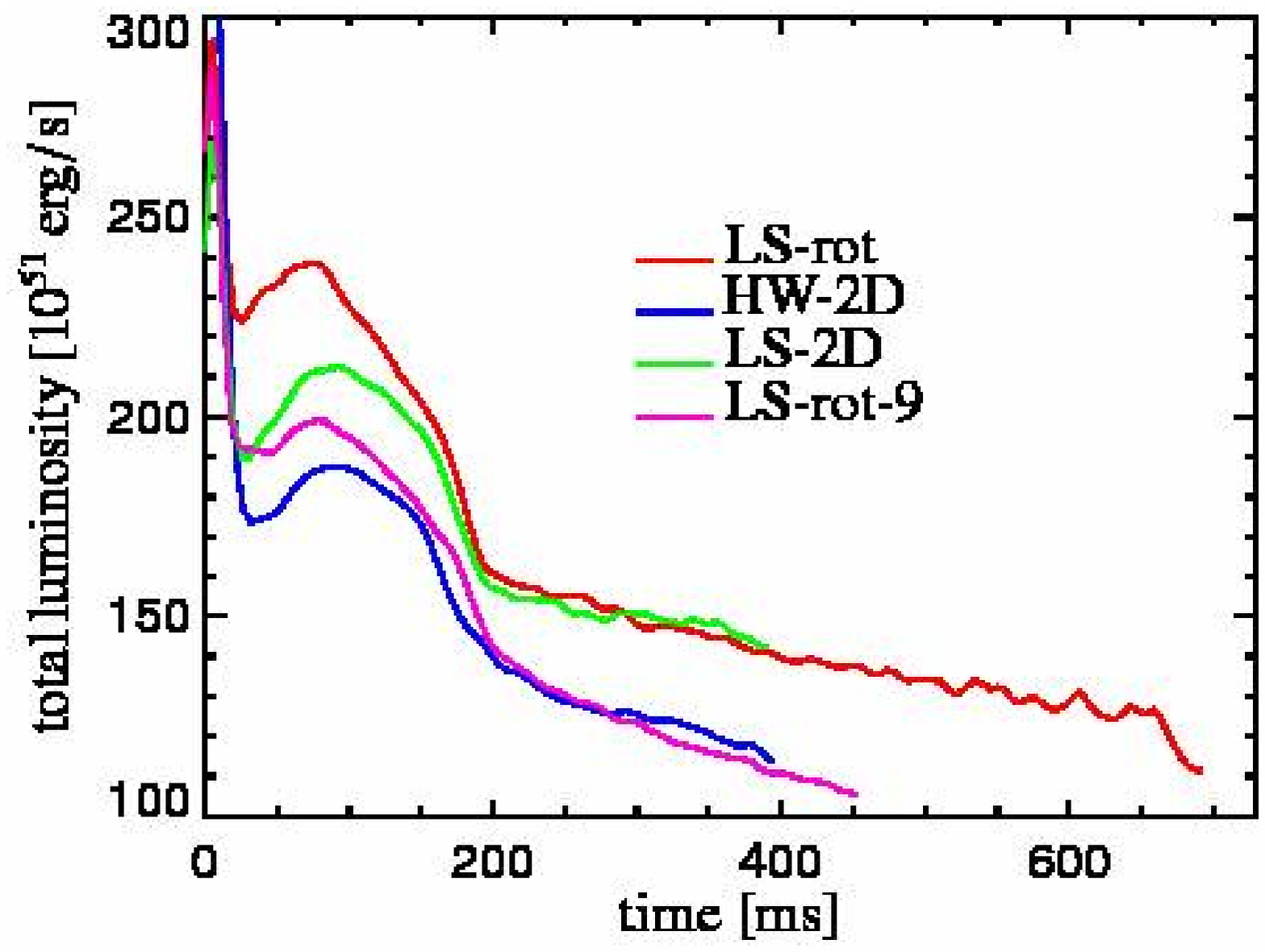}
 \caption{
Luminosities ({\em top}) and mean energies ({\em middle})
of radiated neutrinos in Model~M15LS-rot, and total luminosities
(summed for neutrinos and antineutrinos of all flavors) for
all 2D simulations with the 15$\,M_\odot$ progenitor 
({\em bottom}) as functions of time after
bounce. All quantities are evaluated at a radius of 400$\,$km
for an observer in the rest frame of the
stellar center, and the average energies are defined as ratio of
energy flux to number flux. The decline of the curves near the 
end of the simulation of Model~M15LS-rot signals the onset of the 
explosion, which
leads to reduced accretion luminosities of $\nu_e$ and $\bar\nu_e$
from the nascent neutron star.
\label{fig:rotneutrinos1}}
\end{figure}

\begin{figure*}
\epsscale{1.}
\plottwo{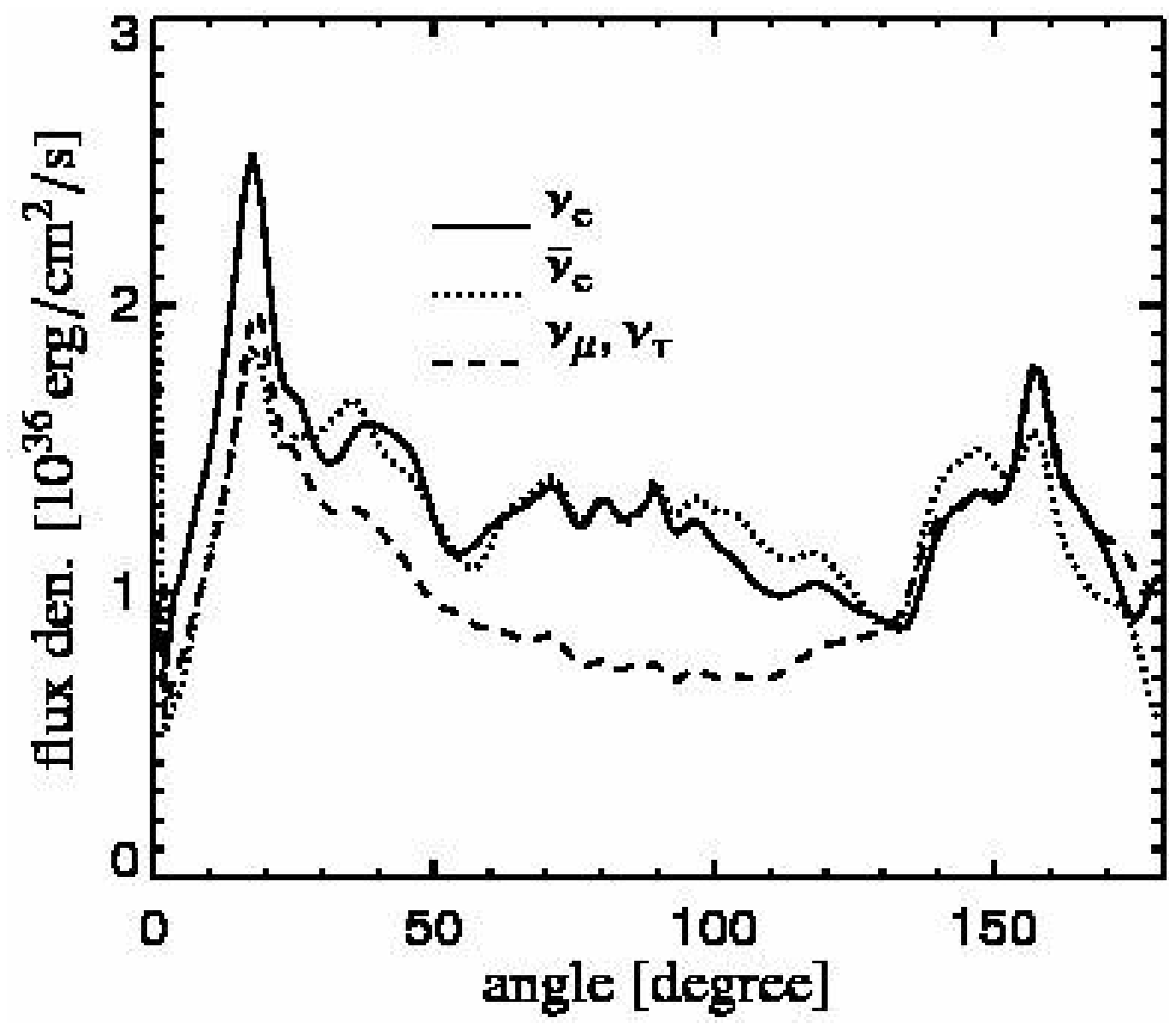}{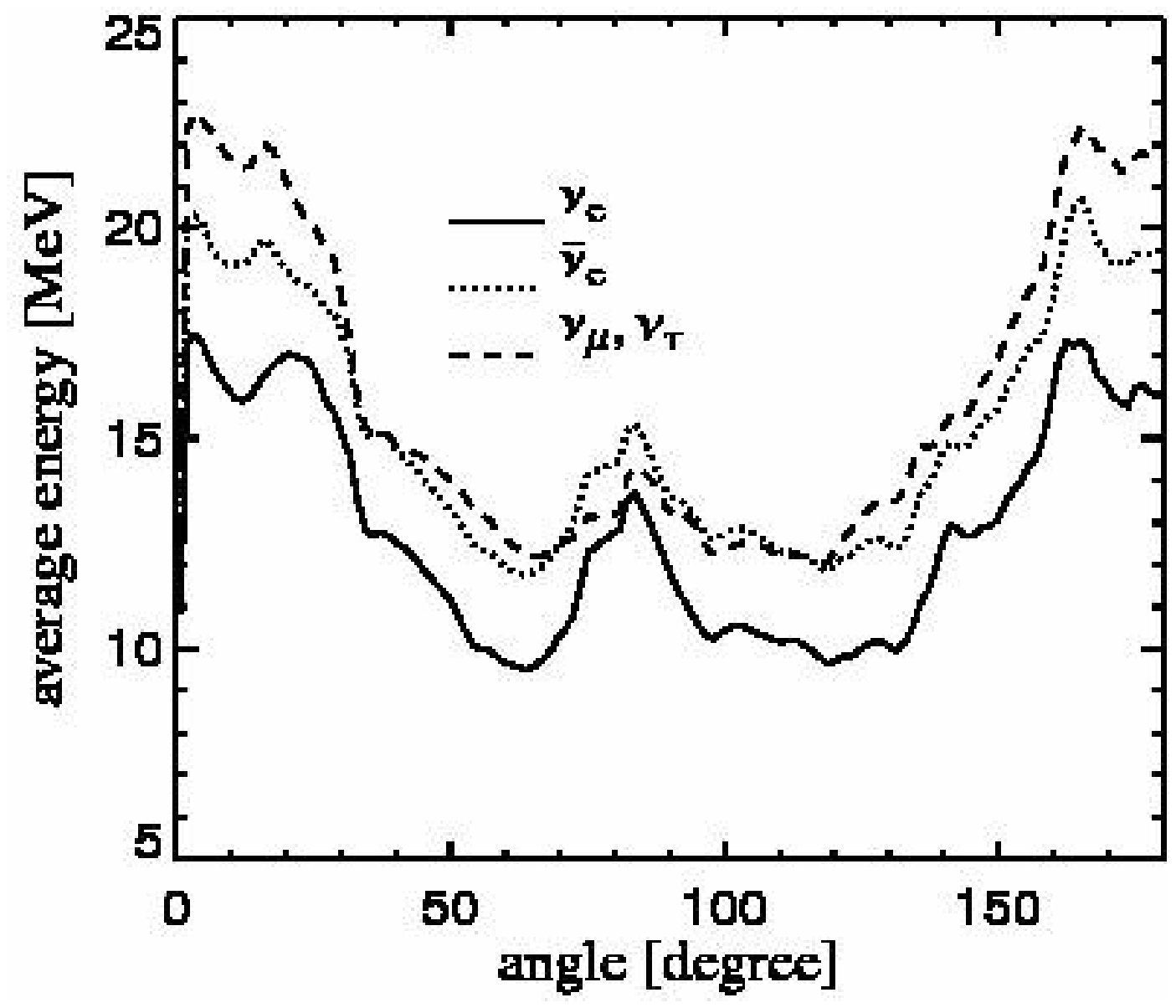}
\plottwo{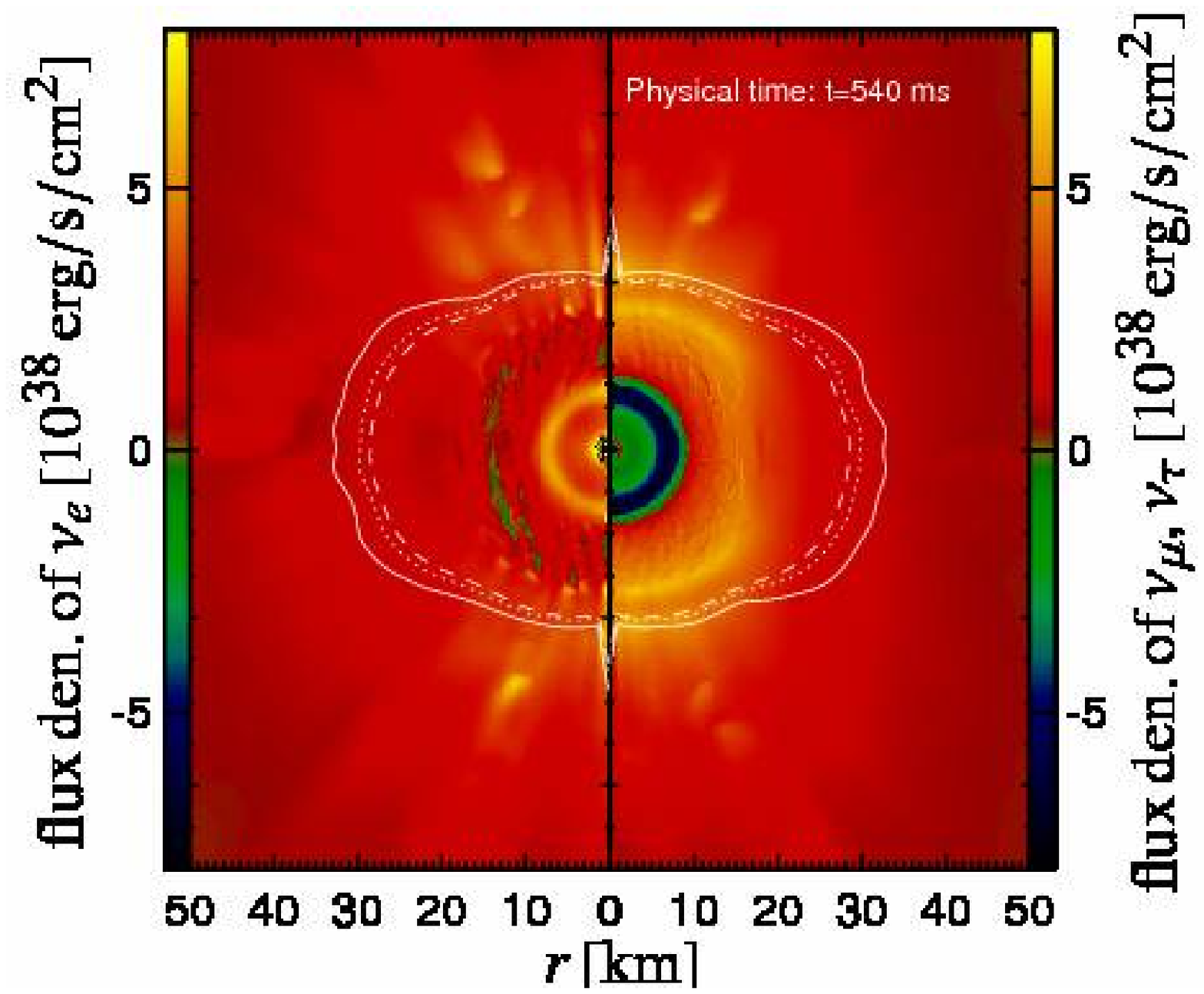}{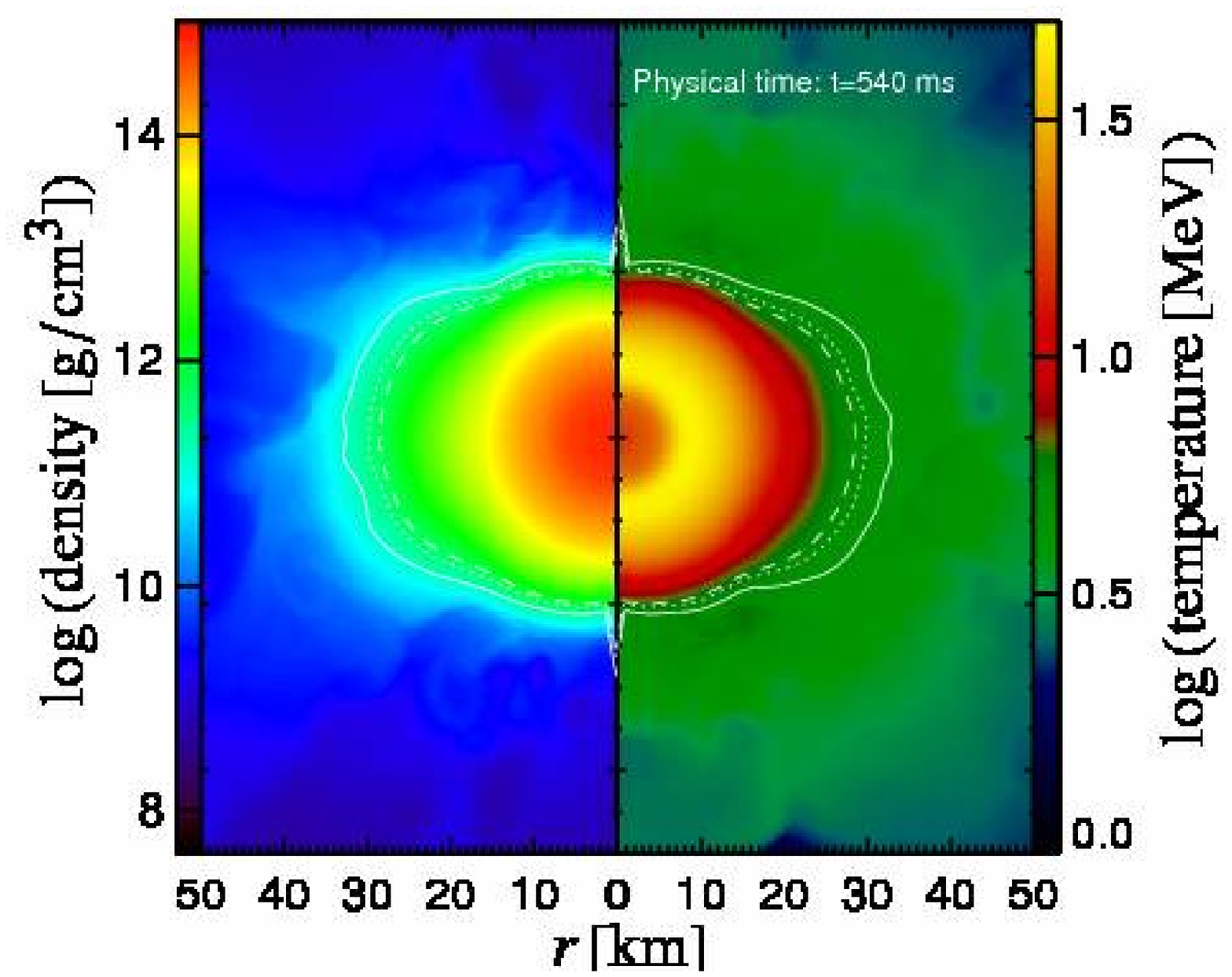}
 \caption{
{\em Top:} Neutrino energy flux densities ({\em left}) and mean energies 
({\em right}) as functions of polar angle
at 540$\,$ms after bounce for Model~M15LS-rot. The quantities
are measured at 400$\,$km and
given in the observer frame, and the average energies are 
defined as ratio of energy flux density to number flux density. 
{\em Bottom, left:} Energy flux densities of electron neutrinos 
and (one kind of) heavy-lepton neutrinos in and around the nascent
neutron star. The mean neutrinospheres,
which are marked by white lines for $\nu_e$ (solid), 
$\bar\nu_e$ (dotted), and $\nu_\mu$ (dashed), are significantly
deformed because of the presence of rotation.
{\em Bottom, right:} Density and temperature in the vicinity of
the forming neutron star.
\label{fig:rotneutrinos2}}
\end{figure*}

In Figs.~\ref{fig:rotneutrinos1} and \ref{fig:rotneutrinos2} we 
present some results on the neutrino emission of our exploding
Model~M15LS-rot. In spite of the deeper effective gravitational 
potential used in this simulation (Case~R of Marek et al.\ 2006), the
luminosities and mean energies of the radiated neutrinos at late
times after bounce ($t \ga 350\,$ms) are slightly lower than those
in Model~M15LS-1D. This is again the influence of rotation, which
compensates some of the consequences of the 
stronger gravity in Model~M15LS-rot, an effect that we have already
discussed in the context of comparing Models~M15LS-rot and M15LS-2D, whose
neutrino luminosities become very similar after 200$\,$ms of post-bounce
evolution (see lower panel in Fig.~\ref{fig:rotneutrinos1}).

The mean energies of the neutrinos radiated from Model~M15LS-rot 
(averaged over all observer directions relative to the rotation axis),
however, do not exhibit the crossing of 
$\left\langle \epsilon_{\bar\nu_e} \right\rangle$ and 
$\left\langle \epsilon_{\nu_x} \right\rangle$ seen in the 1D calculations,
although both mean energies get very close
(compare the middle panel of Fig.~\ref{fig:rotneutrinos1} with the
lower right panel of Fig.~\ref{fig:1dresults}). The reason for
this difference can be understood from the upper right plot of 
Fig.~\ref{fig:rotneutrinos2}, which displays the mean energies of 
radiated neutrinos as functions of polar angle. One can see
that in the vicinity of the poles (at latitudes 
$\theta \la 30^\circ$ and $\theta \ga 160^\circ$) the mean energy
of muon and tau neutrinos is clearly higher than that of electron
antineutrinos, while away from the poles both are very similar and
near the equator their order even reverses. This, as well as the
trough-like shape of the curves (which is also present in the
muon and tau neutrino flux densities, see upper left panel of 
Fig.~\ref{fig:rotneutrinos2}), is a consequence of the rotational
deformation of the neutrinospheric region as visible in the lower 
two plots of Fig.~\ref{fig:rotneutrinos2}. The superimposed
short-wavelength variations of the average radiated energies
and of the neutrino fluxes are caused by local downdrafts of
accreted matter near the poles and close to the equatorial plane.
The neutrino flux production associated with such downdrafts is
visible as bright spots above and around the neutrinospheres
in the lower left panel of Fig.~\ref{fig:rotneutrinos2}. Near
the poles, where the downdrafts penetrate deep inward and become 
very hot, neutrinos of all flavors are produced in the tips of
these flow structures, whereas the lower points of equatorial 
downflows emit only $\nu_e$ and $\bar\nu_e$ (this explains the
absence of a local maximum of the muon and tau neutrino flux 
density around the equator).

Because of rotation, the density and temperature gradients at the
poles (Fig.~\ref{fig:rotneutrinos2}, bottom right) are much steeper
and the radial positions of all neutrinospheres there are extremely 
close together\footnote{In the context of this paper we have adopted 
Eq.~(28) of Buras et al.\ (2006a) for the definition of the energy-averaged
neutrinosphere. This means that the average neutrinosphere coincides with
the ``transport sphere'' where the neutrinos in the bulk of the 
spectrum undergo the transition from
diffusion to free streaming, and not with the ``energysphere'' where
the neutrino production effectively ceases and the neutrinos decouple 
energetically from the stellar matter.}.
Most importantly, the region where the dominant part of the flux
is generated before the flux density gets diluted due to scattering 
losses and the growing distance from the source, is 
significantly below the energy-averaged neutrinosphere for muon
and tau neutrinos (see the yellow areas in the right half
of the lower left plot of Fig.~\ref{fig:rotneutrinos2}). This 
means that before the $\nu_\mu$ and $\nu_\tau$ reach their
neutrinosphere, they have to diffuse through an extended 
scattering atmosphere that acts like a high-energy filter, i.e.,
that is inefficient in producing more of these neutrinos, but 
in which energy transfers in neutrino-nucleon and neutrino-electron
scatterings down-grade high-energy neutrinos in energy space
and reduce the energy flux of heavy-lepton neutrinos
(Raffelt 2001; Keil et al.\ 2003). Moreover, $\nu_e$ and $\bar\nu_e$
in contrast to heavy-lepton
neutrinos are abundantly absorbed and emitted by charged-current
$\beta$-processes and thus are more strongly created in an
extended but cooler and less dense accretion layer (visible, for
example, also by a significantly more extended neutrinospheres at
low latitudes). Both effects in combination are responsible for the 
convergence or even inversion of the mean energies
of the radiated $\bar\nu_e$ and $\nu_x$ in equator-near regions
(for a discussion of this phenomenon in the context of the 
nonrotating 2D models, see Marek et al.\ 2008).
Close to the poles the
extreme steepness of the density gradient and the additional enhancement
of the neutrinospheric fluxes of muon and tau neutrinos and antineutrinos
by convection below the neutrinospheres (see next paragraph)
counteract such a strong filter effect of a scattering
atmosphere. The 1D models (cf.\ Sect.~\ref{sec:1dmodels}) define a
case that is more similar to the equatorial regions of the rotating
model: although the mass infall rate to the nascent neutron star
drops strongly after the first 
$\sim$200$\,$ms of massive post-bounce accretion, the energysphere
of muon and tau neutrinos is considerably below the neutrinosphere
and the density decline below the accretion atmosphere
is sufficiently gradual to establish a thick layer of 
frequent neutrino scatterings between the energysphere and the
neutrinosphere of heavy-lepton neutrinos.

Finally, we point out that the region of convective activity
inside the nascent neutron star can be discerned in the lower left
plot of Fig.~\ref{fig:rotneutrinos2}. In the left half of this
figure, which shows the flux of $\nu_e$, a pattern of green and red 
stripes indicates elongated convective cells that are oriented 
parallel to the rotation axis. This is typical of a situation in
which convection is constrained by the presence of angular momentum
gradients (see the discussion in Buras et al.\ 2006b). The cell structure
becomes visible in this image, because Ledoux convection in the 
newly formed neutron star is driven by negative lepton number gradients and
is therefore associated with variations of the neutron-to-proton ratio.
Due to the short $\beta$-equilibration timescale in the optically 
thick regime, such variations are tightly linked to differences 
in the electron neutrino density as well as flux density. In the 
right half of the lower left image of Fig.~\ref{fig:rotneutrinos2},
one can see that a local maximum of the $\nu_\mu$ flux coincides
with the outer edge of the convective region. This suggests that
convection inside the proto-neutron star causes a noticeable enhancement
in particular of the transport of muon and tau neutrinos, a fact 
that was discussed in detail in Buras et al.\ (2006b).

\section{Neutron star g-mode oscillations}
\label{sec:gmodes}

Recently Burrows et al.\ (2006, 2007b)
discovered in their two-dimensional, Newtonian
core-collapse simulations that 
very late after core bounce the newly formed neutron star can be 
excited to core g-mode oscillations with amplitudes of several 
kilometers. Burrows et al.\ did not obtain neutrino-driven explosions
in their simulations,
but while the shock revealed violent sloshing motions due
to the SASI instability, the anisotropic gas flow around the 
central object remained gravitationally bound and the gas 
continued to be accreted. 
After more than one second of post-bounce evolution, however,
a rapid outward acceleration of the shock set in. For an
11.2$M_\odot$ progenitor, for example, Burrows et al.\ (2007b)
saw this shock revival happening about 1.1~seconds after bounce;
in a 20$M_\odot$ star the outward shock expansion occurred after
1.2~seconds of post-bounce accretion. In the simulations of 
Buras et al.\ (2006b), however, the same 11.2$\,M_\odot$ model
was found to explode about 200$\,$ms after bounce by the 
SASI-supported neutrino-heating mechanism (see Sect.~\ref{sec:model11.2}).
The very late explosions in the runs of Burrows et al.\ were 
powered by the acoustic flux that was sent by large-amplitude 
vibrations of the neutron star into the medium surrounding the 
compact remnant. On travelling outward, the pressure waves steepened
into shocks and dissipated their energy in the medium behind the
supernova shock, thus creating the thermal energy that provided the 
pressure for the shock acceleration.

Burrows et al.\ (2006, 2007b) estimated that the acoustic energy
flux generated by this gravity-wave activity of the forming 
compact remnant dominates the energy deposition by neutrinos in
their models at times later than about one second after bounce.
A sizable fraction of the accretion power is thus converted
by the oscillating neutron star, which acts as a transducer,
to acoustic flux instead of neutrino emission. As
long as anisotropic accretion goes on, the gravity waves are
further excited and the accretor produces a steady
flow of acoustic energy. In order to
provide a dynamically relevant rate of energy input to the
shock and possibly the supernova explosion, however, the
amplitude of the core g-mode oscillations must be sufficiently
large at the surface of the nascent remnant; 
several kilometers seem to be necessary for creating an 
acoustic energy flux of a few $10^{50}\,$erg$\,$s$^{-1}$
(Eq.~(1) in Burrows et al.\ 2007c).

A closer and independent investigation of this interesting
and potentially important phenomenon is highly desirable.
Many questions remain to be answered, for example: Which
decisive factors does the 
excitation of core g-modes due to turbulence and anisotropic
accretion in the SASI layer depend on and how efficient can it
be? How strong is the acoustic coupling between the oscillating
neutron star and its surroundings and how powerful can the
acoustic energy flux become? What is the cause of the sudden
increase of the g-mode activity and of its driving force at 
late post-bounce times
in the simulations of Burrows et al.\ (2006, 2007b)? What are
the numerical requirements for tracking this phenomenon in 
supernova simulations, e.g., is the momentum 
conserving treatment of the gravity source term introduced
by Burrows et al.\ important? What is the influence of the 
grid configuration (choice of coordinates) and of the
grid resolution on the treatment of g-mode excitation and
de-excitation? Does the phenomenon also show up in similar
strength in 3D instead of 2D simulations?
In fact, Yoshida, Ohnishi, \& Yamada (2007) have recently
explored the generation of neutron star g-modes by external SASI-
and turbulence-induced pressure fluctuations. However, they
extracted the boundary condition for a semi-analytic g-mode 
analysis from independent hydrodynamic SASI simulations. While
these latter simulations were performed with Newtonian gravity,
the g-mode excitation of the neutron star was discussed on
the basis of a general relativistic stellar model. 
This approach therefore
lacks consistency in many respects. Also, Yoshida et al.\ 
(2007) left crucial points unaddressed: Does the pressure
fluctuation spectrum vary with different numerical resolution
for the SASI calculations? How strongly
does neutrino bulk viscosity damp the g-mode excitation?
Is the assumed boundary condition appropriate for
representing the coupling of the neutron star core and the 
exterior SASI layers, and how sensitive is the result to changes of
this boundary condition? A serious concern of different sort 
was addressed
in a paper by Weinberg\& Quataert (2008). They calculated the
damping of the primary $l = 1$ g-mode in the core of the 
proto-neutron star by its coupling to higher-order modes,
whose short wavelengths cannot be resolved in numerical 
simulations. They found that the primary mode should saturate
at an energy ($\sim$10$^{48}\,$erg) that is much too low to make 
acoustic power a significant energetic driver of supernova
explosions.

We do not intend to fundamentally challenge here the possibility 
that the acoustic mechanism might work as proposed by Burrows 
et al.\ (2006, 2007b,c), because none of our simulations was
continued to the late times when they see the corresponding 
large-amplitude g-mode oscillations of the neutron star core
in their models. In the following sections we will concentrate
only on two questions: Do neutron star g-mode oscillations
grow to important amplitudes in our simulations and does
the associated acoustic power radiated by the neutron star
contribute to the shock revival seen in our models at any
significant level?

\subsection{Core g-modes in our model runs}
\label{sec:gmodesinruns}

In order to answer these
questions, we have performed an analysis of the gravity-wave 
activity in the region around and below the neutrinosphere. 
A direct quantitative comparison of the
g-mode strength in the neutron star core between our simulations
and those of Burrows et al.\ (2006, 2007b,c) would be desirable
and could be illuminating, but information suitable for this
purpose is hardly available in the published papers. Figure~7
in Burrows et al.\ (2006), which displays a set of time-dependent
$l$-mode amplitudes of the spherical harmonics expansion of
the fractional pressure variations $[P(r,\theta)-\left\langle
P(r,\theta)\right\rangle_\theta]/\left\langle P(r,\theta)
\right\rangle_\theta$ at a fixed radius of 35$\,$km, 
could be considered as a basis for a straightforward comparison.
However, for several reasons the displayed quantity is not the
most appropriate one for diagnosing the presence of 
g-modes in the supernova core and neutron star.
(1) The mentioned figure provided by Burrows et al.\ was not meant
to display a signature of the core g-mode activity but was intended to
reflect the pressure fluctuations in the region between neutron
star and shock. The authors evoked this measure because it is a gauge
of the potential of pressure fluctuations to excite core g-mode
oscillations in the more quiescent interior (A.\ Burrows, private
communication). The primary origins of these ``outer'' pressure
fluctuations are, of course, the SASI and neutrino-driven convection,
not core pulsations. Without a well understood theory of the
excitation mechanism of the large-amplitude low-order g-modes in
the neutron star core, however, it is unclear what the external 
pressure fluctuations can tell about the fluid motions to be 
expected inside the neutron star.
(2) Moreover, 
pressure fluctuations are not an optimal measure of the strength
of core g-mode pulsations, which rely on buoyancy as the restoring
force. This is visible, for example, from the eigenfunctions of 
neutron star core g-modes given by Yoshida et al.\ (2007),
where the pressure-dependent amplitude appears unspectacular while
the Lagrangian displacement amplitude can still be large.
(3) Pressure fluctuations as a function of time at a fixed radius
do not reflect the evolution of the conditions 
at a certain enclosed mass, but they also include
a long-time evolutionary trend due to the
shrinking of the forming neutron star (and due to its growing
centrifugal deformation in the case of rotating models).
In particular at radii close to the boundary layer between the
neutron star and the violently turbulent, SASI-perturbed
postshock region, a gradual increase of the pressure fluctuations
with time does not necessarily signal a rising strength of the g-mode
driving force at the surface of the neutron star core.
Instead, the observed growth of the pressure fluctuations can just be
a consequence of the neutron star contraction, in course of which
the radial position moves out of the neutron star and deeper into 
the stirred SASI layer. The range of radii that is affected by this 
shift depends on the compactness of the neutron star and 
therefore on the softness of the high-density equation of state
and on the depth of the gravitational potential. Since 
we describe gravitational effects by the effective relativistic
potential, our neutron stars are more compact (in particular
with the softer EoS of Lattimer \& Swesty 1991) than those in
the Newtonian simulations of the Tucson group. Comparing
simulations performed with different nuclear equations of state
and different treatments of gravity (Newtonian or relativistic)
is therefore not conclusive when the same radius of evaluation
is picked. Simply for this reason a detailed assessment of the
similarities and differences between our simulations and those
of Burrows et al.\ (2006) is very difficult on grounds of the
limited data published. 

Accordingly, an evaluation of the fractional pressure
variations at different locations in the neutron stars of our
simulations, guided by Fig.~7 in Burrows et al.\ (2006),  
leaves too much room for different interpretations and does
not really illuminate the g-mode activity in the interior of the
nascent neutron star. We therefore decided to present a variety
of alternative quantities whose behavior can better reflect 
nonradial fluid perturbations in different regions of the supernova
core. 

\begin{figure*}
\plottwo{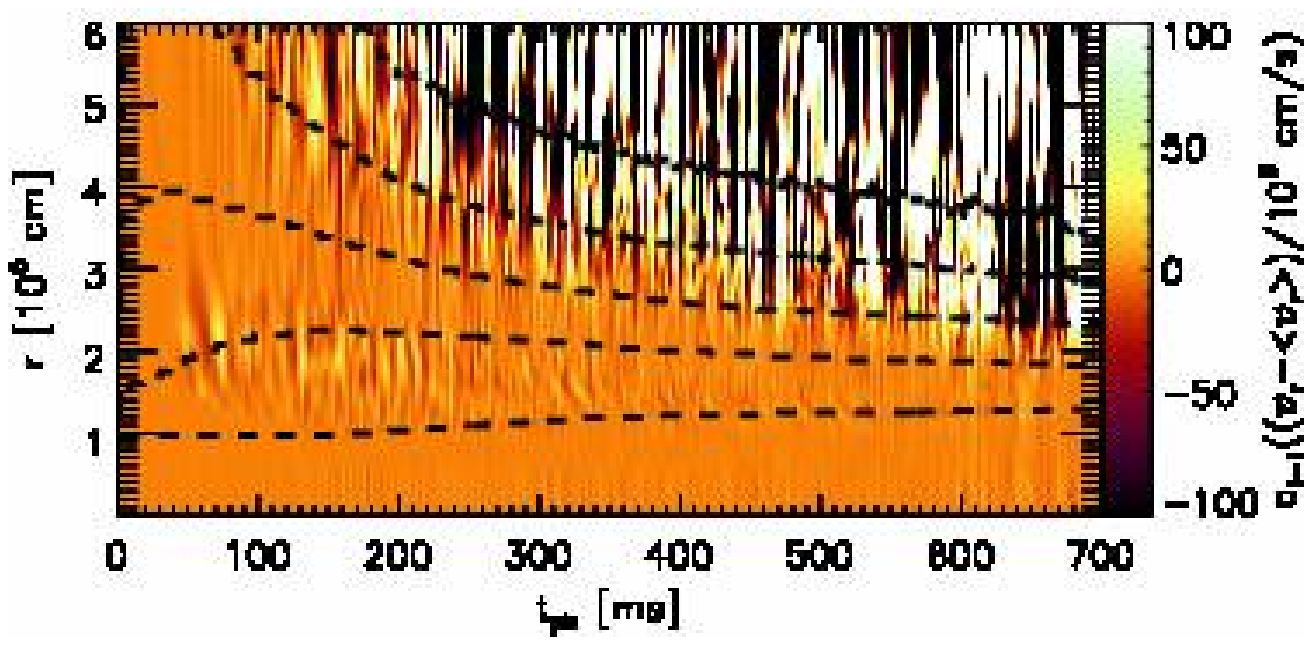}{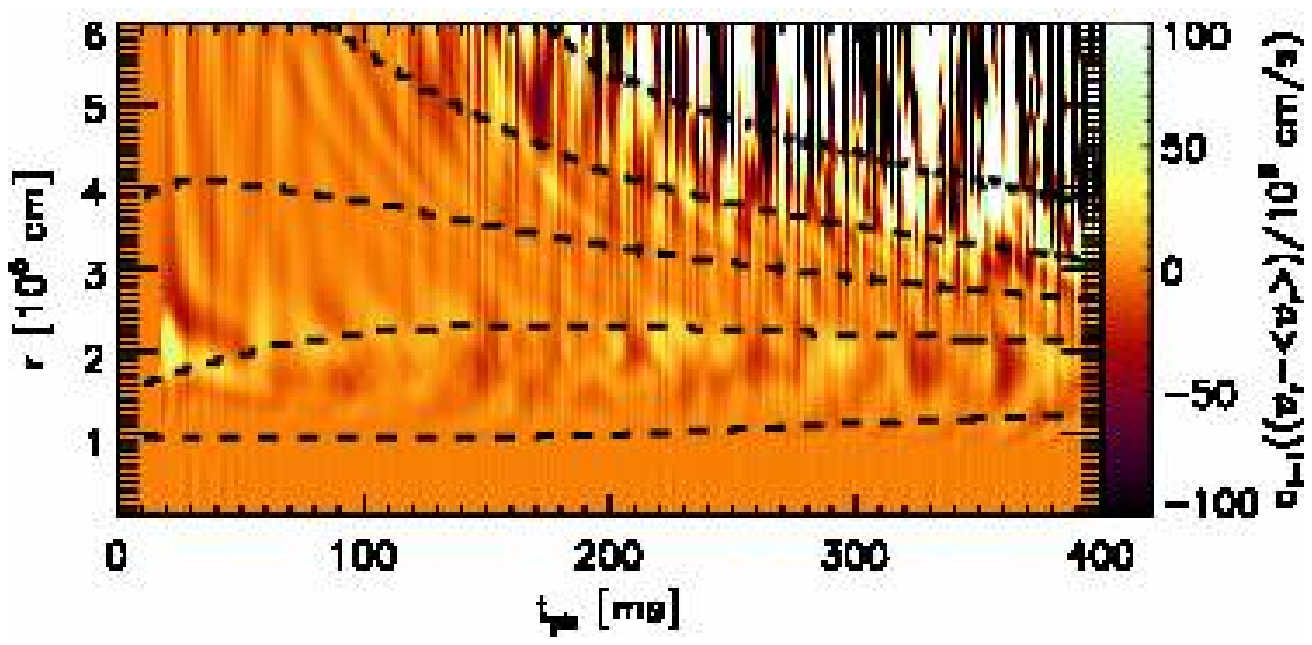}
\plottwo{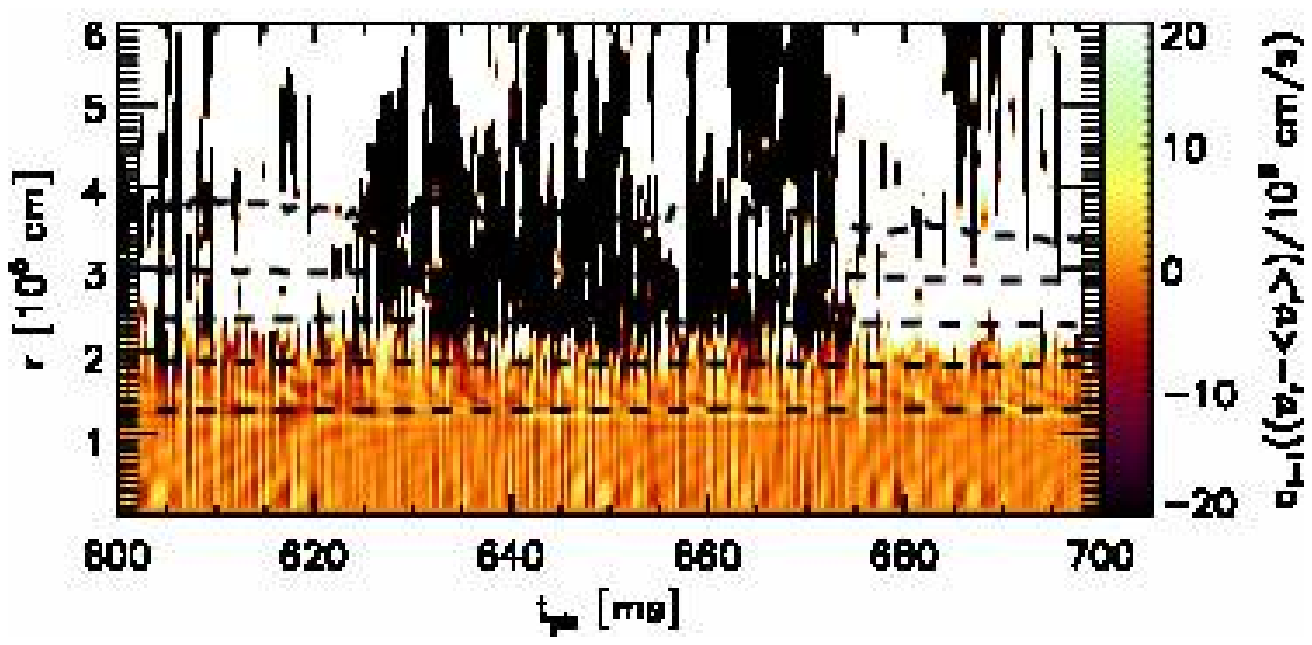}{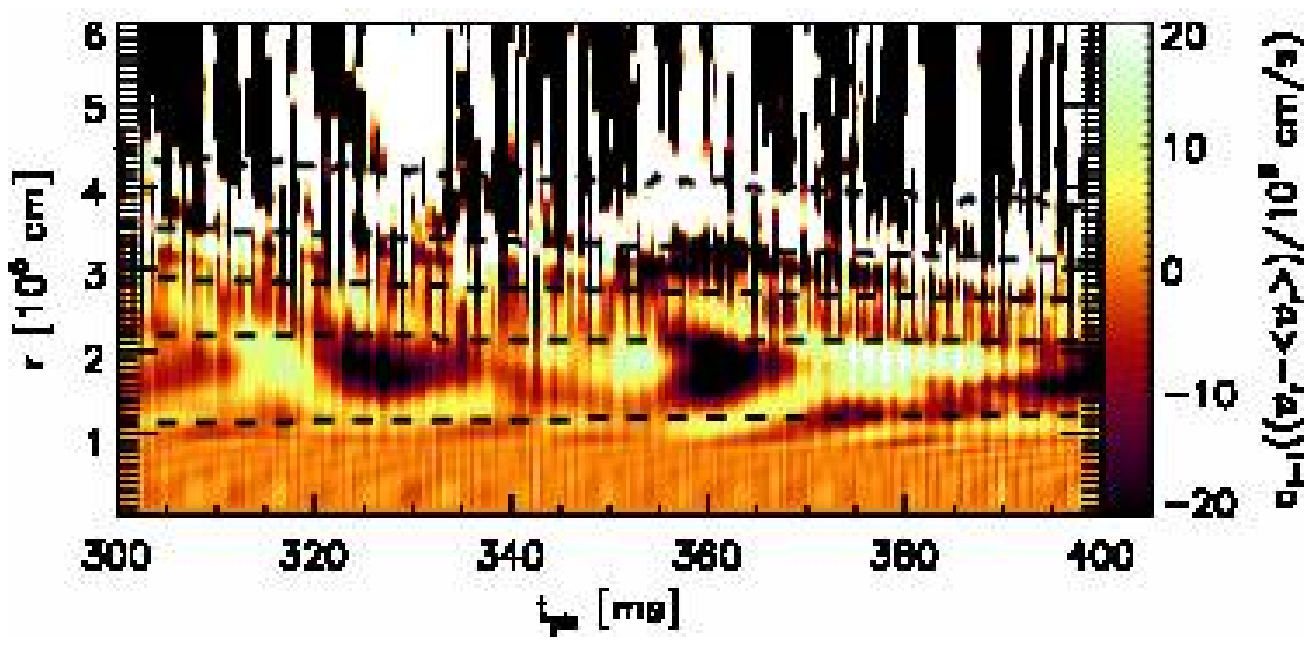}
 \caption{
Amplitude of the $l=1$-mode component of the spherical harmonics
decomposition of the velocity variations normalized by 
$10^6\,$cm$\,$s$^{-1}$,
$(v_r(r,\theta) - \left\langle v_r(r,\theta) \right\rangle_\theta)/
10^6\,\mathrm{cm/s}$, as a function of radius and post-bounce time
in our simulation
of the rotating 15$\,M_\odot$ model M15LS-rot ({\em left}) and in
the corresponding simulation without rotation, M15LS-2D ({\rm right}),
once for the whole post-bounce evolution ({\em top}) and another 
time for the last $\sim$100$\,$ms of both simulations ({\em bottom}). 
In both cases the central region with a radius of 60$\,$km is shown, 
but the color scale for the amplitude is constrained to a 
different range of values (much more narrow than the absolute maxima
and minima) in order to visualize
activity in different regions. The interior of the neutron star at
$r \la 10\,$km is much more quiet than the outer layers where SASI 
and convective motions perturb the surroundings of the proto-neutron 
star and stir g-mode activity in its surface layers. In particular,
there is no sign of any sizable core g-mode oscillations in the 
neutron star core. The dashed lines mark the positions where the
laterally averaged density has values of 
$\left\langle\rho\right\rangle_\theta = 
10^{14}$, $10^{13}$, $10^{12}$, $10^{11}$, and
$10^{10}\,$g$\,$cm$^{-3}$ (from bottom to top).
\label{fig:gmodes}}
\end{figure*}

\begin{figure*}
\plottwo{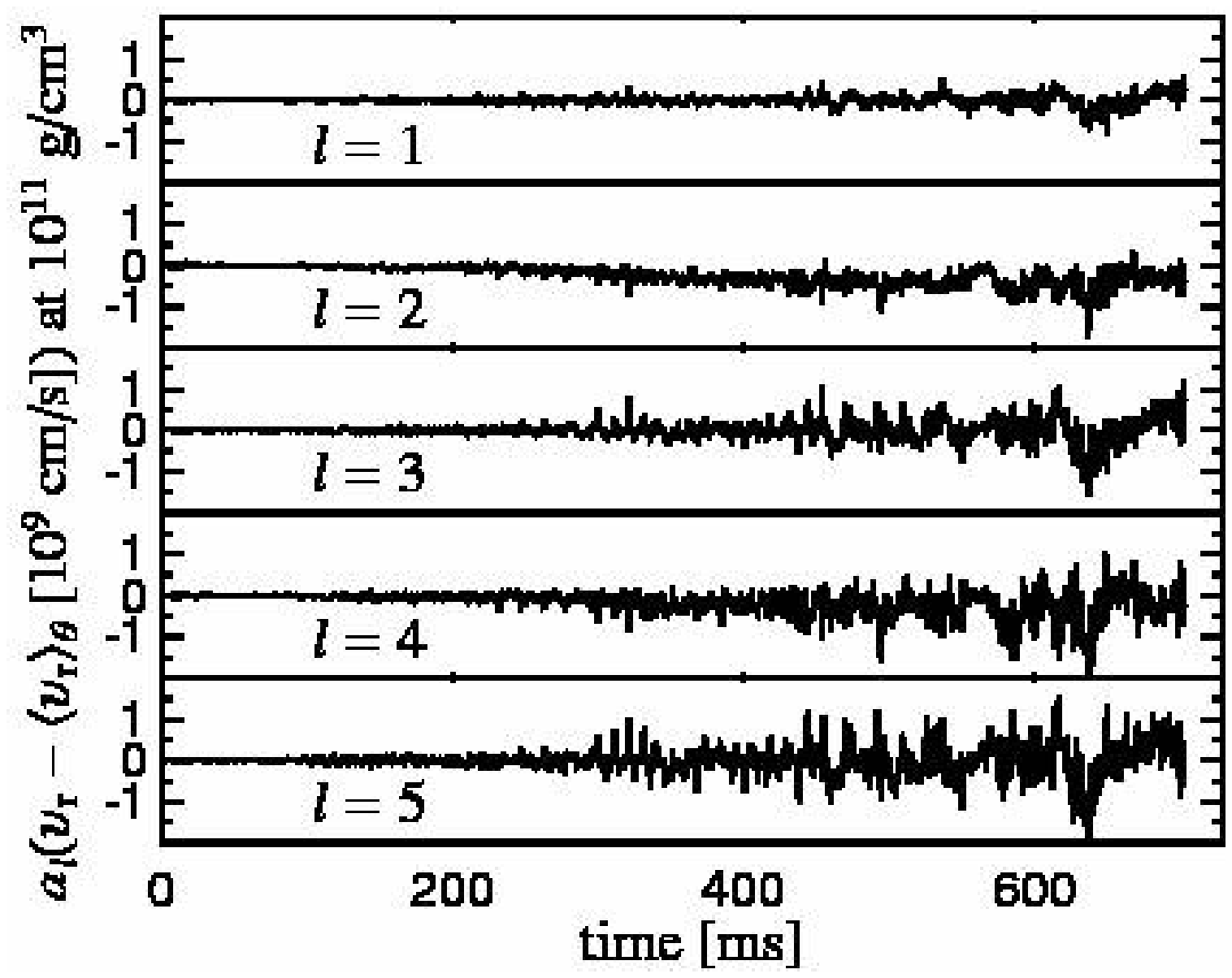}{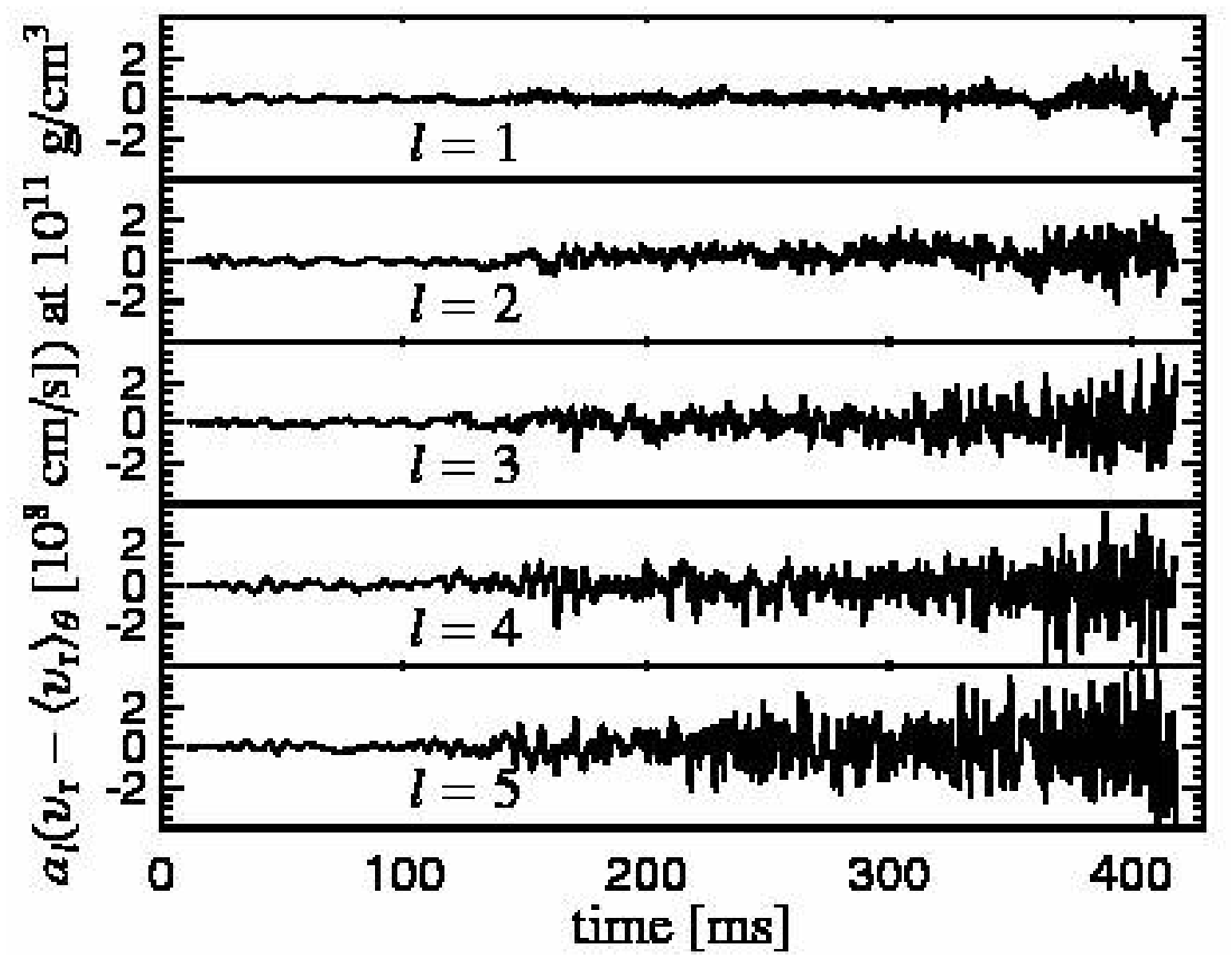}
\plottwo{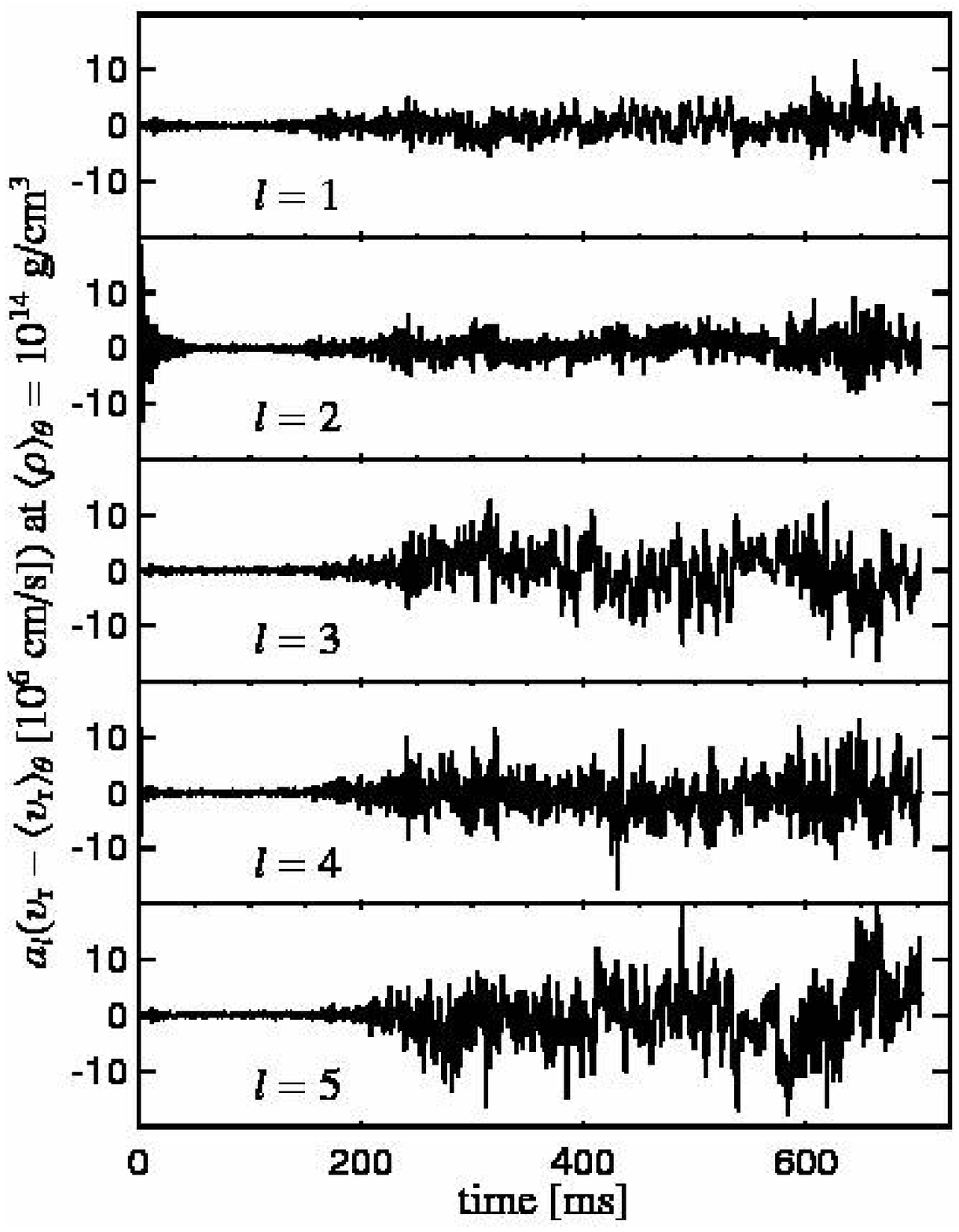}{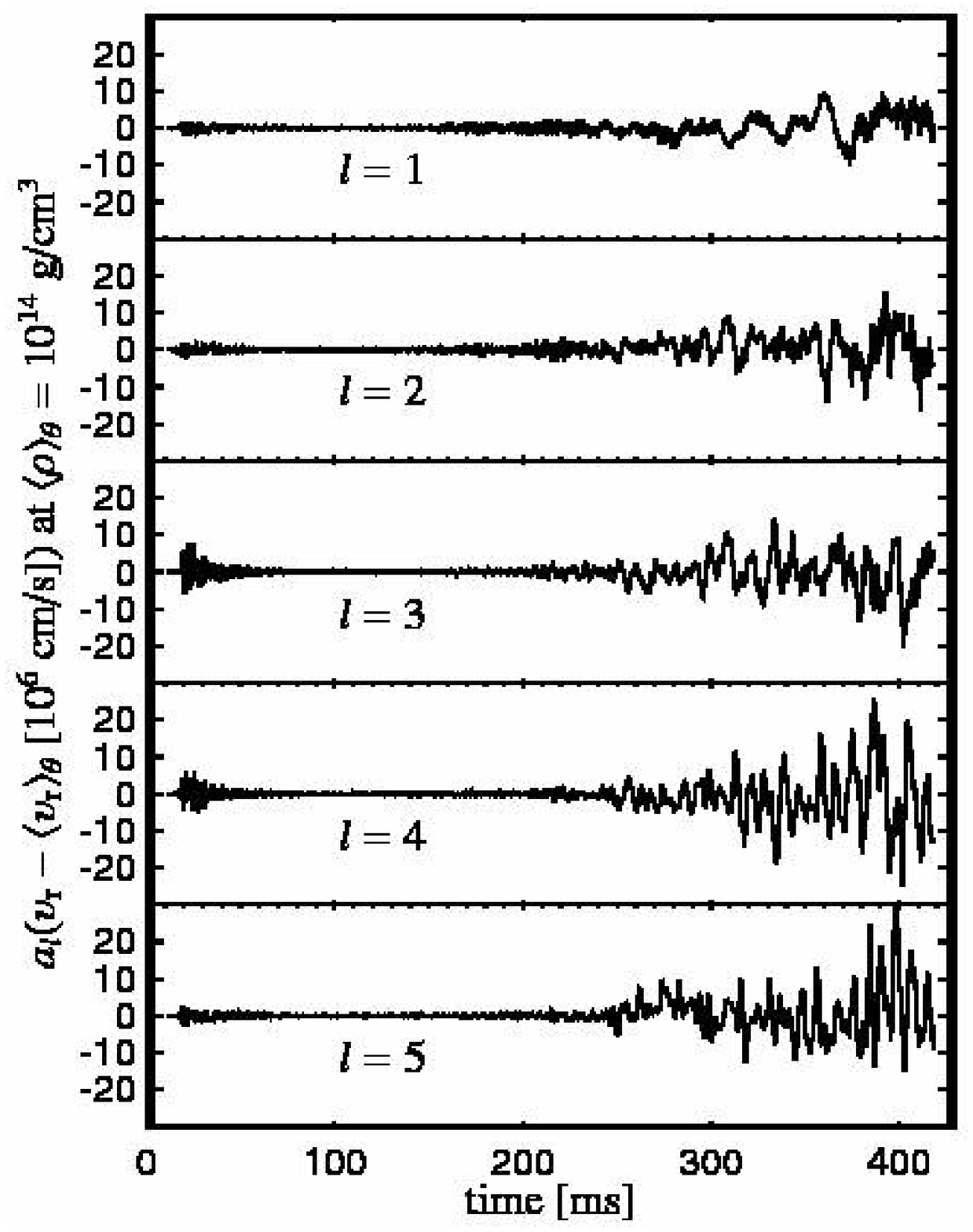}
 \caption{
Amplitudes of the $l=1,\,2,\,3,\,4,\,5$-modes of the spherical harmonics
decomposition of the velocity variations, 
$(v_r(r,\theta) - \left\langle v_r(r,\theta) \right\rangle_\theta)$, 
as functions of post-bounce time at
fixed values of the laterally averaged density (see the dashed lines
in Fig.~\ref{fig:gmodes}) for our Models
M15LS-rot ({\em left}) and M15LS-2D ({\em right}).
The upper panels show the amplitudes at a density of 
$\left\langle \rho\right\rangle_\theta = 10^{11}\,$g$\,$cm$^{-3}$
with a normalization by $10^9\,$cm$\,$s$^{-1}$ (left) and 
$10^8\,$cm$\,$s$^{-1}$ (right), the lower panels at 
$\left\langle \rho\right\rangle_\theta = 10^{14}\,$g$\,$cm$^{-3}$
with a normalization by $10^6\,$cm$\,$s$^{-1}$.
\label{fig:gmodes2}}
\end{figure*}

Figures~\ref{fig:gmodes} and \ref{fig:gmodes2} show the 
time evolution of low-mode spherical harmonics amplitudes (for
$l = 1,...,5$, defined in agreement with Eq.~(\ref{eq:expansion}))
of the velocity perturbations, $v_r(r,\theta)-
\left\langle v_r(r,\theta)\right\rangle_\theta$, in the central
regions of Models~M15LS-rot and M15LS-2D, the former figure 
between the neutron star center and a radius of 60$\,$km, the
latter figure at the positions where the laterally averaged
density has values of
$\left\langle \rho\right\rangle_\theta = 10^{11}\,$g$\,$cm$^{-3}$
and $10^{14}\,$g$\,$cm$^{-3}$. It is obvious that the interior of
the neutron star at high densities, where the second density
location is chosen to be, is much more quiet than the environment 
of the neutron star and its surface-near layers, which are 
represented by the
first value of the density. The stellar plasma in the neutron
star surroundings is stirred by the violent SASI and convective
overturn motions, which instigate g-mode activity in the outer 
layers of the compact remnant. Figures~\ref{fig:gmodes} and 
\ref{fig:gmodes2} reveal that the maximum amplitudes of the 
velocity perturbations deep inside the neutron star are of the 
order of $10^7\,$cm$\,$s$^{-1}$, whereas they are up to a factor
100 higher in the SASI region and adjacent shells. The larger 
values in Model~M15LS-rot compared to Model~M15LS-2D are a 
consequence of the rotational flattened compact remnant
in the former simulation, which causes the radius for
$\left\langle \rho\right\rangle_\theta = 10^{11}\,$g$\,$cm$^{-3}$
to reach deeper into the SASI-perturbed layer above the poles
of the neutron star.

One should note that the contour for an average density of 
$10^{14}\,$g$\,$cm$^{-3}$ is located near the inner edge of 
the convectively unstable shell inside the neutron star, at the
transition of this layer to the convectively stable region 
closer to the center (roughly at $r \sim 10\,$km).
A major fraction of the activity
seen at this location is therefore linked to convection and
convective overshooting and not to
pure gravity waves, but the maximum amplitudes are in the range 
typically found also at positions deeper inside the core. 
In Model~M15LS-2D the dipole amplitude of the
velocity perturbations exhibits a characteristically 
different low-frequency modulation, which
is absent in Model~M15LS-rot (see Fig.~\ref{fig:gmodes2} for
$\left\langle\rho\right\rangle_\theta = 10^{14}\,$g$\,$cm$^{-3}$), 
where it is also more difficult
to discern the convective region inside
the neutron star (Fig.~\ref{fig:gmodes}). This can 
probably be explained by the nature of convection in
rapidly rotating neutron stars, where due to the Solberg-term
in the instability criterion the convective cells are
constrained to tube-like, narrow structures oriented parallel to
the rotation axis and located in regions where the angular momentum 
gradient is
flat (see also Sects.~\ref{sec:postbounce} and \ref{sec:neutrinos}).
The effects of the convection in such an environment are therefore
strongly damped compared to the nonrotating case.

Interestingly, at a given density and time,
all modes from $l=1$ to $l=5$ are similarly strong, with a 
tendency of slightly (roughly a factor of two) higher amplitudes
for $l = 3,\,4,\,5$ (Fig.~\ref{fig:gmodes2}). In the innermost
core of the neutron star (at $r < 10\,$km) the opposite effect 
is present and the difference of the amplitudes is larger:
the amplitudes for $l=1$ and $l=2$ are the clearly dominant
ones during most of the post-bounce evolution, $l=3$ is 
intermediate, and $l=4,\,5$ are significantly weaker.

Consistent with the activity level observed in the velocity
perturbations, we also find very small 
fractional density variations at given positions, 
at most some tenths of a percent in the neutron star interior
and about 10\% just outside of the neutrinosphere. 
There is no sign of any significant coherent movement
of the neutron star core as suggested by Fig.~3 in
Burrows et al.\ (2007c), where the isodensity surfaces show
large relative displacements of their geometrical centers 
instead of being concentric.

\begin{figure*}
\plottwo{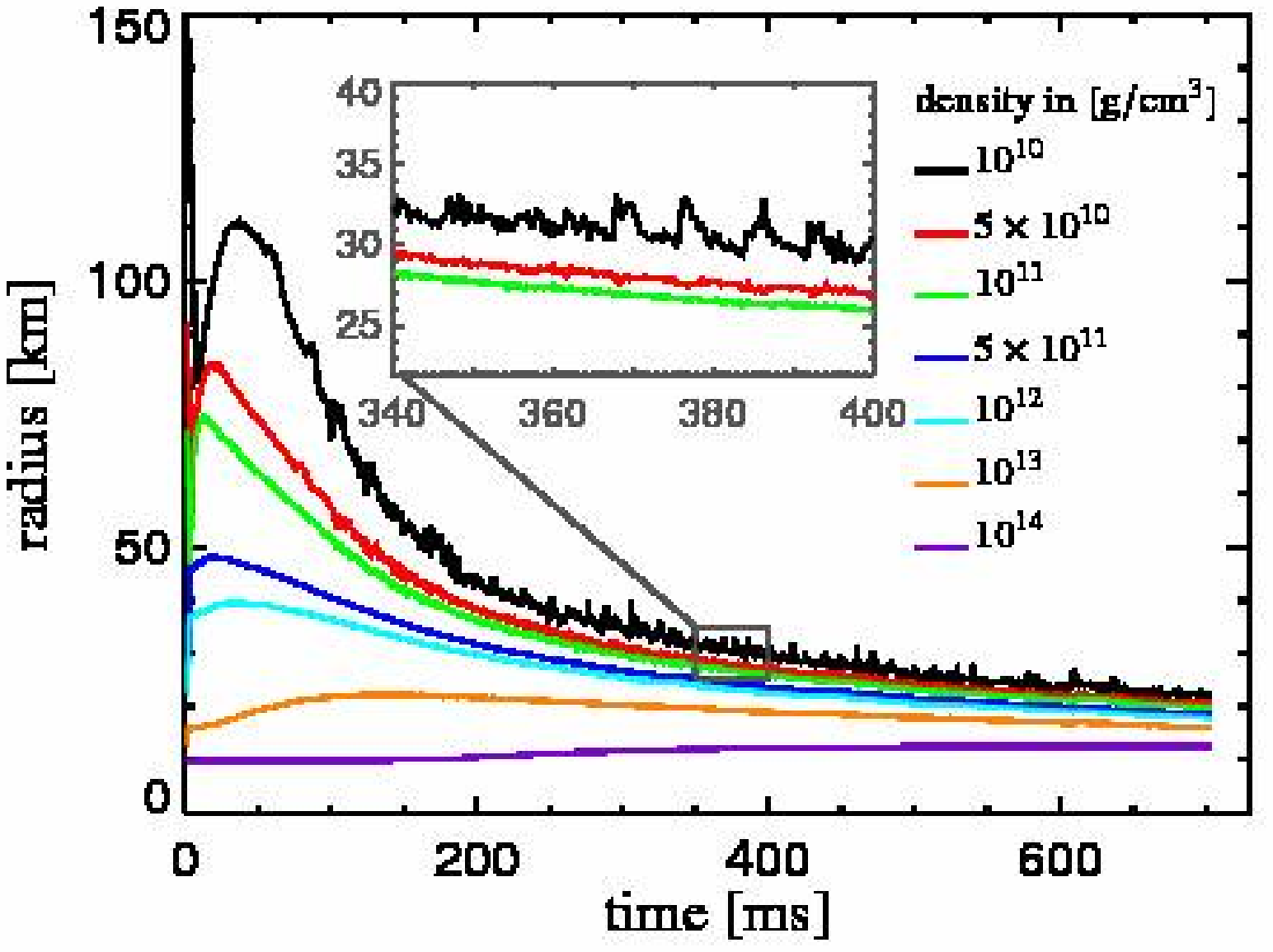}{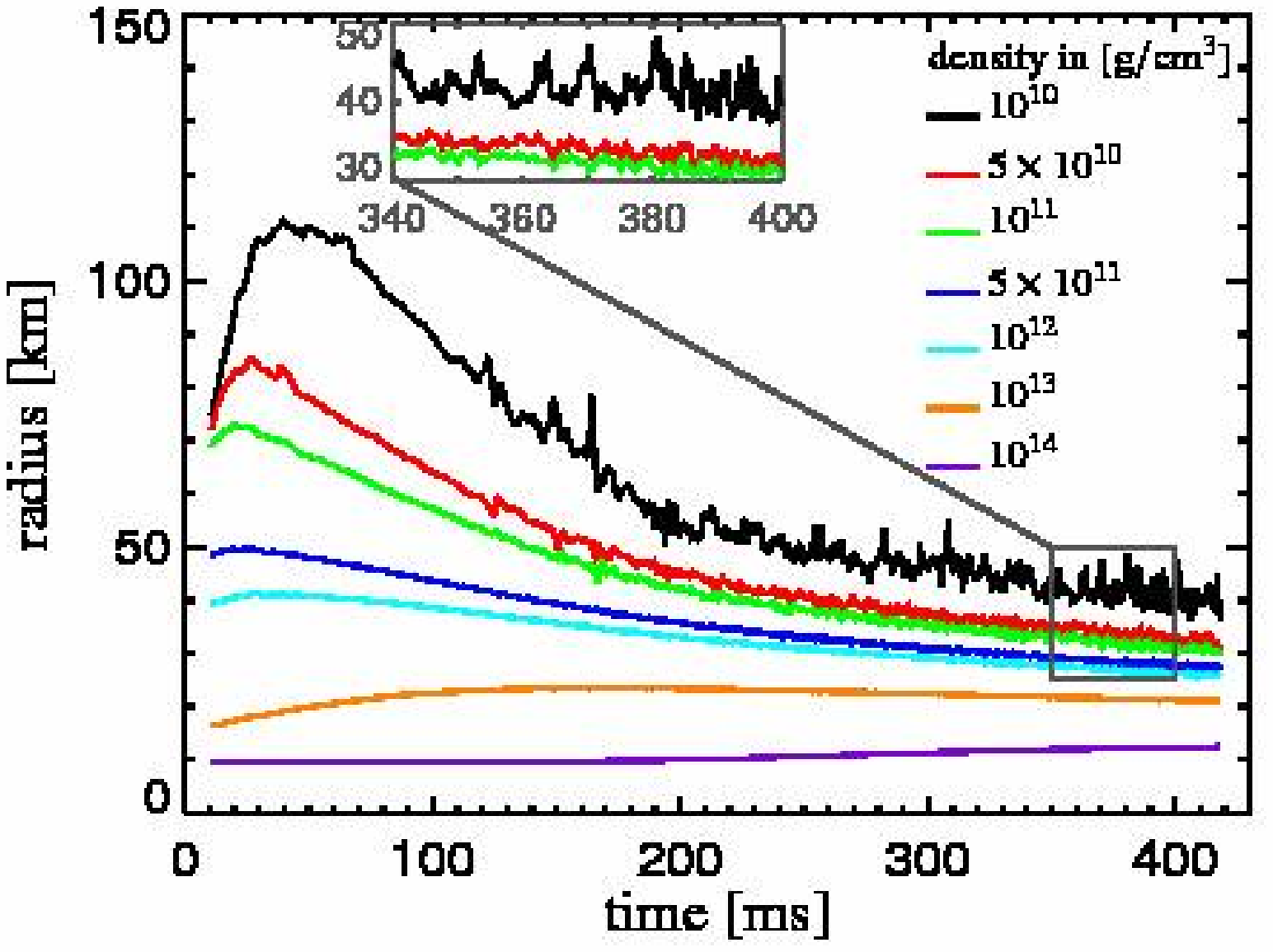}
\plottwo{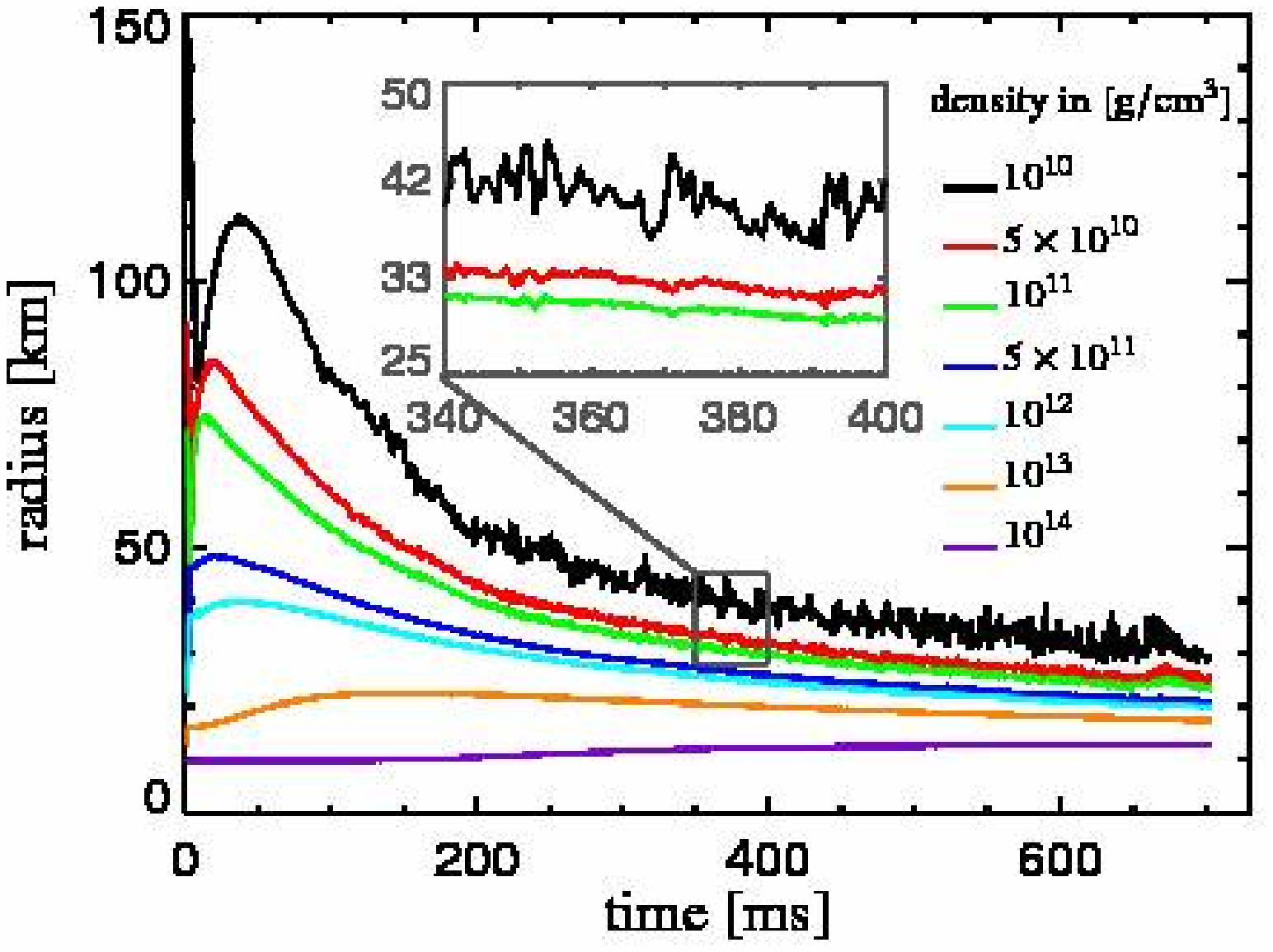}{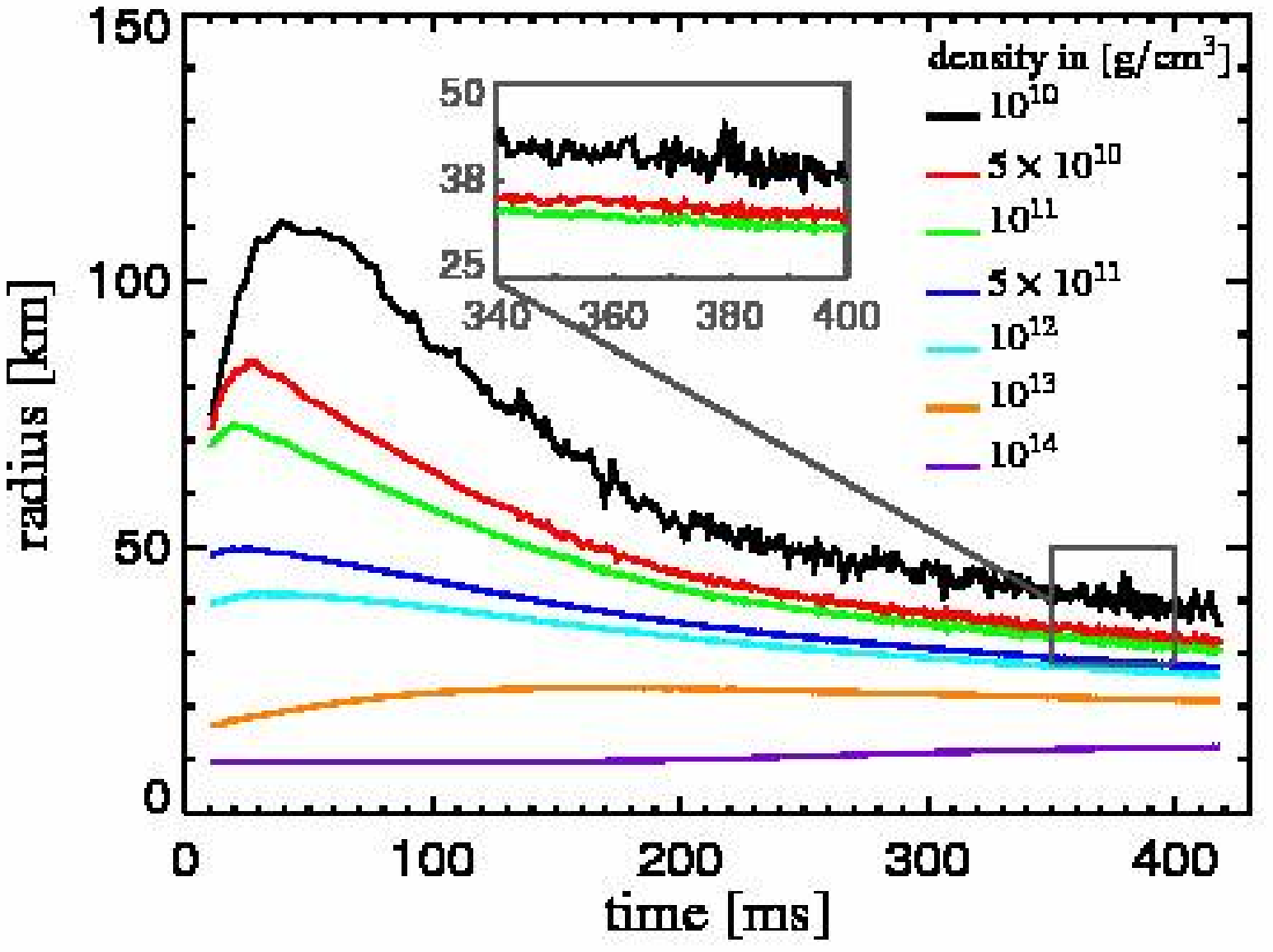}
 \caption{
Radius variations $r_\rho(t)$ as functions of post-bounce time
of the locations for different values
of the local density (as specified in the plots) at a position near
the pole of the spherical coordinate grid ({\em upper panels})
and at a latitudinal angle of 45 degrees ({\em lower panels})
in the nascent neutron star of 
Model~M15LS-rot ({\em left}) and Model~M15LS-2D ({\rm right}).
The displayed variations are indicative for major mass motions and
mass displacements associated for example with convective overturn
or g-mode oscillations. One can see that large amplitudes are present
only in the outermost layers of the neutron star at densities
below some $10^{10}\,$g$\,$cm$^{-3}$ (see also the zoom in the 
inset), where the violent SASI and 
convective activity in the postshock layers around the neutron 
star makes an impact. The interior of the neutron star is 
essentially quiet, in particular there is no sign of any sizable
core g-mode oscillations or pulsational motions of the neutron
star core. 
\label{fig:gmodes3}}
\end{figure*}

In order not to just rely on a potentially misleading
interpretation of local fluctuations of the mentioned 
quantities, we have also evaluated Models~M15LS-rot and M15LS-2D
for the time-dependent variations of the radial positions of 
certain density values in chosen directions (Fig.~\ref{fig:gmodes3}). 
In the case of a coherent translatory movement of some core region,
which for symmetry reasons in the 2D 
simulations has to occur along the $z$-axis of the polar grid, this
would also be a rough measure of spatial shifts of the geometric 
center of isodensity surfaces. Such
variations are not found on the scale of the plots in
Fig.~\ref{fig:gmodes3} for matter at densities above
$\rho > 10^{11}\,$g$\,$cm$^{-3}$, while they are around 0.5--1$\,$km
for densities between $5\times 10^{10}\,$g$\,$cm$^{-3}$ and
$10^{11}\,$g$\,$cm$^{-3}$ and even larger only for lower densities
outside of the neutron star\footnote{It is important to note
that in Fig.~\ref{fig:gmodes3} in contrast to Fig.~\ref{fig:gmodes2},
the evaluation is not performed at the position of a laterally
averaged density value, but at the local density in a chosen
direction. In the case of a rotationally deformed neutron star 
the polar radius for a certain density is significantly smaller
than the radius that corresponds to the same value of the
lateral density average. The former position is therefore 
close the neutron star surface, while the latter position is
deeper in the SASI-perturbed surroundings of the neutron star.}. 
There, however, they originate from
the SASI and convective overturn activity, which also stirs the
neutron star surface layer below. This interpretation is consistent
with the fact that one detects radius variations with the same
behavior and similar or even larger magnitude in 
equator-near regions, where 
violent downflow activity is also observed during the accretion
phase of the stalled shock (Sect.~\ref{sec:results}). 
A direct inspection of the geometrical centers of the 
isodensity surfaces for the density values considered in 
Fig.~\ref{fig:gmodes3} reveals full agreement with the message
and quantitative information
in this figure: the geometrical centers exhibit $z$-displacements
relative to the coordinate center of less than $\sim$0.1$\,$km for
densities above $10^{11}\,$g$\,$cm$^{-3}$, at most 0.2$\,$km
for $\rho = 10^{11}\,$g$\,$cm$^{-3}$, some 0.1$\,$km for 
$\rho = 5\times 10^{10}\,$g$\,$cm$^{-3}$, and 1--3$\,$km only
for $\rho \la 10^{10}\,$g$\,$cm$^{-3}$. Taking all together we 
again conclude that there is no evidence for large-amplitude
core g-modes in the deep interior of the nascent neutron star
in our simulations.

\subsubsection{Acoustic energy input to explosion}
\label{sec:acousticenergy}

Although core g-modes do not seem to play a significant role in 
the discussed models, the surface gravity waves caused by 
the impact of accretion downflows in the outer layers of the 
neutron star could still be an important source of sonic 
waves and secondary shocks, which could contribute energy 
input to the developing explosion. Because of the violently
turbulent and time-dependent conditions in the neutron star 
surroundings, but also because of the possibility to liberate 
or absorb
energy by changes of the composition of the stellar gas, it is 
quite difficult to assess in a quantitatively reliable way 
how important this energy input rate really is. 
Burrows et al.\ (2006, 2007c) therefore refer to a crude 
analytic formula that sensitively depends on the product of
several factors whose values are not well constrained by the
numerical results (see Eq.~(1) in both papers). For checking
the plausibility of their assumptions they compare their
estimated numbers with the available rate of energy
release from gas accretion onto the forming neutron star.

Our approach here is different. Making use of our hydrodynamical 
results, we attempt to estimate the rate of acoustic energy
transfer to the ejected matter by sound waves, 
$\dot E_{\mathrm{sw}}$, as the difference of all other energies that
play a role in the evolution equation for the volume integral 
of the total (i.e., internal plus kinetic plus
gravitational) energy in some region behind the supernova shock.
During the phase of simultaneous accretion and outflow of matter
(stage (2) described in Sect.~\ref{sec:expenergy}) one can write:
\begin{eqnarray}
\dot E_{\mathrm{sw}} &\,\approx\,& 
\partial_t E_{\mathrm{tot}} \nonumber \\
&\,+\,& 
\oint_{S|v_r > 0} \mathrm{d}\Omega\, r^2 v_r
\left ( \varepsilon_{\mathrm{int}} + P + 
\varepsilon_{\mathrm{kin}} - {G M \rho\over r}\right ) \nonumber \\
&\,-\,& 
\zeta\,\int_{V_{\mathrm{g}}}\mathrm{d}V \,\rho \dot q_\nu \,-\,
\oint_{S|v_r > 0} \mathrm{d}\Omega\, r^2 v_r
\rho e_{\mathrm{bind}}^{\mathrm{acc}} \nonumber \\
&\,\equiv\,& \partial_t E_{\mathrm{tot}} \,+\, L_{\mathrm{out}}
\,-\, \zeta\,\dot Q_\nu \,-\, L_{\mathrm{bind}}^{\mathrm{acc}} \ ,
\label{eq:totenergy}
\end{eqnarray}
where for simplicity (and to a very good approximation) we 
assume a spherically symmetric, Newtonian gravitational potential.
In this equation $E_{\mathrm{tot}}$ is the total energy in the
volume of integration, $v_r$ the radial velocity component, 
$\varepsilon_{\mathrm{int}}$ and 
$\varepsilon_{\mathrm{kin}}$ the densities of internal and 
kinetic energy, respectively, $P$ the gas pressure, $\rho$ the
mass density, $\dot q_\nu$ the local rate of neutrino energy
deposition, $\mathrm{d}\Omega r^2$ is the surface element for
the surface integrals, and $\mathrm{d}V$ the volume element for
the volume integrals. The surface integrals are carried out over  
a sphere $S(r)$ at radius $r$, which is located behind the shock
in all parts. The integration includes only material that 
flows through this
sphere with positive radial velocity, a constraint that is 
indicated by $S|v_r > 0$. The volume integration, which also
leads to $E_{\mathrm{tot}}$, has to be performed over the matter 
that fills the volume between the direction-dependent gain radius, 
$R_{\mathrm{g}}(\theta)$, and the radius $r$, {\em and} which
is ultimately able to leave the gain layer through the surface 
$S(r)$ with positive radial velocity. The radius $r$ is chosen
large enough so that the neutrino energy deposition at bigger
radii is negligibly small. If the volume integration
was done for {\em all} matter in the gain layer, i.e.\ over the
whole volume $V_{\mathrm{g}}$, an extra term
would have to be introduced into Eq.~(\ref{eq:totenergy}) to account
for the energy loss associated with matter that is 
advected inward through the gain radius to be accreted onto
the neutron star. Because of the highly turbulent flow around
the gain radius and the corrugated structure of the gain 
surface, it is very difficult to evaluate the 
corresponding surface integral. We therefore avoid this term and
instead introduce the factor $\zeta$ defining the fraction of
the neutrino energy that is deposited in the material swept up 
by the shock and 
ultimately ending up as ejected matter (see Eq.~(\ref{eq:exprate})
and associated discussion). One can relatively easily obtain a 
time-averaged number for $\zeta$ by comparing
the mass accretion rate through the shock and the growth rate of 
the mass below the gain radius. 

The first term on the rhs describes the gain of energy in the
integration volume, the second term the loss of energy by ejecta
mass leaving the surface $S(r)$, the third term the total energy
input to the ejecta by neutrino heating, and the last term
accounts for the fact that energy input by neutrinos (and possibly
other sources like sound waves) does not only have to yield
the positive energy carried by the ejecta, but has to provide
also the energy to overcome the gravitational binding of the
progenitor gas that enters the gain layer through the shock. 
Refering
to the detailed arguments given in Sect.~\ref{sec:expenergy},
we calculate the internal energy always such that
all baryons are assumed to be assembled in iron-group nuclei.
This allows us to ignore in Eq.~(\ref{eq:totenergy}) 
additional terms that describe
the sizable amounts of energy that can be exchanged between
the reservoirs of rest-mass and internal energy by processes
altering the nuclear composition of the medium, i.e.\ by
nuclear burning and photodissociation of nuclei in the accretion
flow and by nucleon recombination in the outflow.
The quantity $e_{\mathrm{bind}}^{\mathrm{acc}}$ in the last term
on the rhs of Eq.~(\ref{eq:totenergy}) is the (negative) 
specific binding energy (computed again
as the sum of the specific internal, kinetic, and gravitational
energies) of matter that flows in accretion funnels from the shock 
through the surface $S(r)$ towards the neutron star. This energy
--- with the nuclear binding energy included as mentioned
before --- is typically around $-$1 to $-$2$\,$MeV per nucleon
during the considered accretion phase. It is roughly constant 
when the gas falls inward, because the energy absorbed from
neutrinos is relatively small when the matter is still located at 
large distances from the neutron star and is also small when the
accretion flows move with high velocities and thus are exposed to
neutrino heating only for a very short period of time.
Therefore its value is not very sensitive to the radius where
it is determined\footnote{Bernoulli's theorem implies that
the constant energy functional includes the specific 
enthalpy $w = (\varepsilon_{\mathrm{int}} + P)/\rho$
instead of the specific internal energy. Thus the constant 
quantity is actually $e_{\mathrm{bind}}^{\mathrm{acc}} + P/\rho$ 
and not $e_{\mathrm{bind}}^{\mathrm{acc}}$. This
difference, however, is not essential for our discussion.}. 

We have performed the evaluation for Model~M15LS-rot in a time
interval around 690$\,$ms after bounce near the end of our 
simulation. At this time the
mass accretion rate through the shock is $\dot M_{\mathrm{acc}}
\sim 0.23\,M_\odot\,$s$^{-1}$ and the gas accretion onto the 
neutron star is still going on. A time-averaged fraction of 
$\zeta\sim\,0.4$--0.5 of the infalling mass is fed back into
the outflow of ejecta, and the conditions
in the region between the gain radius $R_{\mathrm{g}}(\theta)$ 
and the radius $r$ of the sphere $S$ can be assumed as nearly
stationary, in which case $\partial_t E_{\mathrm{tot}} = 0$
is a reasonably good approximation. Nevertheless, there are
still significant time variations in the different terms in
Eq.~(\ref{eq:totenergy}). For a mass outflow rate through $S$
of $\dot M_{\mathrm{out}}\sim \zeta \dot M_{\mathrm{acc}} \sim
0.05$...0.1$\,M_\odot\,$s$^{-1}$ ($\sim \dot M_{\mathrm{gain}}$,
the growth rate of the mass in the gain layer),
we obtain an outflow luminosity of 
$L_{\mathrm{out}}\sim (9...14)\times 10^{50}\,$erg$\,$s$^{-1}$ 
and a flow rate of gravitational binding energy associated with
the infalling and reejected mass of
$L_{\mathrm{bind}}^{\mathrm{acc}}\sim \dot M_{\mathrm{out}}
\left\langle e_{\mathrm{bind}}^{\mathrm{acc}}\right\rangle
\sim -(1.5...3) \times 10^{50}\,$erg$\,$s$^{-1}$ 
(we find an average value of
$\left\langle e_{\mathrm{bind}}^{\mathrm{acc}}\right\rangle 
\sim -1.5\,$MeV per nucleon in
the accretion funnels of Model~M15LS-rot). With a relevant neutrino
energy deposition rate of $\zeta\dot Q_\nu\sim (10...15)\times
10^{50}\,$erg$\,$s$^{-1}$ the rhs of Eq.~(\ref{eq:totenergy})
yields an upper limit for additional energy input to the 
developing explosion
by acoustic waves of $(0.5...2)\times 10^{50}\,$erg$\,$s$^{-1}$.
The acoustic energy flux originating from the violent fluid 
motions caused by the impact of accretion flows near the neutron
star surface is therefore at least a factor
of $\sim$10 lower than the energy deposition by neutrino heating
in our simulations. Until 700$\,$ms after bounce there is no
convincing evidence of any significant energy transfer to the 
shock by pressure waves originating from the neutron star.
Our 15$\,M_\odot$ model M15LS-rot as well as the 11.2$\,M_\odot$ 
progenitor investigated by Buras et al.\ (2006b) and in this paper 
(Sect.~\ref{sec:model11.2})
develop neutrino-powered explosions long before
the acoustic mechanism was found to cause an explosion in the
simulations of Burrows et al.\ (2006, 2007a,c).

\subsubsection{Momentum conservation}
\label{sec:momentum}

\begin{figure*}
\plottwo{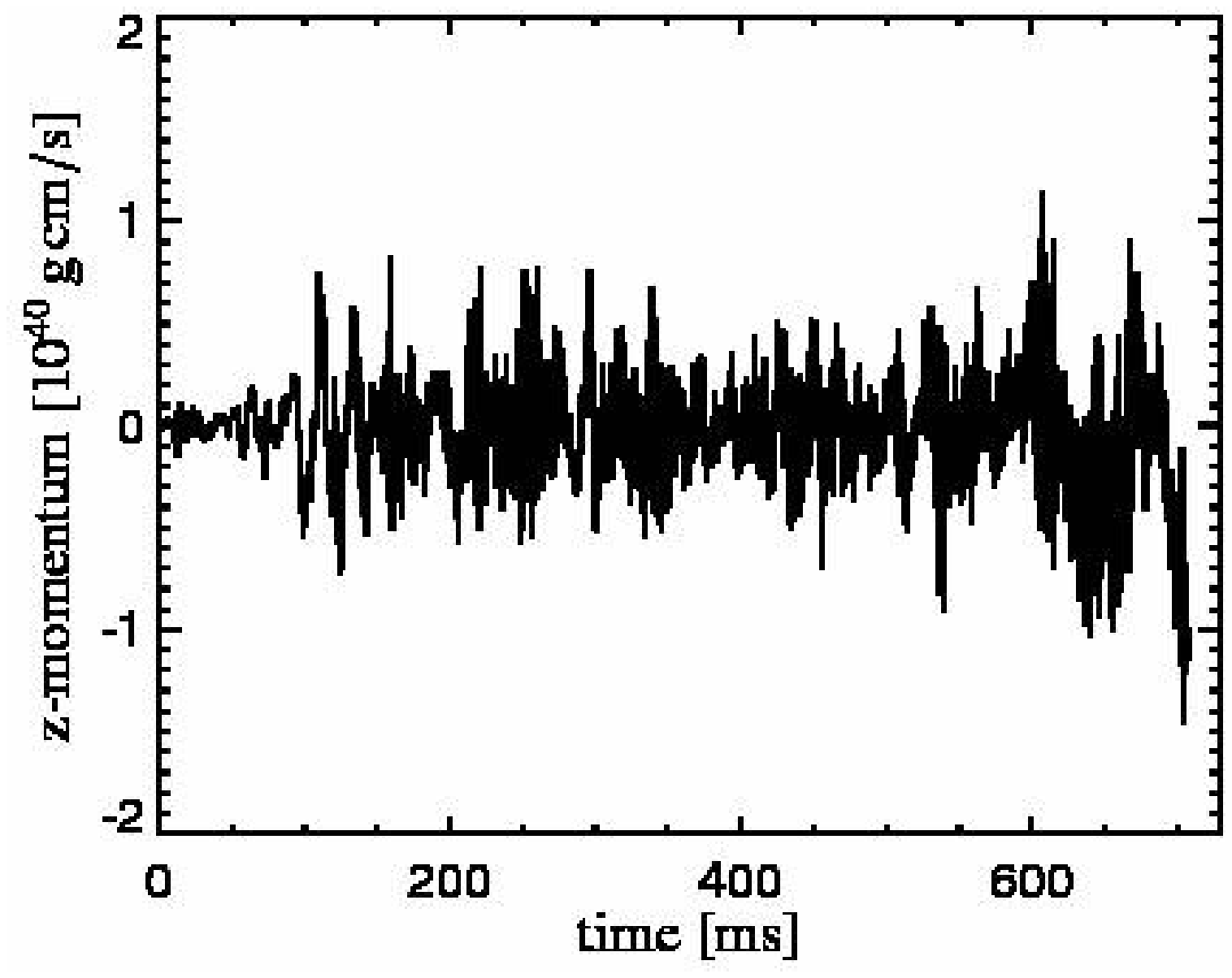}{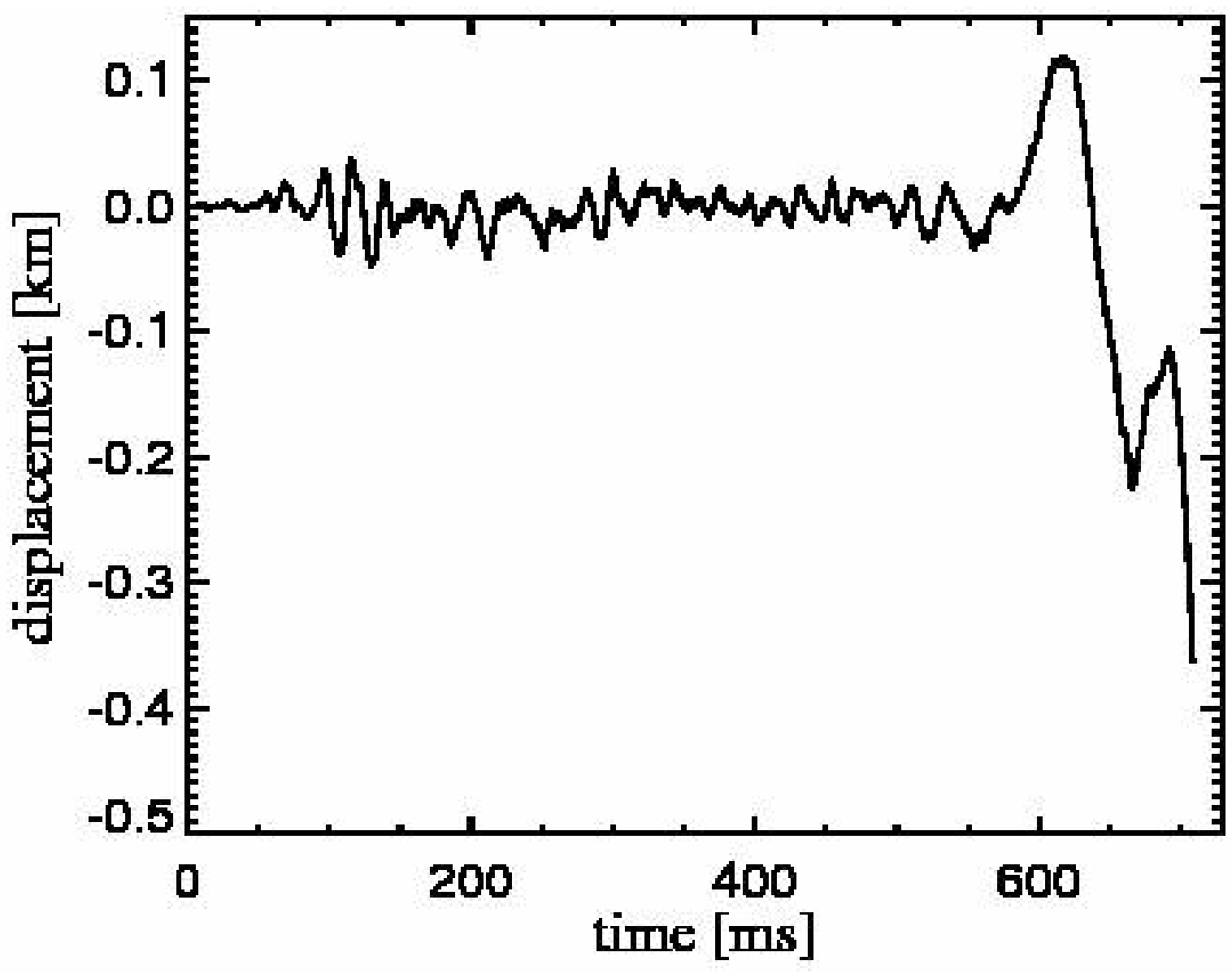}
 \caption{
Momentum conservation in Model~M15LS-rot. The left plot shows the
fluctuating total momentum in the direction of the polar (rotation)
axis for all gas on the grid, the right plot displays the
corresponding displacement of the center of mass from the grid center as
function of post-bounce time. Until near the end of the computed
evolution, this displacement is far less than the radial width of the
innermost grid zone (0.3$\,$km). Only when the postshock gas gains
momentum as the explosion takes off, which happens with more strength
in the southern hemisphere and thus corresponds to a growing
negative momentum value, the $z$-displacement exhibits a clear trend
and becomes slightly larger than the innermost grid zone.
\label{fig:momentumcons}}
\end{figure*}

But is our code actually able to follow the excitation and 
evolution of core g-mode oscillations,
in particular of $l=1$ type, in which case the gas in
the stellar center participates in the motion? Since a few radial
zones in the central $\la$1.7$\,$km of our numerical grid are treated 
in spherical symmetry (to get around too restricted CFL timesteps), 
one might wonder whether a flow pattern with gas motion through
the stellar center can be reasonably well described. Such
a spherical core might create severe perturbations, for example
because the central mass, which cannot move out of the grid center,
absorbs linear momentum and thus may cause a significant violation of
momentum conservation. In Fig.~\ref{fig:momentumcons}, left plot,
the evolution of the total linear momentum of the whole gas on the 
computational grid (about 2$\,M_\odot$, of which about three
quarters are contained in the neutron star) is shown for 
Model~M15LS-rot. The right plot of this figure displays the
corresponding displacement that the center of mass of the gas experiences
relative to the grid center in the direction of the polar grid axis
(which coincides with the rotation and symmetry axis of the 2D model).

The quantities in Fig.~\ref{fig:momentumcons} exhibit variability 
on two timescales, which can be easily discerned. A long-period
modulation with a typical frequency of 30--80$\,$Hz (timescales
of roughly 15--30$\,$ms) is the consequence of the SASI sloshing and
convective overturn in the postshock layer (cf.\ Fig.~\ref{fig:modes}) 
and the corresponding changes of the mass distribution around the
neutron star. This effect is superimposed by high-frequency
fluctuations with periods of only milliseconds, which is the 
typical timescale of small-scale mass motions in the close 
vicinity of the neutron star and of gravity waves in the neutron
star surface and core. Correspondingly, the 
linear momentum along the $z$-axis changes its sign with a 
frequency of several 100$\,$Hz and the associated net 
displacement is tiny. Slightly larger, but still very small is
the displacement caused by the low-frequency variations: the center
of mass of the whole gas wobbles around the grid center with an
amplitude of usually less than 50 meters, which is much smaller
than the radius of our innermost grid cell (300 meters).
The separation of the center of mass from the grid center
grows larger only when the outward shock expansion becomes very
strong and the postshock layer gets inflated with a pronounced
global deformation at $t \ga 600\,$ms (see Fig.~\ref{fig:snapshots}).
A clear trend in one direction seems to be
established towards the end of the computational run (with a
final displacement, however, that is still only insignificantly 
bigger than the first radial grid zone) as a consequence
of the clear dipolar asymmetry of the accelerating postshock gas,
which attains more momentum in the southern hemisphere (bottom right 
panel in Fig.~\ref{fig:snapshots}). 
We observe a very similar behavior in the case of our 11.2$\,M_\odot$ 
explosion model, where the $z$-momentum in the northern hemisphere
dominates when the simulation is terminated. The momentum asymmetry 
of the supernova gas could ultimately imply a sizable recoil 
of the neutron star in the opposite direction, which should drive
the neutron star away from the coordinate center (possibly leading
to numerical problems), but our simulations are not carried on
long enough to be conclusive. In Model~M15LS-rot the velocity 
of the compact remnant is only $\sim 20\,$km$\,$s$^{-1}$ at 
$t = 700\,$ms after bounce, and in the 11.2$\,M_\odot$ model it is
6$\,$km$\,$s$^{-1}$ at 300$\,$ms after bounce; the final neutron 
star kick will be determined only seconds after the onset of the
explosion (see Scheck et al.\ 2006). 

A significant drift of the 
neutron star away from the origin of the grid is likely to lead to 
numerical problems, because the gravity solver as well as the
treatment of the neutrino transport are written for a centrally
condensed mass in a polar coordinate grid. Until the end of our 
simulations, however, neither this problem nor violation of
momentum conservation are a serious concern.

Another conceivable restriction of our simulations might be
connected to our use of a relativistic monopole potential,
whose dominant contribution to the gravitational potential
of the neutron star prohibits an accurate
response of the gravitational field to the nonradial
asymmetry of the mass distribution caused by the SASI 
oscillations, convective mass motions, and gravity waves. 
One should note here,
however, that we still have included the higher multipoles
of the gravitational potential in their Newtonian form,
and thus our treatment of the gravitational effects is certainly
much less constraining and much less approximative than the
so-called Cowling approximation. The latter is widely used for
discussing stellar pulsations in the linear regime, 
although it ignores the influence of the oscillation-induced 
density variations on the gravitational field.

\subsection{Core g-mode tests}
\label{sec:gmodetests}

In order to test whether our code can handle $l=1$ core
g-modes with large amplitudes in the neutron star if they 
were excited (e.g.\ hydrodynamically by pressure variations
around the neutron star surface due to anisotropic accretion flows), 
we artificially instigated a large dipole g-mode
by imposing an $l=1,\,n=1$ perturbation (i.e., we assumed one 
radial node) of the $z$-component of the velocity
field at a time late after bounce (alternatively, we also 
tested an $l=1,\,n=2$ mode with two radial nodes, obtaining
similar results). 
The velocity field was imposed with a constant absolute value
on the matter within a radius of 50$\,$km. It  
was chosen such that the linear momentum associated with this
perturbation was zero (the node was therefore initially located
at 15$\,$km).
The once perturbed model was followed in its evolution with all
(micro)physics being the same as in the long-time supernova runs.
For exploring the response to g-modes with different energies,
we varied the amplitude of the imposed velocity perturbation
by a factor of four, using values of 500$\,$km$\,$s$^{-1}$
and 2000$\,$km$\,$s$^{-1}$, corresponding to kinetic energies
of $3.7\times 10^{48}\,$erg and $5.9\times 10^{49}\,$erg,
respectively.

\begin{figure*}
\plottwo{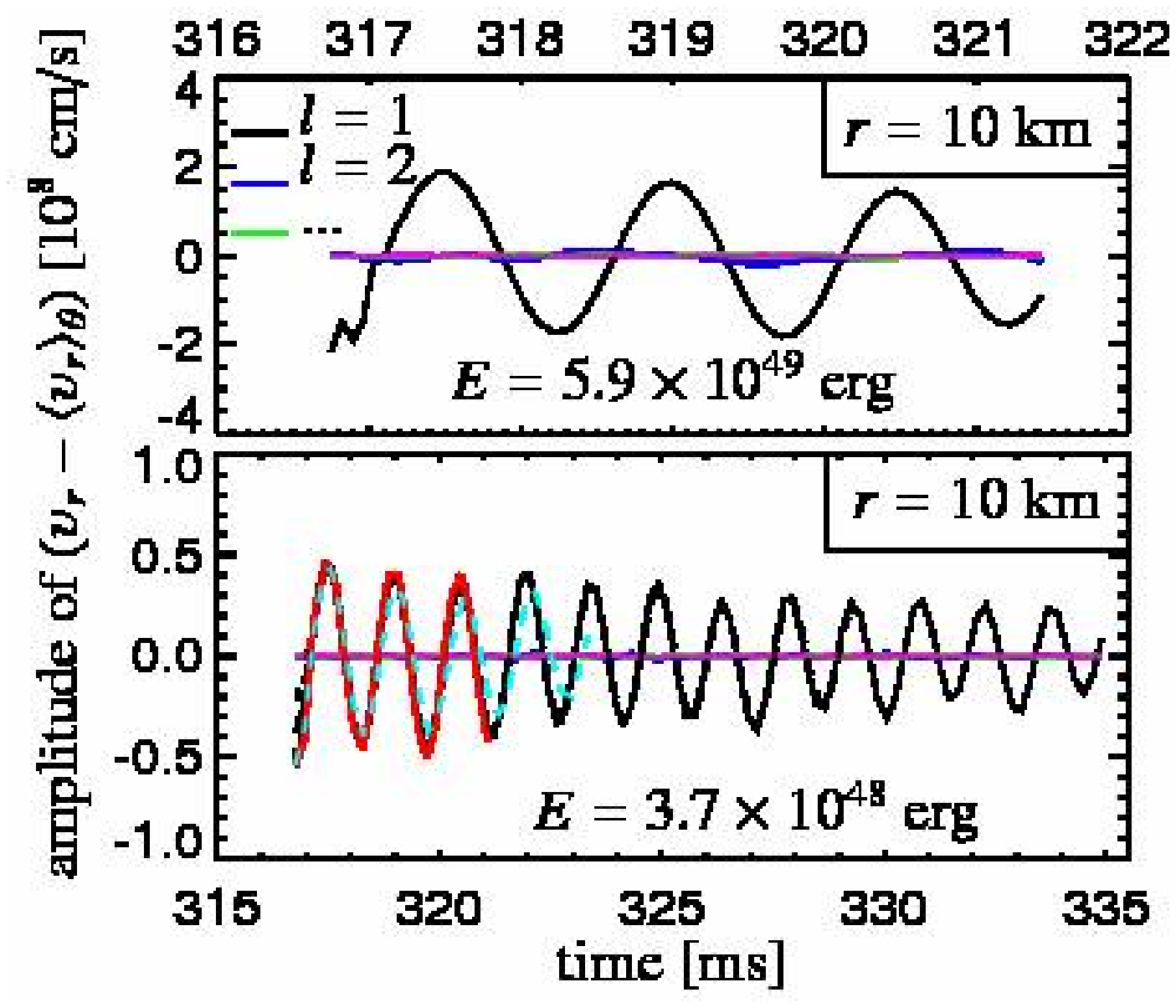}{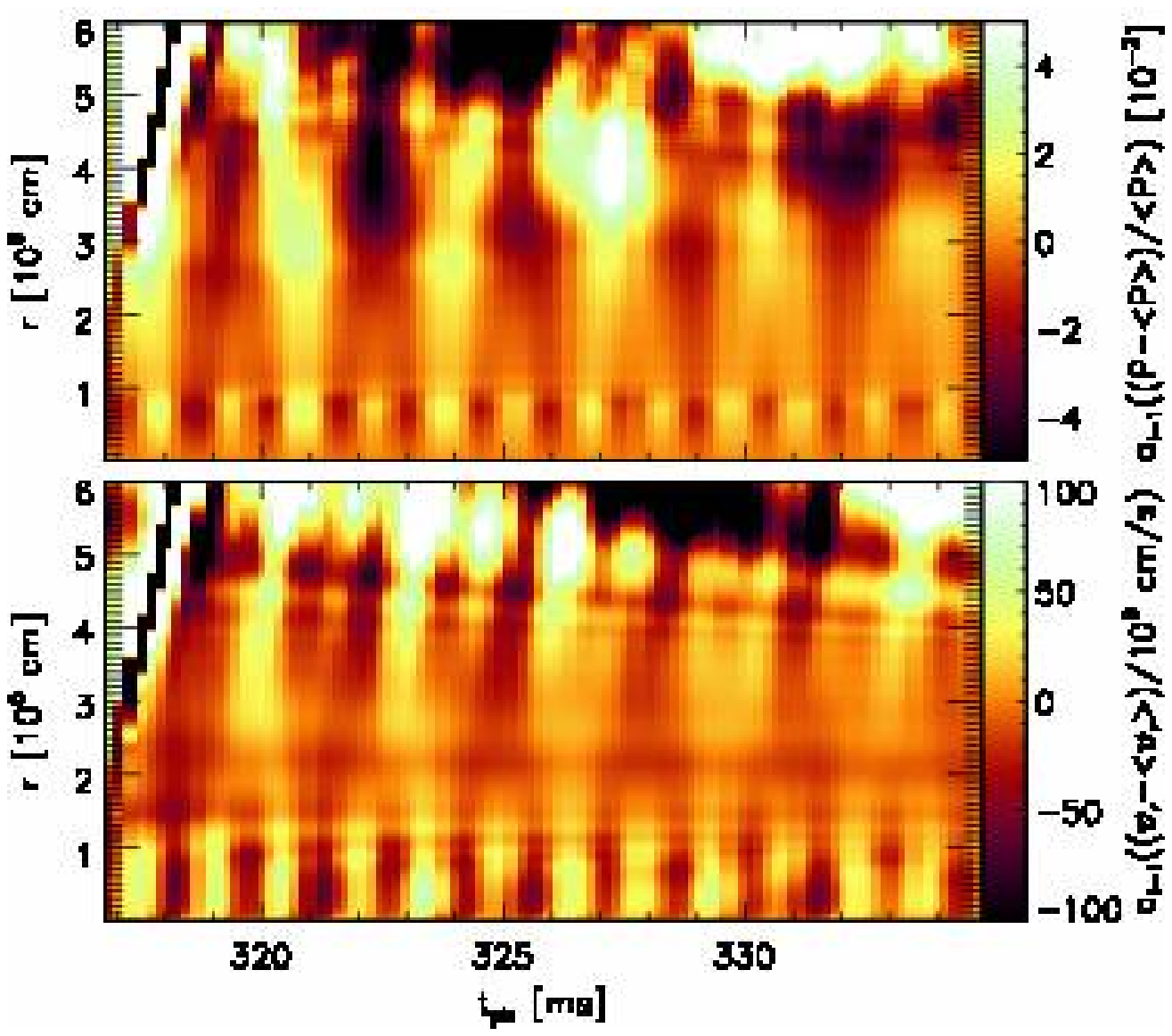}
 \caption{
{\em Left:} Test simulations with artificially excited core g-mode
oscillations of dipole ($l=1$) character in the neutron star (here
in the case of Model~M15HW-2D). Two
different amplitudes were considered for an initially imposed 
velocity field, $5\times 10^7\,$cm/s and $2\times 10^8\,$cm/s,
corresponding to a factor of 16 different kinetic energies (as
indicated in the plot). The time evolution of different spherical 
harmonics components of $(v_r(r,\theta) - 
\left\langle v_r(r,\theta) \right\rangle_\theta)$
is shown at a radius of 10$\,$km. The instigated oscillation
remains basically of dipole character. 
The clear presence of many cycles of the oscillation
demonstrates the ability of our numerical code to follow such
gravity waves, if they are instigated. In the lower
panel three overlapping curves are displayed. The dashed (hardly 
visible) black line gives the result of a 2D simulation with our
standard central
1D-core of $\la$1.7$\,$km radius, the black solid line is the result
with a 1D-core of $\sim$0.8$\,$km radius, and the red bold line
shows a simulation in which the whole star down to the center was
computed in 2D. Neither the frequency
nor the amplitude nor the damping behavior are affected by the 
spherically symmetric treatment of a central region when this 1D
core is as small as chosen. Slightly faster damping and a slowly
evolving frequency shift are observed when the 1D-core is increased
to 3.0$\,$km radius (dashed cyan line).
{\em Right:} The amplitude of the $l=1$ mode shown in the lower 
left panel (with a 1D-core radius of 0.8$\,$km) is displayed as a 
function of time and radius. The upper panel displays the fractional
pressure variations (in percent), $(P(r,\theta) - \left\langle P(r,\theta) 
\right\rangle_\theta)/\left\langle P\right\rangle_\theta$,
the lower panel the velocity variations normalized by $10^6\,$cm,
$(v_r(r,\theta) - \left\langle v_r(r,\theta) \right\rangle_\theta)/
10^6\,\mathrm{cm}$. Note that in both panels the range of values
on the color bar is limited (cutting off the true maxima and
minima of the amplitudes) for visualizing activity in all regions.
Interior
to about 10$\,$km the core oscillates with twice the frequency of the
mantle outside of $r\approx 25\,$km. In the intermediate, convective
layer the gravity waves are damped and a frequency change happens. 
At radii $r \ga 50\,$km one can see the presence of gravity waves 
produced by the violent and chaotic influence of convective
overturn and SASI activity in the layers beween the neutron star 
and the shock.
\label{fig:gmodetests}}
\end{figure*}

Figure~\ref{fig:gmodetests} displays results of two
such tests in the case of Model~M15HW-2D. It shows the time 
evolution of several low-$l$ mode amplitudes of the 
spherical harmonics decomposition    
of the radial velocity field, $(v_r(r,\theta) -
\left\langle v_r(r,\theta) \right\rangle_\theta)$, 
at a chosen radius of 10$\,$km (left panel) and for the 
$l=1$ case in the central
region with 60$\,$km radius (right panel; for comparison we 
provide there also the dipole amplitude of the fractional
pressure variations, $(P(r,\theta) - \left\langle P(r,\theta)
\right\rangle_\theta)/\left\langle P\right\rangle_\theta$). 
One can see that after
an initial, short relaxation (caused by the fact that our chosen
velocity perturbation did not correspond to an eigenfunction) an 
essentially pure dipole mode is present with a clean periodicity.
The amplitudes of the radial velocity variations 
are a factor of 10--20 larger than those that
we find for the core g-modes in our supernova runs 
(see Figs.~\ref{fig:gmodes} and \ref{fig:gmodes2}). 

The test model with the lower g-mode
energy (which is computationally cheaper) 
was run for many cycles and shows a decrease of the
oscillation amplitude at a fixed radius of 10$\,$km
by approximately a factor of two on a timescale
of about 18$\,$ms, but this decay slows down lateron. 
In principle, there can be different reasons for this 
damping. The central 1D region, though small, might brake
the motion because there might be friction and dissipative
losses caused in the 2D flow around the central 1D core. This
possibility, however, seems to be ruled out for the core sizes
chosen and the numerical resolution employed, because test 
runs with a core radius of 0.8$\,$km instead
of 1.7$\,$km and with a full 2D treatment (i.e., no 1D core
and thus painfully tiny computational timesteps)
show the same frequency, amplitude, and damping behavior
(Fig.~\ref{fig:gmodetests}, left lower panel). Only when the
radius of the central 1D core is increased to $\sim$3$\,$km
we can observe a slightly faster damping and a gradually 
decreasing oscillation frequency. 

Another
possible reason for the initial damping is the reconfiguration
of the mode pattern due to the fact that the imposed perturbation
does not correspond to an eigenfunction. This means that 
oscillation energy will be redistributed within several pulsation
periods and therefore amplitude damping is observed
in some regions while the amplitudes in other regions
increase. We find damping in the whole volume of the 
neutron star up to a radius of $\sim$30$\,$km, so that the
motion outside should be amplified. The right panels of
Figure~\ref{fig:gmodetests} indeed reveal such a trend. In contrast,
hardly any energy is transferred from the $l=1$ mode to higher modes
(in spite of unavoidable numerical coupling), whose amplitudes
remain very small in the whole perturbed volume of the
neutron star (Fig.~\ref{fig:gmodetests}, left panel).

Yet another reason for the observed damping, probably  
enhanced by the just mentioned growth of the g-mode
amplitude near the neutron star surface, is the acoustic 
energy flux that is sent by the moving neutron star
into its surrounding medium. Assuming that this
ultimately is the most efficient channel for energy loss
of the ringing neutron star, we can obtain an upper limit 
for the corresponding rate of energy outflow.
Since roughly 75\% of the initial mode
energy are lost within 18$\,$ms, we estimate an acoustic 
energy flux of $\sim$1.5$\times 10^{50}\,$erg$\,$s$^{-1}$, 
comparable to the value stated by Burrows et al.\ (2007c). 
For the g-mode activity and amplitudes observed in our 
supernova runs the energy flux from the 
oscillating neutron star must be expected to be smaller
and thus can only make a small addition to the neutrino energy
deposition behind the shock, in agreement with our conclusions
based on the estimates in Sect.~\ref{sec:acousticenergy}.

The test calculation with the larger excitation amplitude was
computationally expensive because of bigger
oscillation-induced variations of the thermodynamic quantities
that led to shifts of the $\beta$-equilibrium between neutrinos
and the stellar gas. The resulting high neutrino source terms
enforced small timesteps in the neutrino transport.
The test run was therefore
continued only for about three full oscillation cycles, but
the evolution visible in the upper left panel of 
Fig.~\ref{fig:gmodetests} agrees nicely with the run for the
reference case.

These tests give us confidence that our long-time simulations
should be able to trace also large-amplitude core g-modes, 
if an efficient mechanism was driving their excitation. 
We have no reason to suspect that our numerical code might
fail to describe the hydrodynamic driving
due to turbulence and anisotropic accretion that
Burrows et al.\ (2006, 2007a,c)
consider as responsible for instigating the large core g-modes 
in their models (see also Yoshida et al.\ 2007). 
Therefore we are tempted to conclude that 
until the end of our simulations the conditions for such an
efficient excitation of gravity waves in the neutron star 
core do not seem to be present in the supernova center.

\section{Discussion and conclusions}
\label{sec:summary}

We have presented evidence from our 2D neutrino-hydrodynamic 
stellar core-collapse simulations that the neutrino-driven 
mechanism may explain the explosions of progenitor stars in
a wider range of masses. In this work we considered a 
15$\,M_\odot$ progenitor (s15s7b2 of Woosley \& Weaver 1995)
and an 11.2$\,M_\odot$ model (Woosley et al.\ 2002). The former
is a typical representative of supernova progenitors in the
intermediate mass range between somewhat more than 10$\,M_\odot$
and around 20$\,M_\odot$, the latter is more typical of stars near
the lower mass end of supernova progenitors with iron cores.
A neutrino-driven explosion of the 11.2$\,M_\odot$ star was found
before in a 2D simulation by Buras et al.\ (2006b). Results of
a continuation of this run were also discussed here.
Similar to the explosion obtained for
the 11.2$\,M_\odot$ star, the delayed shock revival by neutrino
heating in the 15$\,M_\odot$ model was fostered and enabled
by the presence of a strong, low-mode SASI oscillations of
the postshock layer\footnote{Although our successful explosion 
for the 15$\,M_\odot$ star was obtained in a simulation in 
which we had imposed angular momentum on the progenitor model, our
comparion with the other simulations for this progenitor suggests
that the presence of rotation was not essential for the development
of the explosion (see Sects.~\ref{sec:rotmod9} and 
\ref{sec:nonrot}).}. The SASI modes grow rapidly in the
accretion flow to the neutron star even at conditions where
buoyancy instabilities are damped in the rapidly infalling gas
behind the accretion shock and therefore convection in the gain
layer is initially weak (see Foglizzo et al.\ 2006, Scheck et
al.\ 2008).

The SASI has two crucial consequences, which are favorable 
for the possibility of neutrino-driven explosions. On the one
hand, the violent sloshing motions of the postshock layer
with fast expansion and contraction phases of the shock
create steep entropy gradients in the
postshock flow, which lead to powerful secondary convection and 
thus efficient overturn of the gas in the gain region (for a 
detailed discussion, see Scheck et al.\ 2008). On the
other hand, the SASI also pushes the shock to a larger average
radius. This reduces the mean accretion velocity in the gain 
layer (because the velocity in the infall region ahead of the
shock drops inversely with the square root of the shock 
radius) and thus increases the advection timescale of gas 
from the shock to the gain radius. The accreted matter is
therefore exposed to neutrino heating for a longer time,
a fact that is reflected by a significant increase of the
critical ratio of the advection timescale to the neutrino-heating
timescale. A strong SASI activity has the consequence that this
timescale ratio comes much closer to the value of unity,
signaling favorable conditions for a neutrino-driven explosion.
The combination of these effects indeed turned out to drive the 
15$\,M_\odot$ model (specifically our Model M15LS-rot) 
towards a runaway instability at about 
600$\,$ms after core bounce, while for the 11.2$\,M_\odot$ star
this happened roughly 200$\,$ms after bounce.

Unfortunately, due to the considerable demand of computer time 
with the high numerical resolution needed for converged results, 
we were so far not able to continue our simulations to the phase 
where the
blast is fully developed and the explosion energy can be
determined from our computations. However, several factors are
indicators that the 15$\,M_\odot$ Model M15LS-rot considered here 
just like the investigated 11.2$\,M_\odot$ model 
is in the process of undergoing the transition from collapse
to successful outburst. This is suggested not only by the
rapidly accelerating expansion of the shock
radius near the end of the simulations and by the fact that 
the critical timescale ratio exceeds unity permanently with
rapidly growing value.
It is also suggested by a continuous trend of increasing 
energy in the gain layer. Before Model M15LS-rot was stopped,
a significant and growing gas mass in the gain layer
had obtained a positive specific energy and was thus ready
to become gravitationally unbound and to promote further 
outward propagation of the shock against the gravitational pull
of the neutron star. The mean shock radius has already reached
about 600$\,$km. This is sufficiently large for the temperature 
in the postshock ejecta to be so low that considerable nucleon 
recombination to $\alpha$-particles has set in and for the 
gas swept up by the shock to be so cool that heavy nuclei
do not experience complete dissociation any longer.
The release of nuclear binding energy and the
reduction of dissociation energy allow for a higher pressure
behind the shock and support the shock expansion. Last but not
least, the shock has entered the progenitor layer
(at about 500$\,$km) where silicon is not yet burned to iron,
and its most extended parts reach the shell where
oxygen is still unburned (at $\sim$700$\,$km). 
According to Bruenn et al.\ (2006) and Mezzacappa et al.\ (2007),
shock-initiated nuclear burning and the reduced ram pressure
in these layers will assist the ongoing strong neutrino heating
in driving the explosion. For all these facts we do not see any 
reason to suspect that the outward acceleration of the shock 
near the end of our simulated Model M15LS-rot might break down 
again.

The energy of such SASI-supported, neutrino-driven explosions
is not determined at the moment when the runaway sets
in. Instead, accretion of gas towards the gain radius and the
neutron star can proceed at the same time as shock acceleration
sets in due to the expansion of rising, neutrino-heated, buoyant
gas. This characterizes the generically multi-dimensional nature
of the developing explosion. The ongoing accretion continuously
channels fresh material, which is swept up by the shock, to the 
region near the gain radius, where neutrino heating is strongest.
Roughly half of the gas penetrates to the cooling layer below
the gain radius and ultimately settles onto the neutron star,
the other half absorbs with high efficiency energy from the 
neutrinos radiated by the neutron star. This gas turns around 
and rises again, driven by buoyancy forces, and contributes to
the energy of the explosion. Our estimates show (and hydrodynamic
simulations by Scheck et al.\ 2006 have demonstated)
that simultaneous accretion and shock acceleration can go on
for many hundred milliseconds and therefore the final explosion
energy can only be determined by computations that cover a much
longer time evolution than we were able to follow in our current
simulations. Our analytic considerations, based on the 
situation at the beginning of the explosion, reveal that 
the 11.2$\,M_\odot$ model can be expected to develop an
explosion energy of several $10^{50}\,$erg, while due to the
higher mass accretion rate the canonical
value of $10^{51}\,$erg is well 
in reach of the considered 15$\,M_\odot$ star. The final 
explosion energy should increase sensitively when the initial 
expansion of the shock is relatively slow and when the progenitor
maintains an appreciable mass accretion rate for a longer time.

Because the onset of the discussed SASI-supported 
neutrino-driven explosions for 11.2 and 15$\,M_\odot$ stars
is considerably delayed, the neutrino-heated gas mass in the 
gain layer at the time when the blast takes off is fairly low
and therefore a powerful gas ejection has to rely on a 
longer lasting phase of simultaneous accretion and outflow.
The possibility of such stable accretion with downflows from
one (or more) direction(s) and gas outflow
in the other direction(s) was indeed seen in the simulations of 
Scheck et al.\ (2006). 
The significance of this generically multi-dimensional phenomenon
in powering the explosion to completion and in achieving the 
necessary energies contradicts the concept of supernovae being
energized by a spherically symmetric neutrino-driven wind as
suggested by Burrows \& Goshy (1993). This has been realized 
recently also by Burrows et al.\ and was advocated by them
in the context of the acoustic explosion mechanism (Burrows
et al.\ 2007c), which requires the accretion of 
gravitational/mechanical energy, and in the context of 
magnetohydrodynamical explosions (Burrows et al.\ 2007a), which
require the accretion of differential kinetic energy.

The strong shock acceleration, which we found to begin about
200$\,$ms after bounce in the 11.2$\,M_\odot$ progenitor and at
600$\,$ms after bounce in the investigated 15$\,M_\odot$ model,
is associated with a large dipolar deformation of the shock.
This is a clear indication for the dominance of the lowest
nonradial SASI mode at the onset of the runaway. It suggests that
the developing explosion is likely to be very asymmetric,
similar to the artificially initiated neutrino-driven explosions
that were computed by Scheck et al.\ (2006) and led to 
neutron star kick velocities (with maximum values of more than
1000$\,$km$\,$s$^{-1}$) in agreement with measurements. Also
the large-scale anisotropies and mixing processes that can be
found in many supernovae and that seem to determine the
appearance of their gaseous remnants might be a consequence
of the pronounced asymmetry of the developing blast waves in our
simulations (see Kifonidis et al.\ (2006) for
studies of the long-time evolution of anisotropic neutrino-driven
explosions).

We compared 2D models with and without rotation and 
with a stiff and a soft nuclear equation of state
Moreover, we tested the influence of
different prescriptions of the effective relativistic
gravitational potential in our approximation to fully relativistic
calculations. We found that a nascent neutron star that contracts
faster and is more compact, which typically happens in the case of
a softer high-density equation of state
(or a stronger gravitational potential), has a 
favorable influence on the SASI-supported neutrino-driven
mechanism. This is a consequence on the one hand of a more rapid 
growth of low SASI modes, amplified by the release of 
gravitational binding energy, and on the other hand of higher 
luminosities and mean energies of the emitted neutrinos 
during the accretion phase and correspondingly stronger neutrino 
heating in the gain layer. The latter also supports the development 
of violent SASI and convective activity in the postshock region.
We stress that the EoS properties that lead to a faster
compactification of the nascent neutron star during its post-bounce
accretion do not need to be linked to the EoS characteristics at
core bounce or to the value of the incompressibility modulus
of symmetric nuclear matter at saturation density. The evolutionary
changes in the interior of the remnant could also trigger a 
softening of the EoS by a reduction of the adiabatic index, e.g.,
as a result of a phase transition.

Rotation, in contrast, can be diagnosed to have the opposite
effect. Although centrifugal support leads to a larger
average shock radius, a longer advection timescale of the matter
falling through the gain layer, and a reduced binding energy of 
accreted matter, these helpful effects are more than compensated
by lower luminosities and mean energies of the neutrinos radiated
by the significantly more extended and cooler rotating 
proto-neutron star. The overall neutrino heating and heating
efficiency found in rotating models is considerably
lower and the neutrino heating timescale in the gain layer
correspondingly longer. The critical ratio of advection timescale
to heating timescale therefore signals less favorable conditions
for an explosion in comparison to a nonrotating model with 
otherwise the same physics. This pessimistic judgement of the
role of rotation is opposed by the assessment of Yamasaki \& 
Yamada (2005), who found by linear analysis that sufficiently
rapid rotation leads to a reduced critical neutrino luminosity
for shock revival in the polar direction. Simulations for longer
postbounce times are needed to obtain conclusive information 
about this possibility.

Although we clearly see the presence of gravity-waves
in the newly formed neutron star, we cannot detect any important 
influence on the dynamical evolution of the
beginning supernova explosion. On the one hand, the amplitudes 
of core g-mode oscillations in our supernova runs remain very
small. On the other hand, the estimated flux
of acoustic power associated mainly with the considerable surface
gravity-wave activity
is dwarfed by the rate of the neutrino energy deposition.
The small amplitudes of core g-modes found in our models are
not in disagreement with the calculations of Burrows
et al.\ (2006, 2007b,c), who discovered violent neutron 
star core motions only at very late times ($t \ga 1\,$s) 
after core bounce, which we are unable to reach in our 
simulations.

We convinced ourselves by numerical tests with artificially
instigated large-amplitude core g-mode vibrations that our
supernova code is well able to capture this effect even 
in the case of sizable core displacements (of order one
kilometer and more).
Moreover, we do not see any reason to suspect that our numerics
could be unable to track the g-mode excitation by pressure
fluctuations due to anisotropic accretion and turbulence in
the SASI layer, in particular since we make sure to have good
numerical resolution inside the neutron star as well as 
in the region between the
neutron star and the shock. Both facts together make us
confident that we should observe large-amplitude core pulsations
if the physical conditions were present to drive their excitation.

The 11.2$\,M_\odot$ and 
15$\,M_\odot$ simulations discussed in the present work suggest that
--- at least in this mass range of progenitor stars ---
the SASI-aided neutrino-driven mechanism can lead to explosions 
significantly earlier than the acoustic mechanism proposed by
Burrows et al.\ (2006, 2007b,c), a possibility that is not challenged 
by the latter authors. 
Naturally, our simulations, which had to be terminated at latest 
$\sim$700$\,$ms after bounce, do not allow us to make any statement
about the neutron star g-mode activity
at even later times (with the potential to pump additional energy 
into the already launched blast), nor can we exclude a potentially
important role of the vibrating neutron star as an acoustic energy
source for triggering the explosion in cases where the 
neutrino-driven mechanism is too weak and the explosion is as
delayed as observed by Burrows et al.\ (2006, 2007b,c).

Neutrino-driven explosions have recently been also found
in 2D simulations with sophisticated, energy-dependent
neutrino transport by
Bruenn et al.\ (2006) and Mezzacappa et al.\ (2007). 
Their calculations for 11$\,M_\odot$ and 15$\,M_\odot$ stars,
which in contrast to ours were Newtonian and were performed
with a flux-limited neutrino diffusion scheme,
revealed the initiation of an explosion at the time 
when the inner edge
of the oxygen layer accretes through the shock. The authors
reported that explosions were only obtained when they used
an alpha network
of nuclei for following composition changes by
nuclear burning in the collapsing stellar layers, but not when 
they applied a ``flashing'' treatment with instantaneous conversion
of the nuclear abundances from non-NSE to NSE. 
They diagnosed that effective energy release by oxygen burning 
in the immediate vicinity of the shock, which in their case 
happens only with the alpha network,
assists the explosion, in particular in case of a weak 
shock. Their papers do not provide the information that would
allow us to conduct a detailed comparison of the composition 
evolution in their simulations and in ours.
However, we emphasise that although we do not use a nuclear 
reaction network, we still follow
composition changes in the non-NSE regime by
converting oxygen to silicon and silicon to iron when the
corresponding burning temperatures are reached due to 
compressional (or shock) heating in the infalling stellar 
layers. In our calculations for the 11.2$\,M_\odot$ and 
15$\,M_\odot$ progenitors a variety of important indicators
signal the onset of the blast and strong outward shock 
acceleration already long before the accretion
shock has reached the inner boundary of the
oxygen-rich layer. This, of course, does not mean that oxygen
burning in the collapsing layers has no bearing on the possibility
of an explosion, but it means that the inauguration of the blast
is not correlated with oxygen combustion happening at the shock.
Independent of whether a more accurate network treatment of nuclear 
reactions and of the transition to NSE instead of our
simplified description can make a difference for the
shock propagation or not, the simulations by Bruenn et al.\ (2006) 
and Mezzacappa et al.\ (2007) suggest that any such improvement
is likely not to disfavor explosions of the investigated stars.
In this sense the fundamental agreement of the outcome of these 
different simulations, in spite of various differences in 
numerical aspects and input physics, may be considered as
very encouraging.

Certainly, our current simulations can only be suggestive
for how neutrino heating and hydrodynamic instabilities in
collapsing stellar cores can collaborate to initiate the supernova
explosion. The simulations must be continued to later
times for determining the explosion properties, and definitely
more calculations are needed to obtain a clear understanding how
the shock revival depends on the core structure of the 
progenitor stars and on the physics that plays a role in the
supernova core, e.g., the neutrino opacities, equation of state, 
rotation, the depth of the gravitational potential, and also
stochastic elements like chaotic fluctuations of the SASI
strength in the nonlinear regime. Also the influence of the 
numerical resolution has to be more closely studied.
Our simulations cannot assess all the interesting questions 
involved. The results as presented here should only be seen as
an indication that the neutrino-driven mechanism is a viable
possibility for driving supernova explosions of progenitor stars
significantly more massive than 10$\,M_\odot$.

The positive trend towards a
SASI-supported, neutrino-driven runaway late after bounce
might be linked to a (relatively) soft equation of state for 
neutron star matter, like the EoS of Lattimer \& Swesty (1991)
used by us. Our set of calculations for the considered
15$\,M_\odot$ star, although including also a model with a 
significantly stiffer nuclear EoS, 
does not provide an answer to the question
how robust the success of the SASI-supported neutrino-heating 
mechanism is to variations of such an important physics ingredient.
Systematic long-time simulations for different
nuclear equations of state are needed to explore the differences
at times later than covered by our present calculations. 
Also the exact moment when the explosion sets in can well
be sensitive to various uncertain
aspects like the amount of rotation in the stellar core,
the possible influence of amplifying magnetic fields, unsettled
details of the subnuclear equation of state and of neutrino-matter
interactions, neutrino oscillations, the missing treatment
of lateral neutrino fluxes in our simulations,
and the approximative description of the effects of 
relativistic gravity. Even stochastical variations at some fairly
low energy level (${\cal{O}}(10^{49})\,$erg in the gain layer)
seem to be sufficient to foster an explosion at an earlier time.
Simulations in full relativity are certainly
desirable, in particular for stellar core collapse with rotation,
because in this case the optimal choice of the effective 
gravitational potential is not obvious (see Marek et al.\ 2006).
Moreover, supernova modeling in three spatial dimensions must
be the ultimate goal, because the growth rate and properties of  
the hydrodynamic instabilities in 3D must be expected to differ
from the axially symmetric 2D case, and new degrees of
freedom may play a non-negligible and potentially helpful 
role, for example spiral waves ($m\neq 0$ modes) and triaxial 
instabilities (see, e.g., Blondin \& Mezzacappa 2007, 
Ott et al.\ 2007, Yamasaki \& Foglizzo 2007, Iwakami et al.\ 2008).

In the light of our present results, supernova explosions 
appear to be an {\em accretion instability} rather than an
aspherical wind as described by Burrows et al.\ (2007c).
The SASI and convective instabilities, and in particular the
simultaneous presence of accretion and outflow of neutrino-heated
gas, which drives the shock expansion, are crucial ingredients of
the explosions described in this paper and constitute the 
explosions as a generically multi-dimensional phenomenon. 
The gas outflow during the main phase of shock revival and
the build-up of the explosion energy are fed by neutrino-heated,
accreted gas. Only much later, after the accretion
has ceased, does the gain radius retreat to the neutron star
surface and the cooling layer between neutrinosphere and gain 
radius shrinks to a very narrow region or disappears completely. 
This marks the 
onset of the neutrino-driven baryonic wind phase, in which the
dilute gas outflow from the nascent neutron star is determined
solely by the conditions at the surface of the hot, compact
remnant and not by the (accretion) properties of the dying star.

\acknowledgments
  We are very grateful to R.~Buras, K.~Kifonidis, B.~M\"uller, E.~M\"uller, 
  and M.~Rampp for their input to various aspects of the reported project,
  to R.~Johanni for his support in parallelizing our code,
  and to S.~Woosley for data of his progenitor model. We thank
  A.~Burrows, C.D.~Ott, and an anonymous referee for their numerous
  suggestions of how to improve our manuscript and to extend our
  investigations.
  This work was supported by the Deutsche Forschungsgemeinschaft
  through the Transregional Collaborative Research Centers SFB/TR~27
  ``Neutrinos and Beyond'' and SFB/TR~7 ``Gravitational Wave Astronomy'', 
  the Collaborative Research Center SFB-375 ``Astro-Particle Physics'',
  and the Cluster of Excellence EXC~153 
  ``Origin and Structure of the Universe''
  (\url{http://www.universe-cluster.de}). The computations were
  performed on the IBM p690 of
  the John von Neumann Institute for Computing (NIC) in J\"ulich,
  on the national supercomputer NEC SX-8 
  at the High Performance Computing Center Stuttgart (HLRS) under
  grant number SuperN/12758, on the IBM p690 of the Computer Center 
  Garching (RZG), on the sgi Altix 4700 of the Leibniz-Rechenzentrum (LRZ)
  in Munich, and on the sgi Altix 3700 of the MPI for Astrophysics.
  We also acknowledge support by AstroGrid-D, a project funded by
  the German Federal Ministry of Education and Research (BMBF) 
  as part of the D-Grid initiative.

\clearpage
%
%

\begin{deluxetable}{l|crrrr}
\tablecolumns{6}
%
\tablecaption{ Investigated 15$\,M_\odot$ models.
\label{tab:models}}
\tablehead{ \colhead{Model}  &
            \colhead{Dimension} &
            \colhead{$N_r$\tablenotemark{a}} &
            \colhead{$N_\theta$\tablenotemark{b}} &
            \colhead{$N_\epsilon$\tablenotemark{c}} &
            \colhead{$\Phi_{\mathrm{eff}}$\tablenotemark{d}} }
\startdata
M15LS-1D   & 1D              & 600--960 & 1    & 17 & new \\
M15HW-1D   & 1D              & 600--750 & 1    & 17 & new \\
M15LS-rot  & 2D \& rotation  & 400--850 & 128  & 17 & old \\
M15LS-2D   & 2D              & 600--960 & 192  & 17 & new \\
M15HW-2D   & 2D              & 600--700 & 192  & 17 & new \\
M15LS-rot9 & 2D \& rotation  & 400--500 & 128  &  9 & new \\[-5pt]
\enddata
\tablenotetext{a}{Number of radial grid points, increasing with time.}
\tablenotetext{b}{Number of lateral grid points from pole to pole.}
\tablenotetext{c}{Number of energy grid points for the neutrino transport.}
\tablenotetext{d}{Effective relativistic potential; ``old'' corresponds to
definition introduced by Rampp \& Janka (2002),
which is equivalent to Case~R of Marek et al.\ (2006);
``new'' corresponds to Case~A of Marek et al.\ (2006).}
\end{deluxetable}

\clearpage
%

%

%

%

%

%

%

%

%

%

%

%

%

%

%

\clearpage

\clearpage

\end{document}